\documentclass[sigconf]{acmart}

\usepackage[makeroom]{cancel}
\usepackage{pifont}
\usepackage{marvosym}

 \usepackage[justification=centering]{caption} 
 \usepackage[caption=false]{subfig}
\usepackage{color}
\usepackage{multirow}
 \usepackage{graphicx}
\usepackage[linesnumbered, vlined, ruled,algo2e]{algorithm2e} 
\usepackage{booktabs}
\usepackage{amssymb}
\usepackage{amsfonts}
\usepackage{amsmath}
\usepackage{mathtools}

\usepackage{amsthm}
\usepackage{soul}
\usepackage{etex}
 \usepackage{relsize}
\usepackage{epstopdf}
\usepackage{multirow}
\usepackage{url}
\usepackage{tablefootnote}
\usepackage{setspace}
 \usepackage{hyperref}  
 \hypersetup{ 
    colorlinks,%
    citecolor=black,%
    filecolor=black,%
    linkcolor=black,%
    urlcolor=black,%
}

\newcommand{\eat}[1]{}
\newcommand{\RED}[1]{  {\color{red}{#1}}} 
 
\newcommand{\BLUE}[1]{  {\color{blue}{#1}}}

\newcommand{\stitle}[1]{\vspace{0.2cm}\noindent\textbf{#1}} 
\newcommand{\sstitleM}[1]{\vspace{0.1cm}\noindent${{#1}}$}

\DeclareMathOperator*{\argmax}{arg\,max}

  \DeclareMathOperator*{\Mod}{mod}

\renewcommand{\qed} {\hfill$\blacksquare$}

\definecolor{dark-gray}{gray}{0.2}

\newcommand{\ses}{{SES}\xspace}
\newcommand{\bsc}{{ALG}\xspace}
\newcommand{\bnd}{{INC}\xspace}
\newcommand{\hor}{{HOR}\xspace}

\newcommand{\horb}{{\mbox{HOR-I}}\xspace}
\newcommand{\rand}{{RAND}\xspace}
\newcommand{\stat}{{TOP}\xspace}

\newcommand{\meetupc}{{\textsf{Meetup}}\xspace}
\newcommand{\music}{{\textsf{Concerts}}\xspace}
\newcommand{\zip}{{\textsf{Zip}}\xspace}
\newcommand{\uni}{\textsf{{Unf}}\xspace}
\newcommand{\norm}{{\textsf{Nrm}}\xspace}

\newcommand{\maxAss}{{\mathsf{getBetterAssgn}}}
\newcommand{\popAss}{{\mathsf{popTopAssgn}}}
\newcommand{\topAss}{{\mathsf{getTopAssgn}}}

\newcommand{\E}{\ensuremath\mathcal{E}}
\newcommand{\A}{\ensuremath\mathcal{A}}
\renewcommand{\S}{\ensuremath\mathcal{S}}

 \renewcommand{\L}{\ensuremath\mathcal{L}}
 \renewcommand{\C}{\ensuremath\mathcal{C}}
 \newcommand{\T}{\ensuremath\mathcal{T}}
 \renewcommand{\U}{\ensuremath\mathcal{U}}
 
 \newcommand{\M}{\ensuremath\mathcal{M}}
 \newcommand{\B}{\ensuremath\mathcal{B}}

 \newcommand{\W}{\ensuremath\mathcal{W}}

\newcounter{prob}
\newenvironment{myprob}[1][]
{\refstepcounter{prob}\par\setlength{\leftskip}{0pt}\setlength{\rightskip}{0pt}\medskip\noindent\ignorespaces
\textbf{#1}}
{\medskip\par}

\newcounter{theorcnt}
\newenvironment{mytheor}
{\refstepcounter{theorcnt}\medskip\setlength{\leftskip}{6pt}\setlength{\rightskip}{0pt}\par\noindent\ignorespaces 
   \textbf{Theorem~\thetheorcnt.}}
{\smallskip\par}

\newenvironment{myproofSk}
{\medskip\setlength{\leftskip}{0pt}\setlength{\rightskip}{0pt}\par\noindent\ignorespaces 
   \textsc{Proof Sketch.}}
{\medskip\par}

\newcounter{corollcnt}
\newenvironment{mycoroll}
{\refstepcounter{corollcnt}\medskip\setlength{\leftskip}{6pt}\setlength{\rightskip}{0pt}\par\noindent\ignorespaces 
   \textbf{Corollary~\thecorollcnt.}}
{\smallskip\par}

\newcounter{claimcnt}

\newcounter{remarkcnt}
\newenvironment{myremark}
{\refstepcounter{remarkcnt}\vspace{0.2cm}\setlength{\leftskip}{0pt}\setlength{\rightskip}{0pt}\par\noindent\ignorespaces 
   \textbf{Remark~\theremarkcnt.}}
{\smallskip\par}

\newcounter{proposcnt}
\newenvironment{myproposition}
{\refstepcounter{proposcnt}\smallskip\setlength{\leftskip}{6pt}\setlength{\rightskip}{0pt}\par\noindent\ignorespaces 
   \textbf{Proposition~\theproposcnt.}}
{\smallskip\par}

\newcounter{lemmcnt}
{\medskip\par}

\newcounter{ex}
\newenvironment{myExample}
{\refstepcounter{ex}\medskip\setlength{\leftskip}{0pt}\setlength{\rightskip}{0pt}\par\noindent\ignorespaces 
   \textbf{Example~\theex.}}
{\medskip\par}

\newcommand{\myFontP}[1]{{\fontfamily{ppl}\selectfont #1}} 

\newcommand{\DEF}[1]{\textit{\myFontP{\textbf{#1}}}}

\newcommand{\tline}  {\specialrule{0.8 pt}{0pt}{1pt}}		 
\newcommand{\bline}  {\specialrule{0.8 pt}{1pt}{0pt}}		
\newcommand{\dline}  {\specialrule{0.4 pt}{0pt}{2pt} \specialrule{0.4 pt}{0pt}{4pt}}		

\SetKw{Logand}{and}	
\SetKw{Logor}{or}
\SetKw{Lognot}{not}	

\SetKw{mybreak}{break}	
\SetKw{mycontinue}{continue}	

\SetKwInput{KwVar}{Variables}
\SetKwInput{KwPar}{Parameters}

\SetKwData{myEOF}{EOF}	
\SetKwData{myfalse}{false}
\SetKwData{mytrue}{true}
\SetKwData{mynull}{null}
\SetKwData{myvalid}{\texttt{valid}}
\SetKwData{myfeasible}{\texttt{feasible}}
\SetKwData{myupdated}{\texttt{updated}}
\SetKwData{mynotupdated}{\scalebox{0.9}[1]{ \texttt{prtl  \!\! \!\!\!updated}}}

\SetKwData{myempty}{empty}
\SetKwBlock{myblock}{beginnikos}{end} 
\SetAlgoInsideSkip{} 
\SetAlgoSkip{} 
\SetVlineSkip {0.5mm} 
\SetAlgoCaptionSeparator{.} 

\newcommand{\mycomment} [1] 	{\tiny{\textcolor{dark-gray}{\myFontP{{#1}}}}} 
\SetKwComment{Comment}{\mycomment{/\!\!/}}{}
 \DontPrintSemicolon
\SetAlgoCaptionLayout{small}
\SetProcNameSty{small}
\SetProcArgSty{small}

\graphicspath{{figures/}{plots/}}

\usepackage{times}%

\begin{document}

 \title{ \vspace*{-10pt}Attendance Maximization for Successful \linebreak Social Event Planning}

\titlenote{{\mbox{This paper appears in 22nd Intl.\ Conf.\ on Extending Database Technology (EDBT~2019)}}}



\author{Nikos Bikakis}
\affiliation{%
  \institution{Athens Univ.\ of  Econ.\ \& Bus.\ }
   \country{Greece}
} 
\author{Vana Kalogeraki}
\affiliation{%
  \institution{Athens Univ.\ of  Econ.\ \& Bus.\ }
   \country{Greece}
} 
\author{Dimitrios Gunopulos}
\affiliation{%
  \institution{University of  Athens}
   \country{Greece}
}

\newcommand{\alt}[2]{
#1
}


\begin{abstract}

Social event planning has received a great deal of attention in recent years where various entities, such as event planners and marketing companies, organizations,  venues, or users in Event-based Social Networks, organize numerous social events (e.g., festivals, conferences, promotion parties). 
Recent studies show that ``attendance'' is the most common metric used to capture the success of social events, since the number of attendees has great impact on the event's expected gains  (e.g., revenue, artist/brand publicity).
In this work, we study the   \textit{Social Event Scheduling} (\ses) 
problem which   aims at identifying and
assigning social events to appropriate time slots, so that the number of events attendees is maximized. 
%
We show that,  even in highly restricted instances, the \ses problem 
is NP-hard to be approximated  over a factor. 
%
To solve the \ses problem, we design three efficient and scalable  algorithms.
%
These algorithms exploit several novel schemes that we design.
We conduct extensive experiments using several real and synthetic datasets, 
and demonstrate that the proposed algorithms 
perform on average half the computations compared to 
the existing solution and, in several cases, are 3-5 times faster.
%

%
\end{abstract}

%

\maketitle

 \hypersetup{ 
    colorlinks,%
    citecolor=black,%
    filecolor=black,%
    linkcolor=black,%
    urlcolor=black,%
pdftitle={Attendance Maximization for Successful Social Event Planning},
pdfauthor={Nikos Bikakis, Vana Kalogeraki, Dimitrios Gunupulos},
pdfsubject={Social Event Planning Arrangement Organization  Profit, Attendance Maximization},
pdfproducer={pdfbikTeX-1.1},
pdfcreator={bikTeX},
pdfkeywords={Social Event Planning, Social Event Arrangement, Social Event Organization  SES problem, Social Event Mining, 
profit maximization, attendance maximization, user preferences, meetup, Event-based Social Networks, EBSN, event recommendation, social networks,
assignment problem, matching problem, event organizers, event planning, marketing, event scheduling, nphard, knapsack, np-hard,
approximation, optimization, large data, big data, data science, data mining, greedy, user habits, user behavior, spatiotemporal, 
given a set of events a set of time periods and a set of users
our objective is to determine how to assign events on the time periods
so that the maximum participant enrollment is achieved, 
the Social Event Scheduling SES problem,  
which schedules a set of social events considering 
user preferences and behavior, events spatiotemporal conflicts, and competing events, 
in order to maximize the overall number of participants.
}
}
\section{Introduction}
\label{sec:intro}

\begin{figure*}[ht]
\vspace{-20pt}
\centering
\scriptsize
\hspace{-10pt}
 \subfloat{
\setlength{\tabcolsep}{4pt}
\begin{tabular}[t]{cc}
{{\textbf{Event}}} & {{\textbf{Location}}} \\ \cmidrule{1-2}
 $e_1$ & Stage 1 \\
$e_2$  & Stage 1 \\
 $e_3$  & Room A \\ 
$e_4$  & Stage 2 \vspace{4pt} \\ 
\multicolumn{2}{c}{{\small{\textbf{(a) Candidate Events}}} }
\end{tabular}
}
\hspace{11pt}
%
\subfloat {
\begin{tabular}[t]{c}
$t_1 = \langle$Friday 8--11pm$\rangle$  \vspace{4pt}\\ 
$t_2  = \langle$Sat.\ 6--9pm$\rangle$ \vspace{8pt} \\
\small \textbf{(b) Time Intervals} 
\end{tabular}
}
\hspace{8pt}
\subfloat {
\begin{tabular}[t]{c}
$c_1$ $\langle$Friday 6--9pm$\rangle$, $t_{c_1}=t_1$ \vspace{4pt} \\
$c_2$ $\langle$Sat.\ 8--10pm$\rangle$, $t_{c_2}=t_2$ \vspace{8pt} \\ 
\small \textbf{(c) Competing Events}\\
\end{tabular}
}
\hspace{15pt}
 \subfloat{
\setlength{\tabcolsep}{4pt}
\begin{tabular}[t]{ccccccccc}
 & \multicolumn{4}{c}{{{\textbf{Event Interest}}}} &\multicolumn{2}{c}{\hspace{-2pt}{{\textbf{Comp.\ Ev.\ Interest}}}} &\multicolumn{2}{c}{{{\textbf{Activ.\ Prob.\ } }}} \\ \cmidrule(lr){2-5} \cmidrule(lr){6-7} \cmidrule(lr){8-9}
{{\textbf{User}}} & {$e_1$} & {$e_2$}& $e_3$ &$e_4$ &  \hspace{13pt}$c_1$ & \hspace{-9pt}$c_2$& \hspace{5pt}$t_1$& $t_2$\\ \cmidrule{1-9}
 $u_1$ & 0.9 &   0.3  & 0  & 0.6 & \hspace{13pt} 0.8  & \hspace{-9pt}  0.3 & \hspace{5pt}0.8 & 0.5 \\
$u_2$ & 0.2 &  0.6  &  0.1  & 0.6 & \hspace{13pt} 0.4  &\hspace{-9pt}  0.7 &\hspace{5pt} 0.5 &  0.7 \vspace{5pt}\\
\multicolumn{9}{c}{{\small{\textbf{(d) Users}}}} \\
\end{tabular}
}
\vspace{-5pt}
\caption{Running example \normalfont{(4 Candidate events, 2 Intervals, 2 Competing events, 2 Users)}}
\vspace*{-0pt}
\label{fig:exaple}
 \end{figure*}

%
%
%
%
%
%
 

 The event planning industry has grown enormously in the past decade, with
large \textit{event planning} and \textit{marketing} companies (e.g., MKG, GPJ),
organizing and managing a variety of social events  (e.g., \mbox{multi-themed} festivals,  promotion parties, conferences).
In addition to companies, social events are also organized by \textit{venues} (e.g., theaters, night clubs), \textit{organizations} (e.g., ACM, TED), as well as \textit{users} in \textit{Event-based Social Networks} (e.g., Meetup,  Eventbrite, Plancast). 

The \textit{Event Marketing Benchmark Report 2017},\footnote{\href{https://www.certain.com/blog/certain-presents-the-event-marketing-benchmark-report-spring-2017}{www.certain.com/blog/certain-presents-the-event-marketing-benchmark-report-spring-2017 \vspace{0pt}}}
where marketing 
\mbox{decision-makers}  from large organizations  participate, indicates that   
``\textit{attendance}'' \textit{is the most common metric used to measure the success of social 
events}, since the number of attendees has a great influence on the event's expected gains
(e.g., revenue,  artist publicity).
%
Therefore,  achieving maximum attendance is the \textit{organizers first challenge}, as also indicated in the    \textit{Event Marketing Trends 2018}  study.%
\footnote{\href{https://welcome.bizzabo.com/event-marketing-2018}{https://welcome.bizzabo.com/event-marketing-2018}\vspace*{-20pt}}



Examples of events organization include large festivals
and conferences where a large number of (multi-themed) events are organized
over several stages and sessions attracting several thousands
of people.
For example, \textit{Summerfest Festival} 
has performances from over 800 bands, attracting more than 800K people each year. Beyond music concerts, numerous 
multi-themed events take place,  ranging from art-makings and theatrical performances to fitness activities and parties. 
%
In such scenarios, successful event planning is extremely challenging, since various factors need to be taken into account, such as the 
\textit{large number of events} and available \textit{time slots}, 
the \textit{diversity of events' themes} and \textit{user interests}, 
the presence of \textit{overlapped events}, 
the \textit{available resources} (e.g., available stages), etc.

 Assume the following scenario. On Monday two events are  scheduled to take place during a festival: (1)  a \textit{rock concept} from 19:00 to 22:00; and 
(2) a \textit{fashion show} between 19:00 and 21:00. 
Additionally, from 18:00 to 20:00 a music concert of a \textit{rock singer} is taking place in a nearby (competing) venue. 
Consider that Alice enjoys listening to \textit{rock music}, and is a \textit{fashion lover}.
Although Alice is interested in all three events, she is only able to attend one of them.

In this work, we study the \textit{Social Event Scheduling} (\ses) \linebreak problem \cite{ses}.
\textit{Given a set of {candidate events}, a set of {time intervals} and
a set of {users}, \ses   assigns events to time intervals,  
in order to {maximize the overall number of participants}.}
The assignments are determined by considering  several events' and users' factors, such as
{user preferences} and {habits}, events' {spatiotemporal conflicts}, etc.

%
%
%
%



Recently, several studies have been published
examining the problem of assigning \textit{users} (i.e., participants)  to a set of \textit{pre-scheduled events} 
in Event-based Social Networks  \cite{Li2014,She2015a,She2016,She2015,Tong2015,She2017,ChengYCGW17}.  
The objective in these works is to find the \textit{\mbox{user-event} assignments that maximizes the satisfaction of the users}.
Here, in contrast to existing works, we study  a substantially different problem. 
Briefly, instead of assigning users to events, we assign \textit{events} to \textit{time intervals}.
The objective here is to find the \textit{\mbox{event-time} assignments that maximize the number of event's attendees}.
More or less, the \ses problem studies the ``satisfaction'' (e.g., revenue, publicity) of the entities 
involved in event organization  (e.g., organizer, artist, sponsors, services' providers).
In other words, \ses is an ``\textit{organizer-oriented}'' problem, while the existing works 
are ``\textit{participant-oriented}''.
Overall, the objective, the solution and the setting of the \ses problem are substantially
different from the related works.

\alt{}{Furthermore, the problem studied here has common characteristics with the {Generalized assignment} (GAP) and {Multiple knapsack} (MKP) problems \cite{Martello1990}. 
However, a major difference of \ses compared to GAP (and MKP) is that in \ses the 
the utility (i.e., expected attendance)
of assigning an event (resp.\ item) to an interval (resp.\ bin) is defined w.r.t.\ the other events assigned to this interval (see Section~\ref{sec:anal} for more details).}

The \ses problem was recently introduced in \cite{ses}  where a greedy algorithm was proposed. 
In the proposed solution, in each assignment selection, the algorithm recomputes (i.e., updates) 
the scores for a large number of assignments. 
Additionally, in each selection the algorithm has to examine (e.g., check for validity) all the assignments. 
The aforementioned result to poor performance of this solution.
In this work, we design three efficient and scalable algorithms which are implemented on top of the 
following novel schemes. 
First, we propose an \textit{incremental updating} scheme in which a reduced number of score 
computations are performed in an incremental manner. 
Further, we design an \textit{assignment organization} scheme which significantly reduces the number of 
assignments that are examined. 
Finally, an \textit{assignment selection policy} is proposed, minimizing the 
impact of performing a part of the required score computations, on the quality of the results.
In our extensive experiments, we illustrate that the proposed algorithms 
perform about half the computations and, in 
several cases, are 
3-5$\times$ faster compared to the method proposed in~\cite{ses}.
 

Further, examining the theoretical aspects of the \ses problem,
we study its approximation, 
showing that even in highly restricted instances, it is NP-hard to be approximated over a factor 
larger than $(1- \epsilon)$.



\stitle{Contributions.} 
The  main contributions of this work are summarized as follows:
%
%
$(1)$~We show that the \ses problem is {\mbox{NP-hard} to be approximated over a factor larger than} $(1- \epsilon)$. 
$(2)$~We design {three efficient and scalable approximation algorithms}.
%
These algorithms outperform  the existing algorithm by exploiting a series of  schemes  that we develop. 
\eat{Particularly:
$(i)$~we propose an \textit{incremental updating scheme} in which a
reduced number of computations is performed in an incremental  manner;
$(ii)$~we introduce an \textit{assignment selection policy}
which minimizes the impact of performing a reduced number of computations, 
on the quality of the results; and
$(iii)$~we devise an \textit{assignment organization} that {takes into account} 
the incremental updating scheme, to reduce the problem search space.
 }
%
 $(3)$~We conduct an detailed experimental analysis using several real and synthetic datasets.

\alt{}{
\begin{table}[]
\centering
\caption{Common  notation}
\label{tab:notation}
\scriptsize
\begin{tabular}{cl}
\tline
\textbf{Symbol} & \textbf{Description}\\ \dline
$\theta$ & Number of available resources \\ \rowcolor{gray!10}	
$\T, t$  & Set of time intervals,  a time interval\\
$\E, e$  & Set of candidate events, a candidate event\\ \rowcolor{gray!10}	
$\ell_e$ & Location of  $e$\\
$\xi_e$ & Number of resources required for $e$\\ \rowcolor{gray!10}	
$\C, c$ & Set of competing events, a competing event \\
$t_c$ & Time interval associated with competing event $c$\\ \rowcolor{gray!10}	
$C_t$ & Set of  competing events associated with  $t$\\ 
$\S$  & Event scheduling \\\rowcolor{gray!10}	
$\alpha_e^t$ & Assignment of $e$ at $t$\\
$t_e(\S)$ &   The time interval that $\S$ assigns to $e$\\ \rowcolor{gray!10}	 
$E(\S)$ & Set of events that $\S$ assigns time intervals \\ 
$E_t(\S)$ & Set of events that $\S$ assigns to $t$ \\ \rowcolor{gray!10}	
$T(\S)$ &  Set of time intervals that $\S$  have assigned events \\ 
$\U, u$ & Set of users, a user \\ \rowcolor{gray!10}	
$\mu_{u, h}$ & Interest of   $u$ over (candidate or competing) event $h$\\
$\sigma_u^t$  & Probability of $u$ participating in a social activity at $t$\\  \rowcolor{gray!10}	
$\rho_{u, e}^t$  &  Probability of $u$   attending $e$   at $t$ \\
$\omega_{e}^t$  & Expected attendance of $e$ at $t$ \\  \rowcolor{gray!10}	
 
 $g_{e}^t$  & Score (i.e., utility gain) of assignment  $\alpha_e^t$ \\ 

$\Omega({\S})$ & Total utility  for $\S$ \\  \rowcolor{gray!10}	
$\Phi$ & Score bound  \\ 
 \bline
\end{tabular}
 \end{table}
}


 \section{Social Event Scheduling Problem}
\label{sec:prob}

%
In this section we first define the \textit{Social Event Scheduling} 
(\ses) problem; and then we study its approximation.
In what follows, we present a simple example that
introduces the main entities involved in \ses problem.

  \begin{myExample} [\myFontP{\textit{Running Example}}]
\label{ex:intro}
Figure~\ref{fig:exaple} outlines our running example
  involving four \textit{{candidate events}}  ($e_1$--$e_4$), two \textit{{time intervals}} ($t_1$,~$t_2$), 
two \textit{{competing events}} ($c_1$, $c_2$), and \textit{{two users}} 
\mbox{($u_1$, $u_2$).}

%
 
%
%
%

The \textit{location} of each \textit{candidate  event} is presented in Figure~\ref{fig:exaple}a 
We  notice that both $e_1$ and $e_2$ are going to be hosted at  {Stage~1}. 
Hence, these events cannot be scheduled to take place during the same time period. 
 Figure~\ref{fig:exaple}c presents the \textit{competing events} along with the
time periods during which these are scheduled to take place. 
For example, $c_1$ is schedule to take place on Friday between 6:00 and 9:00pm \mbox{(at a nearby competing venue)}. 
Further, in Figure~\ref{fig:exaple}b we observe that there is the candidate \textit{time
interval} $t_1$ defined on the same day between 8:00 and 11:00pm.
Thus, due to overlapping time periods, a user cannot attend both
$c_1$ and a candidate event that will be possibly scheduled to take
place during $t_1$.

Finally, 
 Figure~\ref{fig:exaple}d shows, for each user, the \textit{interest values} \linebreak
 \mbox{(i.e., affinity)} for the  events, as well as the 
\textit{{social activity probability}}  (e.g.,~based on user habits) during the time periods defined by the two intervals.
For example, $u_1$ has high social activity probability (equals to $0.8$) at $t_1$, 
since Friday night $u_1$ does not work and usually goes out and participates in social activities.  
 
 %

\end{myExample}
\vspace{-2pt}

%
%

\subsection{Problem Definition}
\label{sec:def}
\alt{}{Table~\ref{tab:notation} summarizes the most common notation and their
description.}

In our problem, we assume an \DEF{organizer}  (e.g., company, venue) managing the events' organization.  
Each organizer  possesses a number of  \DEF{available resources} $\theta \in \mathbb{R}^+$. 
These are abstractions used to refer to staff,  materials, budget or any other means related to event organization.

Further, let $\T$ be a set of \DEF{candidate time intervals}, 
representing time periods that are available for organizing events.

\vspace{2pt}
Assume  a set  $\E$  of available events to be scheduled, referred as \DEF{candidate  events}.
Each $e \in \E$ is associated with a \DEF{location} $\ell_e$ 
representing the place (e.g., a stage) that is going to host the event.  
Further, each event $e$ requires a specific amount of resources $\xi_e \in \mathbb{R}^+_0$  for its  organization, referred as \DEF{required resources}. 

%



\vspace{4pt}
An \DEF{assignment} $\alpha_e^t$ denotes that the candidate event $e \in \E$ is scheduled to take place at $t \in \T$.
An event \DEF{schedule} ${\S}$ is a set of assignments, where there exist no two assignments referring to the same event. 
Given a schedule $\S$,  we denote as $\E(\S)$ the set of all candidate events that  are scheduled by ${\S}$, i.e., $\E(\S)=\{ e \mid a_{e}^t \in \S\}$; 
 and $\E_t(\S)$ the candidate events that  are scheduled  by ${\S}$  to take place at $t$ 
 \linebreak
 (i.e.,~assigned to $t$).
Further, for a candidate event $e \in \E(\S)$, we denote as
 $t_e(\S)$ the time interval on which $\S$ assigns $e$.


\vspace{4pt}
A \textit{schedule} $\S$ is said to be \DEF{feasible} 
if the following constraints are  satisfied: 
$(1)$~$\forall  t \in \T$ holds that 
$\nexists  e_i, e_j \in \E_t(\S)$ with  $\ell_{e_i} = \ell_{e_j}$ 
(\DEF{location constraint}); and 
$(2)$~$\forall t \in \T$ holds that 
${\underset{\forall e \in \E_t(\S)}{ \!\! \!\!\!\! \!\! \sum} \!\!\!\!\!\!\!\! \xi_e} \leq \theta$ 
(\DEF{resources constraint}).
In analogy, an \textit{assignment} $\alpha_e^t$ is said to be \DEF{feasible} if the aforementioned constraints hold for $t$.
Further, we  call \DEF{valid assignment}, 
an assignment $\alpha_e^t$ when the assignment is \textit{feasible}
and $e \notin \E(\S)$.


\vspace{4pt}
 Let $\C$ be a set of \DEF{competing events}, with $\C \cap \E = \varnothing$. 
As competing events we define events that have already been scheduled by third parties, 
and will possibly attract potential attendees of the candidate events.  
Based on its scheduled time, each competing event 
$c \in \C$ is associated with a time interval $t_c \in \T$. 
\alt{
}
{This association implies a ``conflict" between the competing event $c$ and 
and a possible candidate event that will take place at the time interval $t_c$. 
Note that,  at each time interval, a user can attend at most one event. }
Further, as $\C_{t}$ we denote  the competing events that are associated with 
the time interval $t$.

%

 \vspace{4pt}
Consider a set of {users} $\U$, for each \DEF{user} $u \in \U$ and event ${h \in \E \cup \C}$, 
there is a function ${\mu \colon \U \times (\E \cup \C) \to [0, 1]}$, 
denoted as  $\mu_{u,h}$,  that models the \DEF{interest} of user  $u$ over $h$. 
The interest value (i.e., affinity)  can be estimated by considering a large number of  factors 
(e.g.,   preferences, social connections). 
%

Moreover,  we define the 
 \DEF{social activity probability} $\sigma_u^t$, representing the {probability of user} $u$ participating in a social activity at~$t$. 
%
 This probability   can be estimated by examining the user's past behavior (e.g., number of check-ins). 

 \vspace{4pt}
Assume a user $u$  and a candidate event $e \in \E$ 
that is scheduled by $\S$ to take place at  time interval $t$; 
$\rho_{u,e}^t$ denotes the  \DEF{probability of} $u$  \DEF{attending} $e$  \DEF{at} $t$.
Considering the \textit{Luce's choice theory} \cite{Luce1959}, 
the probability $\rho_{u,e}^t$ is influenced  by  the
social activity probability $\sigma$ of $u$ at $t$, and
the interest $\mu$ of $u$ over $e$, $\C_t$ and $\E_t(\S)$.
%
We define the \textit{probability of} $u$  \textit{attending} $e$  \textit{at} $t$  as: %
%

{ \vspace{-2pt}
\small{
 \begin{equation}
\rho_{u,e}^t =  \sigma_u^t \:  \dfrac{\mu_{u,e}}{
{\underset{\forall c \in \C_t}{ \sum}\mu_{u,c}} +
{\underset{\forall p \in \E_t(\S)}{ \sum} \! \!  \! \! \mu_{u,p}}} 
 \label{eq:att}
  \end{equation}
}
\vspace{-0pt}}

\noindent
 Furthermore, considering all users $\U$,  we define the 
\DEF{expected  attendance}    for an event  $e$ scheduled to take place at   $t$ as:

\vspace*{-2pt}
{ \small{
 \begin{equation}
     \omega^t_{e} =  \underset{\substack{\forall u \in  \U}}{ \sum} \:\:  \rho_{u,e}^t 
\label{eq:user}
  \end{equation}
}}
\vspace{-0pt}

The \DEF{total utility} for a schedule $\S$, denoted as $\Omega(\S)$,
is computed by considering the expected attendance over all scheduled events:

\vspace{-3pt}
{ \small{
 \begin{equation}
     \Omega(\S) =  
    \underset{\substack{ \forall e \in  \E(\S) }}{ \sum} \:
    \omega^{t_e(\S)}_{e}  
\label{eq:total}
 \end{equation}
}}


\vspace{2pt}
\noindent
The \textit{Social Event Scheduling} (\ses)  problem is defined as follows: \footnote{Several of the problem's involved  factors  (e.g., user interest, activity and attendance probability)  can be  computed using event-based mining methods, e.g.,
\cite{ZhangWF13,Zhang2015,XuZZXCL15,Liu2012,Pham2015,MinkovCLTJ10,YQVHNH18,DuYMWWG14,Boutsis15}.  
However, this is beyond the scope of this work.}


\begin{myprob}[Social Event Scheduling  Problem (\ses)]\textbf{.}
Given an positive \textit{integer}~$k$, a set of candidate \textit{time intervals} $\T$; 
a set of \textit{competing events} $\C$;
a set of \textit{candidate events} $\E$; and a set of  \textit{users} $\U$;
our goal is to find a \textit{feasible schedule} $\S_k$ that determines how to assign $k$ candidate events such that the \textit{total utility} $\Omega$ is maximized; i.e.,  
${\S_k = \argmax \Omega({\S})}$ and  $|\S|=k$. 
\end{myprob}

\vspace{2pt}

Note that, by performing  trivial modifications to the algorithms proposed here, 
additional factors and constraints to those defined in \ses, 
can be easily handled.
For example, include event's organization cost/fee (to define a ``profit-oriented'' version of the \ses problem),  associate events with duration, or  
define weights over the users
\mbox{(e.g., based on their influence)}.


\eat{
\stitle{Remark.}
A simple ``profit-oriented" version of the \ses problem may also be defined
by associating each event with   an organization cost and a fee.
These factors can be directly included in the \ses problem.
In this setting, the total utility will determine the  expected scheduling profit. 
Also, several additional  constraints  may be included, 
e.g., associate  events with duration, or  
a set of time intervals during which 
the event's location can host the event, etc.
}

%


%

\alt{}
{Without loss of generality, in  the \ses definition we have assumed that $|\E| \ge k$ 
and there exists at least one feasible assignment for $k$  events. 
}

\subsection{Approximation Hardness}
\label{sec:anal}

Here,  we show that even in highly restricted instances the \ses problem is  NP-hard to be approximated over a factor.
 Therefore, \ses does not admit a \textit{Polynomial Time Approximation Scheme} (PTAS).


 \vspace{1pt}

 \begin{mytheor}
\label{th:apprx}
There exists an $\epsilon > 0$ such that it is \mbox{NP-hard} to approximate 
the   \ses problem 
 within a factor larger than $(1- \epsilon)$.   Thus, \ses does not admit a PTAS.
  \footnote{Due to lack of space,   we only include proof sketches, 
while in simple cases, the proof sketch is also omitted.}
\end{mytheor}


\begin{myproofSk}
In our proof we reduce the 3-Bounded \linebreak
3-Dimensional Matching  problem (3DM-3) \cite{Kann91}  to a restricted  instance of \ses.
The  following is an instance of the \mbox{3DM-3} problem.
Given a set $T \subseteq X \times Y \times Z$, with $|X|=|Y|=|Z|=n$,  
 $|T|=m$ and with each element of $X \cup Y  \cup Z$ appearing at most $3$ times as a
coordinate of an element of $T$. 
A matching in $T$ is a subset $M \subseteq T$, such that no elements in $M$ agree in any coordinate.  
%
In our proof, we exploit the following result: 
\cite{Kann91}  showed that in \mbox{3DM-3} there exists an $\epsilon_0>0$ such that it is \mbox{NP-hard} to decide whether 
an instance has a matching of size $n$ or if every matching has size at most $(1 - \epsilon_0)n$.

Consider  the following \textit{associations between} 3DM-3 {and} \ses:
edges $g$ in $T$ to time intervals; and
elements in $X, Y, Z$ to candidate events, with required resources $\xi=1$. 
Let the aforementioned candidate events form a set $E_1$ (i.e., $|E_1|=3n)$.

Further, in the proof, we consider the following \textit{restricted instance of} \ses:
$(1)$~The \textit{available resources} are three.
$(2)$~There are no  \textit{location constraints}.
$(3)$~There is only one \textit{competing event} in each time interval.
$(4)$~The \textit{social activity probability}  is the same for each user and time interval.
$(5)$~The \textit{users} are as many as the candidate events.
$(6)$~There is a set $E_2$ that contains $m-n$ additional (w.r.t.\ $E_1$) candidate events, with ${\xi=3}$. 
Thus, the candidate events $\E$ in the restricted instance is $\E = E_1\cup E_2$, with $E_1 \cap E_2=\varnothing$. 
$(7)$~Regarding the \textit{interest function} we  assume  two disjoint sets of users $|U_1|=3n$ and $|U_2|=m-n$, a well as the following: 
$(7a)$~Each  user $u_1 \in U_1$ likes only one event $e_1 \in E_1$ (as a result,  each $e_1$ is liked only by one user $u_1$), with ${\mu_{u_1, e_1} =0.25}$.
$(7b)$~
Regarding the \textit{competing events} and the users $U_1$ we have the following.
Fix a positive constant $\delta < \frac{1}{12}$.
Let $u_p \in U_1$ the user that  likes the event $e_p$, 
where $e_p$ corresponds to the element $y_p$ in \mbox{3DM-3}.
Then, if  $y_p$ is {included in the edge} $g_t$ (i.e., $y_p \in g_t$), 
the interest of $u_p$ in the competing event $c$ 
that appears in the interval $t$ (which in \mbox{3DM-3} {corresponds to edge}~$g_t$) 
is  $\mu_{u_p, c}=0.25(0.75-\delta)/(0.25+\delta)$ and $0.75$, otherwise. 
  $(7c)$~Each  user $u_2 \in U_2$ likes only one event $e_2 \in E_2$ (as a result,  each $e_2$ is liked only by one user $u_2$), with $\mu_{u_2, e_2}=0.75$.
$(7d)$~{For each competing event $c$ and user $u_2\in U_2$, we have $\mu_{u_2, c}=0$.} 

We can verify that, for each matching in \mbox{3DM-3}, we can obtain 
a schedule in \ses  with  total utility $3(0.25+\delta)$, 
by assigning 3 events of $E_1$ in a same interval.
Then, if 3DM-3  has a matching of size $n$, we can verify that 
the total utility in \ses is $3n (0.25+\delta)+m-n$.
Otherwise, if every matching has size at most $(1-\epsilon_0)n$,
 the total utility in \ses is
\mbox{$ 1-\frac{ \epsilon_0- 12 \delta\epsilon_0 }{ 12 \delta +3} <
 1-\frac{1 }{3}\epsilon_0 $.
 \:\:\:\:\:\:\:\:
\qed}
\end{myproofSk}



\vspace{-2pt}
\section{Algorithms}
\label{sec:methods}

%


%
%


%

\ses is  known to be strongly NP-hard, even in highly restricted instances \cite{ses}.
 Due to its hardness, it is  computationally prohibitive to find an optimal solution 
even in small problem sizes. 
Particularly, in the worst case, we have to  enumerate an exponential number of possible assignments,   
where   each assignment requires always $|\U|$ computations.
For example, the greedy algorithm proposed in \cite{ses}, in several cases in our experiments, took more than 5 hours   to solve the problem in the default parameters setting, while more than 31 hours in larger settings. 
To this end, to cope with the hardness of the \ses problem we design three efficient and scalable approximation algorithms which 
perform about half the computations
and, in several cases, are 3-5 times faster 
compared to the method proposed in \cite{ses}.

\subsection{Existing Solution}
\label{sec:bcsicde}

Here, we outline the previously proposed algorithm.
First, we define the {assignment score}.
%
Given a schedule $\S$ and  an assignment $\alpha_{r}^t$, 
as  \DEF{assignment score} (also referred  as  \DEF{score}) of an assignment $\alpha_{r}^t$, denoted as $\alpha_{r}^t.S$, 
we define the \textit{gain} in the expected attendance by including $\alpha_{r}^t$ in $\S$. 
The assignment score (based on Eq.~\ref{eq:user}) is defined~as: 
 
\vspace*{-4pt}
{ \small{
 \begin{equation}
  \label{eq:gain}
 \alpha_{r}^t.S \,  \,  = \!  \! \underset{\substack{ \forall e_j \in  \\ \E_t(\S)\cup \{  r \}}}{ \sum} \!\!\!\!\!\!
  \omega'^{\, t}_{e_j}  
  \, - \,
  \underset{\substack{ \forall e_i \in  \\ \E_t(\S) }}{ \sum} \!
  \omega^{t}_{e_i}  
  \end{equation}
   }}
 \vspace{0pt}

\noindent 
Given a set of assignments, the term  \textit{\myFontP{\textbf{top assignment}}} refers to the assignment with the largest score. 

 In \cite{ses},  a  simple   greedy algorithm is outlined, referred here as \bsc.
 This method starts by initially generating assignments between all pairs of events and intervals. Then, in each iteration, the assignment with the largest score (i.e., top assignment) is selected. After selecting an assignment, a subset  of the  assignment's scores need to be updated. 
Recall that, the assignment's score is defined w.r.t.\ the events assigned in the assignment's interval (Eq.~\ref{eq:gain}).
Hence, when an assignment $\alpha_e^t$ is selected,  
then the scores of the assignments referring to interval  $t$  need to be recomputed (updated).
The time complexity of \bsc is 
$O(|\U||\C|+|\E||\T||\U|+k|\E||\T|+ k|\E||\U| - k^2|\T|-k^2|\U|)$; 
and the space complexity is $O(|\E||\T|)$.

  \begin{myExample} [\myFontP{\textit{\bsc Algorithm}}]
\label{ex:bsc}
Based on our running example, Figure~\ref{fig:bscEx} outlines the execution of   the \bsc algorithm.  
In this, as well as in the rest of the examples, we assume that $k=3$. 
That is, three out of four events have to
be scheduled. 
Each row represents the selection of a single assignment. 
Rows include the assignment scores (Eq.~\ref{eq:gain}), as well as 
the selected assignment (presented in bold red font) and the assignments that have to be updated after the selection. 
Initially \mbox{(i.e., \ding{172} selection)}, the algorithm selects the assignment with 
the largest score (i.e., $\alpha_{e_4}^{t_2}$).
Thus, after this selection  the assignments referring to $e_4$ have to be omitted (marked with /), and the assignments referring to $t_2$ have to be updated.
After the second selection, the algorithm has to update only $\alpha_{e_3}^{t_1}$ since 
$\alpha_{e_2}^{t_1}$ is no longer feasible (marked with $\times$) due to location constraint. 
Note that,  for the sake of simplicity, the resources constraint has been omitted from the running example. 
Finally,  the schedule contains $\alpha_{e_4}^{t_2}$, $\alpha_{e_1}^{t_1}$ and $\alpha_{e_2}^{t_2}$.
 \end{myExample}

   \begin{figure}[t]
  \vspace*{-0pt}
  \hspace{-3pt}{
\tiny
\setlength{\tabcolsep}{3.0pt}
\begin{tabular}[t]{lcc ccc ccc c ccc}
& $\alpha_{e_1}^{t_1}$  & $\alpha_{e_2}^{t_1}$ & $\alpha_{e_3}^{t_1}$ & $\alpha_{e_4}^{t_1}$ & $\alpha_{e_1}^{t_2}$ & $\alpha_{e_2}^{t_2}$ & $\alpha_{e_3}^{t_2}$ & $\alpha_{e_4}^{t_2}$ & &\hspace{3pt}\scriptsize{\textbf{Select}} & \hspace{-0pt} \scriptsize{\textbf{Update}} \\\cmidrule{2-9} \cmidrule{11-11}\cmidrule(lr){12-12}
\rowcolor{gray!30}	

\hspace*{-0pt}{\small\ding{172}} &0.59  & 0.52 &0.10  &  \cancel{0.64}&  0.53 & 0.57 & 0.09  & \RED{\textbf{0.66}}&\hspace*{-1pt} $\rightarrow$ \hspace*{-4pt}&\hspace{3pt}$\alpha_{e_4}^{t_2}$ &\hspace{0pt}$\alpha_{e_1}^{t_2}$ $\alpha_{e_2}^{t_2}$ $\alpha_{e_3}^{t_2}$\vspace{2pt} \\
\hspace{-0pt}{\small\ding{173}}  &  \RED{{\textbf{0.59}}}  &  {0.52} & {0.10}  &--&\cancel{0.34}& 0.16 & 0.03  & --  & \hspace*{-1pt} $\rightarrow$ \hspace*{-4pt}&\hspace{3pt}$\alpha_{e_1}^{t_1}$ & \hspace{0pt} $\alpha_{e_3}^{t_1}$ \vspace{2pt} \\\rowcolor{gray!30}	
\hspace{-0pt}{\small\ding{174}} &  --  & \xcancel{0.52}&0.05 & -- &  -- &   \RED{{\textbf{0.16}}}&{0.03}   & -- &\hspace*{-1pt} $\rightarrow$ \hspace*{-4pt} &\hspace{3pt}$\alpha_{e_2}^{t_2}$ & \hspace{-0pt}  --\\

\end{tabular}}
\vspace{-4pt}
\caption{\bsc algorithm example} 
  \label{fig:bscEx}

 \vspace{8pt}
 \end{figure}

\vspace{-4pt}
\subsection{Incremental  	{Updating} Algorithm \textnormal{{(\bnd)}}}
\label{sec:bnd}

 The \bsc algorithm proposed in \cite{ses}, has the following  shortcoming:  
$(1)$~each time   \bsc selects an assignment, it has to 
recompute (i.e.,~update)   from scratch all the scores for all the 
assignments associated with the selected assignment's interval.
This process is referred to {assignment updating} or simply  as {updates}; and%
\alt{}{Note that, for each score calculation (Eq.\ \ref{eq:gain}),  $|\U|$  computations are performed.}
$(2)$~in  each step, \bsc has to examine (and traverse) all the available assignments 
in  order to perform its tasks (e.g., select assignment, perform updates). 



Considering the aforementioned issues,   we design  the 
\textit{Incremental  Updating algorithm} (\bnd). 
Regarding the first issue, \bnd exploits an \textit{incremental updating scheme}, 
performing  {incremental assignment updates}. 
Incremental updating allows \bnd, to provide the same solution as \bsc, 
while, in  each step, \bnd performs only a part of the updates (i.e., score computations). 
Regarding the second issue, \bnd attempts to reduce the number of assignments that should be examined  in each step, i.e., search space.
To this end, we devise an \textit{assignment organization}
that takes into account 
the   incremental updating scheme. 
In several cases in our experiments,  \bnd \textit{is more than three times faster} than the existing algorithm.

Essentially, \bnd follows a similar assignment selection process to \bsc, 
selecting  the top assignment in each step, in a greedy fashion. 
However, in \bnd  the assignments' update process has been designed based on  the introduced incremental updating scheme.



\subsubsection{Incremental Updating}
 \label{sec:bscheme}
\hfill \break

 \vspace{-8pt}

\alt{
}
{This section introduces the \BLUE{bound-based} update scheme, 
which is adopted by \bnd to reduce the number of assignments updates.}
%
%
In the proposed scheme,   the updates are computed in an incremental  manner, 
where after each assignment selection only a part of the updates  are performed.
As a result,  during the  algorithm execution,  some of the assignments may not be up-to-date.

An {assignment} is denoted as \DEF{updated}, 
if its score has been computed by considering all the (previously) selected assignments, 
and \DEF{not updated} otherwise.
%
In analogy, a set of assignment is referred as   \textit{updated}, when
all its assignments are updated, and \DEF{partially updated}, otherwise.


The basic idea of our scheme is that we can determine a subset of the not updated assignments  that have to be updated before each selection. 
First we show that, from the available assignments $\A$,  we can find a set $\B \!\!  \subseteq\!  \A$  which includes the next algorithm selection $\chi$.
 Then, we also show that the not updated assignments included in $\B$ 
are the only not updated assignments that have to be updated in order to find $\chi$.

In order to specify $\B$, we use a numeric \textit{bound} $\Phi$. 
As shown next, \scalebox{0.93}[1]{the value of $\Phi$ is  the score of  the 
\mbox{\textit{top}, \textit{updated} and \textit{valid} assignment of $\A$.}}

\vspace{2pt}

\begin{myproposition}
\label{prop:bound}
Let  $\Phi$ be the score of the \textit{top}, \textit{updated} and \textit{valid} assignment of the available assignments $\A$.
Then, the next selected assignment $\chi$
is one of the assignments  that  in $\A$ have score larger or equal to $\Phi$; 
i.e., $\chi \in \B$, where ${\B={\{\alpha_e^t \in \A \mid a_e^t.S \geq\Phi\}}}$.
\end{myproposition}

\vspace{2pt}


\alt{
\begin{myproofSk}
First, we show that {the score of a not updated assignment 
is always larger or equal to the score of  the assignment resulted by its update}. 
Note that the proof for this is not trivial for arbitrary numbers of candidate and competing events. 
Based on the aforementioned, the not updated assignments of $\A$, having score
lower than $\Phi$,   also have score lower than $\Phi$ in $\A'$, 
where $\A'$ be the set of assignments resulting from $\A$ by updating its not updated assignments.
Further, the score of each updated assignment of $\A$  remains the same  in $\A'$.
So, both the updated and the not updated assignments of $\A$ have scores 
lower than $\Phi$; their scores in $\A'$ also remain  lower than $\Phi$.
Thus, the Proposition~\ref{prop:bound} holds.
\qed
 \end{myproofSk}
}{}


Based on Proposition~\ref{prop:bound}, since $\chi$ is included in $\B$, we can easily verify that,  $\chi$   is the \textit{top} assignment of $\B'$, 
where $\B'$ results from $\B$ by updating its not updated assignments.
Thus, in order to find $\chi$, we have to update the not updated assignments of $\B$.
Based on the aforementioned, the following corollary describes  the incremental updating process.




\begin{mycoroll} 
\label{cor:up}
In each step, in order to select the next assignment,
only the not updated assignments having score larger or equal to $\Phi$ have to be updated.
\end{mycoroll}

%

%


\alt{}{
\stitle{\bsc vs.\ \BLUE{Bound-based} Updates.}
In what follows, we outline some main differences between the 
update process adopted in the  \bsc, and the \BLUE{bound-based} approach.
Recall that, after selecting an assignment $\alpha_e^t$, 
\bsc updates all the  assignments referring to $t$. 
On the other hand, in the \BLUE{bound-based} approach, 
after selecting an assignment, a subset of the ``affected'' assignments is updated.
As a result,  through  the execution of the algorithm, some of the  
assignments are up-to-date,  while others are not.
Further, in the \BLUE{bound-based} approach, it is possible that, 
after selecting $\alpha_e^t$, 
assignments referring to several intervals will be updated, while the interval $t$ will may not be included.
Finally, similar to \bsc, when an assignment $\alpha_e^t$ is updated, then $\alpha_e^t$ remains up-to-date until a new assignment referring to $t$ is selected.

\stitle{\BLUE{Bound-based} Update Process.}
In the following, we point out a few issues arising in \BLUE{bound-based} update process. 
The existence of both updated and not updated assignments, 
impose  its categorization to updated or not updated, based on its update statues. 
%
We should note that, the running bound may not be next selected assignment. 
This occurs in cases where one of the assignments that will 
be updated 
has a new score with larger value than the running bound. 
%
}


  \begin{myExample} [\myFontP{\textit{{Incremental Updating} Scheme}}] 
\label{ex:bnd}
 Figure~\ref{fig:bndEx} illustrates the utilization of the incremental updating scheme.
For clarity of presentation, we omit the assignment scores since these are the same as in Example~\ref{ex:bsc}. 
To better understand the procedure,
in each row the assignments are presented in descending order, based on their score.  
The +/- notation is used to denote that the assignment  is \textit{updated}, or \textit{not updated}, respectively. 
After   the first selection,  $\Phi$ is equal to $\alpha_{e_1}^{t_1}.S$ (i.e.,~top, updated and valid assignment), and all the assignments referring to $t_2$ change  to \textit{not updated}. 
Further, since all the not updated assignments have score lower than  $\Phi$, none of the assignments have to be updated.
Then (\ding{173} selection), after selecting $\alpha_{e_1}^{t_1}$, all the assignments become not updated; so $\Phi$ becomes unavailable.
Next, the algorithm updates $\alpha_{e_2}^{t_2}$ and sets $\Phi$ equal to its score (0.16). 
%
In the last selection, since the current $\Phi$ is larger than the scores (0.10 and 0.9)
 of the not updated  assignments $\alpha_{e_3}^{t_1}$ and $\alpha_{e_3}^{t_2}$,
  the algorithm does not have to update it.
Compared to the \bsc algorithm (Example~\ref{ex:bsc}) which performs four updates, 
our scheme performs only one. 
 \end{myExample}

 \begin{figure}[t]
\vspace*{-7pt}
\def\arraystretch{1.9}
\hspace*{-2pt}
\tiny
\setlength{\tabcolsep}{1.0pt}
\begin{tabular}[t]{lccc ccc ccc c  cc}
&\multicolumn{8}{c}{\scriptsize{Assignments Sorted by Score} \tiny{({{``$+$" / ``$-$''  : \textit{Updated}/\textit{Not updated} }}})}& &\hspace{2pt}\scriptsize{\textbf{Select}}&\hspace{-0pt}\scriptsize{$\Phi$} &\hspace{-3pt}\scriptsize{\textbf{Update}} \\\cmidrule(r){2-9} \cmidrule(){11-11}\cmidrule(lr){12-12}\cmidrule(lr){13-13}
\rowcolor{gray!30}	

\hspace*{-1pt}{\small\ding{172}}&   \RED{\scalebox{0.9}[1]{${\boldsymbol\alpha_{e_4}^{t_2+}}$}} & \scalebox{0.9}[1] {\cancel{$\alpha_{e_4}^{t_1+}$}} & \scalebox{0.9}[1]{$\alpha_{e_1}^{t_1+}$} & \scalebox{0.9}[1]{$\alpha_{e_2}^{t_2+}$} & \scalebox{0.9}[1]{$\alpha_{e_1}^{t_2+}$}& \scalebox{0.9}[1]{$\alpha_{e_2}^{t_1+}$} & \scalebox{0.9}[1]{$\alpha_{e_3}^{t_1+}$} &\scalebox{0.9}[1]{ $\alpha_{e_3}^{t_2+}$} & \hspace*{-3pt} $\rightarrow$ \hspace*{-3pt}& \hspace{0pt}\scalebox{0.9}[1]{${\alpha_{e_4}^{t_2}}$}&\hspace{-0pt}\scalebox{0.9}[1]{$\alpha_{e_1}^{t_1}.S$}  & -- \vspace{3pt}\\ 

\hspace*{-1pt}{\small\ding{173}} &  -- & \scalebox{0.9}[1]{--} & \RED{\scalebox{0.9}[1]{$\boldsymbol\alpha_{e_1}^{t_1+}$} }& \scalebox{0.9}[1]{$\alpha_{e_2}^{t_2-}$  }& \scalebox{0.9}[1]{\cancel{$\alpha_{e_1}^{t_2-}$}  }&  \scalebox{0.9}[1]{{$\alpha_{e_2}^{t_1+}$} }& \scalebox{0.9}[1]{$\alpha_{e_3}^{t_1+}$ }& \scalebox{0.9}[1]{$\alpha_{e_3}^{t_2-}$ }&\hspace*{-3pt} $\rightarrow$ \hspace*{-3pt} & \hspace{0pt}\scalebox{0.9}[1]{${\alpha_{e_1}^{t_1}}$}& \hspace{-2pt} $\varnothing$ &\hspace{-2pt}\scalebox{0.9}[1]{$\alpha_{e_2}^{t_2}$} \vspace{3pt}\\ \rowcolor{gray!30}	

\hspace*{-1pt}{\small\ding{174}} &  -- &--    &--&\scalebox{0.9}[1]{\RED{$\boldsymbol\alpha_{e_2}^{t_2+}$}  } &\scalebox{0.9}[1]{--}&\scalebox{0.9}[1]{\xcancel{$\alpha_{e_2}^{t_1+}$}  }&  \scalebox{0.9}[1]{$\alpha_{e_3}^{t_1-}$} & \scalebox{0.9}[1]{$\alpha_{e_3}^{t_2-}$ }& \hspace*{-3pt} $\rightarrow$ \hspace*{-3pt}&\hspace{0pt}\scalebox{0.9}[1]{${\alpha_{e_2}^{t_2}}$}&\hspace{0pt}\scalebox{0.9}[1]{$\alpha_{e_3}^{t_2}.S$}& --\\ 

\end{tabular}
\vspace{-4pt}
\caption{{Incremental  updating} scheme example}
  \label{fig:bndEx}
\vspace{12pt}
 \end{figure}

%


  { \subsubsection{\mbox{Assignments Organization over Incremental Updating}}
\label{sec:interv-org}
\hfill \break

\vspace{-18pt}
In each step,
the algorithm needs to examine and traverse all the available assignments, in order to perform the following   main tasks: 
$(1)$~select the top assignment;
$(2)$~perform  updates;
and $(3)$~maintain the  bound.
In order to accomplish these tasks, for each of the available assignments, 
the algorithm  has to perform numerous computations.
 Indicatively, it has to check  validity constraints, compare scores, 
consider bounds and possibly compute the new assignment score, etc.

Given the above, we introduce an \textit{interval-based assignment organization}  that incorporates with our incremental updating scheme.
This organization attempts to reduce the number of assignments that are accessed and examined, i.e., \textit{search space}.
Using our   organization,  
in most cases in our experiments, \bnd \textit{examines slightly 
more than  half assignments}  compared to the existing algorithm.


  \stitle{Search Space Reduction in Assignment Updates.}
Here, we describe how we reduce the assignments that should be examined 
in order to perform the updates. 
An interval-based assignment organization allows to access at the interval-level, 
only the assignments that should be examined for update. 
Adopting this organization  in a simple (not incremental) updating process, like in \bsc, 
in each step, the algorithm  needs only to access  the assignments of one interval in order to performs the updates.
%
On the other hand, in the incremental updating setting,  
several intervals  become {partially updated}  during the execution.
In this scenario, in order to identify the assignments that need to be updated, we have  
to examine all the assignments included in partially updated intervals.
As a result,  in our setting,  a simple interval-based organization  will not be effective, 
since it will allow to skip accessing only the   updated intervals.


 Beyond ignoring only the updated intervals, 
in order to further reduce the search space, 
we have to be able to identify (and skip) the partially updated  
intervals whose assignments are not going to be updated.
In our organization scheme, this  is addressed by defining a score over each interval. 
Particularly, for each  interval $t \in \T$, a value $M_t$ is defined, 
where $M_t$ is equal to the  score of the \textit{top}, \textit{updated} and \textit{valid} assignment of interval $t$. 
Exploiting $M_t$, we can directly identify the partially updated intervals 
that have to be accessed through the updating process. 
Particularly, it is easy to verify that we have to access all the partially updated intervals $t \in \T$ for which $M_t \leq \Phi$.

\alt{}{Note that, these bound, correspond to the information ``stored'' for each interval 
in order to  select the top  assignment and compute $\Phi$  
(as described in  the previous paragraphs and presented in Algorithm~\ref{algo:bnd}).

To sum up, the proposed interval-based organization which is enriched with bounds, allows us, in each step, to reduce the number of assignments that should be examined by the \bnd algorithm. 
}

\alt{}{Considering the increased number of assignments through the execution of the algorithm, 
the number of    updated intervals is reduced. 
Thus, this interval-organization is not effective in \BLUE{bound-based} scheme, since 
there are cases where the search area cannot be (significantly) reduced.%
}

%
%
\alt{}{Considering the proposed approach, 
there are cases where the search space, instead of including 
all the available assignments, includes only $|\T|$ of them.}%
%


 \stitle{Search Space Reduction in Assignment Selection and Bound \linebreak Maintenance.}
The  organization described so far allows to reduce the search space 
during the assignment updating process. 
However, after performing the updates, the algorithm has to accomplish also the tasks 
of selecting the next assignment and maintaining the bound $\Phi$, which in turn enforce 
the examination of all the available assignments.
In what follows, we show how  to perform all the tasks without examining any further 
assignments beyond the ones examined during the updating process. 

The intuition is that, in each step, only a subset of the assignments is  updated,
while the rest remain the same as in the previous step, referred here as \textit{static}. 
Therefore, it is reasonable to assume that  the algorithm is able to accomplish all of its tasks by utilizing ``information'' previously captured from the static assignments $\W$.
So, 
if this ``information'' is known, then, after performing the updates, we can ignore  $\W$ (i.e., avoid access). 
%
As shown next, this ``information'' can be captured by two numeric values $I_\chi$ and $I_\Phi$ determined from $\W$. Briefly,  $I_\chi$  is exploited to 
specify the next selected assignment $\chi$,
and $I_\Phi$ to compute the new $\Phi$.



\begin{myproposition}
\label{prop:search}
Given a set of static assignments $\W$. Let  $I_\chi$ and $I_\Phi$ be the scores of the \textit{first} and the \textit{second larger
 {top},  {updated}} and {\textit{valid}} assignment of  $\W$, respectively.
Then, if $I_\chi$ and $I_\Phi$  are know, the algorithm can ignore $\W$.%
 \end{myproposition}
 
 \vspace*{2pt}
In our interval-based scheme implementation, 
the static assignments $\W$ correspond to a set of static intervals 
$\T_\W \subseteq \T$. 
Both $I_\chi$~and~$I_\Phi$   can be  be directly computed based on  the values of  $M_t$  of the static intervals $\T_\W$. 
Particularly, 
$I_\chi={\underset{\forall t \in \T_\W}{ \max} M_t}$ and 
$I_\phi= \!\!\!\!\!\!{\underset{\forall t \in \{\T_\W \backslash t_\chi\}} { \max} 
\!\!\!\!\! M_t}$,  where $t_\chi$ is the interval of $I_\chi$%
\alt{.}
{\footnote{Note that, there is a special case in which Proposition~\ref{prop:search}
does not hold in our interval-based scheme implementation. 
Briefly, this case arises, when, after selecting the assignment $a_{e_s}^{t_s}$,   the value 
$z={\underset{\forall t \in \T \backslash t_s}{ \max} M_t}$ corresponds to the event $e_s$.
In this case, the algorithm has to access the (static or updated)  interval $h$,
 in which $M_h =z$.  
Then, if $h$ is a static interval, the proposition does not hold.}.}
Therefore, based on Proposition~\ref{prop:search}, in each step, the algorithm needs to access only intervals that have been updated (i.e.,~subset of partially updated intervals).

\stitle{Assignment Organization   Summary.}
To sum up, the presented organization allows:
$(1)$~the reduction of the assignments that are examined during the updating process; and 
$(2)$~skipping the examination of any further assignments beyond the updated ones. 

}


%

%

\begin{algorithm2e}[t]
\scriptsize
\SetInd{0.4em}{0.8em}
\setstretch{0}

\caption{\bnd ($k$, $\T$, $\E$, $\C$, $\U$)}
\label{algo:bnd}
\KwIn{
$k$: number of scheduled events;  \: 
$\T$:  time intervals;  \linebreak
$\E$: candidate events;  \:  
$\C$: competing events;  \:  
$\U$: users;
}
\KwOut{$\S$: feasible schedule containing $k$ assignments}
 \KwVar{\mbox{$\L_i$: assignment list for interval $i$;  $\Phi$:   bound; 
$\M$: top, valid and updated assign list}
}

\vspace{2pt}

 $\S \gets \varnothing;$\:\: 
$\M  \gets \varnothing;$\:\:  
$\L_i \gets \varnothing,$\:   
$\L_i.U \gets \myupdated$\:
$1 \leq i \leq |\T|$

 \ForEach(\Comment*[f]{\mycomment{{generate assignments}}}){$(e, t) \in  \E \times \T$}{
 compute $\alpha_e^t.S$;  \:  \:  $\alpha_e^t.U$ $\gets$  \myupdated;
 \Comment*[r]{\mycomment{{{by Eq.\ \ref{eq:gain}}}}}

insert $\alpha_e^t$ into $\L_t$
  \Comment*[r]{\mycomment{{{initialize assignment lists}}}}


$\M_{[t]} \gets  \maxAss(\M_{[t]},  \alpha_e^t)$
\hspace{5pt}\scalebox{0.8}[1]{
\Comment*[r]{\mycomment{{{initialize $\M$ with the top assignment from each interval}}}}}
 
}

\While{$|\S|<k$ }{

 	 $\alpha_{e_p}^{t_p}\gets$  $\topAss(\M)$
		  \Comment*[r]{\mycomment{{{select the top, valid \& updated assignment}}}}

 insert $\alpha_{e_p}^{t_p}$ into $\S$  
		  \Comment*[r]{\mycomment{{{insert into schedule}}}}

  remove $\alpha_{e_p}^{t_p}$ from $\L_{t_p};$  \: \: 
$\L_{t_p}.U \gets  \mynotupdated;$ 
		  \Comment*[r]{\mycomment{{{update $\L_{t_p}$ status}}}}


$\alpha_{e}^{t}.U$ $\gets$  \mynotupdated, $\forall \alpha_{e}^{t} \in \L_{t_p}$

\ForEach(\Comment*[f]{\mycomment{{update $\M$ list based on selected assignment}}}){$\alpha_{e}^{t} \in \M$}{

\uIf{$t=t_p$} {
$\alpha_{e}^{t} \gets  \varnothing$
\Comment*[r]{\mycomment{{{i.e., \: $\alpha_{e}^{t}.S \gets  -\infty$}}}}
}
	 
\ElseIf{$e=e_p$}{
$\alpha_{e}^{t} \gets$ $\topAss(\L_i)$
}

}


	$\Phi \gets$  score of top $\M$ 
	 \Comment*[r]{\mycomment{{{set   bound}}}}

\For(\Comment*[f]{\mycomment{{update assignments}}}){$i=1$ \emph{\KwTo} $|\T|$}{

  \If(\Comment*[f]{\mycomment{{check   for updates}}})
  {$\L_i.U = \mynotupdated$ \Logand $\M_{[i]}.S \leq  \Phi$ }{
\ForEach
{$\alpha_{e}^{t} \in \L_i$}{

\uIf{$\alpha_{e}^{t}$ is not \myvalid}{ 
remove $\alpha_{e}^{t} $  from $\L_i$
 }
 
 \ElseIf
{ $\alpha_{e}^{t}.U=\mynotupdated$ \:  \Logand \:  $\alpha_{e}^{t}.S \geq \Phi$ }
{ 

compute new $\alpha_{e}^{t}.S;$  \: \: $\alpha_{e}^{t}.U \gets \myupdated$; \Comment*[r]{\mycomment{{{by Eq.\ \ref{eq:gain}}}}}


$\M_{[i]} \gets \maxAss(\M_{[i]},  \alpha_{e}^{t})$ \Comment*[r]{\mycomment{{{update top assignment}}}}

$\Phi \gets \maxAss(\Phi, \alpha_{e}^{t}.S)$ \Comment*[r]{\mycomment{{{update   bound}}}}

}
}

\lIf{all $\alpha_{e}^{t} \in \L_i$  is \myupdated}
{$\L_i.U=\myupdated$;}
  } 
}
  }

 \Return $\S$\;
 \end{algorithm2e}


\subsubsection{\bnd Algorithm Description \& Analysis}  
\hfill \break
%
%
%

\vspace*{-8pt}

\stitle{Algorithm Description.}
 Algorithm~\ref{algo:bnd} describes the \bnd algorithm; \bnd receives  
the same inputs as \bsc. 
Additionally,  \bnd employs  $|\T|$ lists, with each list $\L_i$ filled with the assignments  of  interval $i$.
Further, each assignment $\alpha_e^t$ and list $\L_i$, use a flag $U$ 
(denoted as $\alpha_e^t.U$) to define its update status.
Finally, the algorithm uses a  list $\M$ that holds the top, valid and updated assignments of each interval.
 %
Initially, like \bsc, \bnd calculates the scores 
for all possible assignments (\textit{loop in line} 2). 
At the same time, 
the assignments are inserted into the corresponding list $\L_i$ (\textit{line}~3-4). 
Note that, the \textit{getBetterAssgn} function returns the assignment with the larger score.

Then, at the beginning of each iteration (\textit{line} 6), the algorithm selects the top 
assignment  from $\M$,
 and inserts it into schedule (\mbox{\textit{lines}~7-8}).
 %
Also, the algorithm has to revise the 
information related to update status (\textit{lines}~9-10). 
%
%
%
After the assignment's selection phase, the algorithm performs score updates. 
Initially, the   bound $\Phi$ is defined by the top updated assignment (\textit{line}~16). 
Then, the algorithm traverses the  lists $\L_i$, 
using as upper bound the $\M_{[i]}.S$, to identify the   lists that have to be checked for updates (\textit{line}~18). 
From the verified lists, the algorithm performs  incremental updates (\textit{lines} 22-23), 
updating also $\M$ and $\Phi$ (\textit{lines} 24-25).


\stitle{\bnd vs.\ \bsc Solution.}
The following proposition states that both    \bnd    and \bsc,
return   the same solutions.

\begin{myproposition}
\label{prop:bnd}
\bnd and \bsc always return the same solution.
\end{myproposition}

\vspace{-0pt}

\stitle{\scalebox{0.94}[1]{Complexity Analysis.}} 
\alt{
%
\scalebox{0.94}[1]{The cost in the first loop (\textit{line}~2)
 is $O(|\E| |\T||\U|)$.}
Note that, each assignment score  (Eq.~\ref{eq:gain}) is computed in $O(|\U|)$.
 Then, the second loop (\textit{line}~6) performs $k$ iterations.
 The overall cost for the {$\topAss$ operation (\textit{line}~7) is} $O(\sum_{i=0}^{k-1}|\T|)$.
 Further, the loop in  \textit{line}~11 performs $|\T|$ iterations, 
 whereas in the worst case, in $|\T|-1$ of these iterations, 
\bnd performs a $\topAss$  operation which costs  {${O(|\E|-(i+1))}$}, with $0 \leq i \leq k-1$.
Thus, the overall cost for this $\topAss$ operation is  $O(\sum_{i=0}^{k-1}(|\T|-1)(|\E|-i-1))$.
%
Next (\textit{line}~17), in the worst case,  
all the not updated assignments are updated (same as \bsc).
Note that, in the \textit{Best case}, \bnd does not perform any computations 
for assignment updates,  while in \bsc, in every case, the cost for updates is  $O(|\U|\sum_{i=0}^{k-2}|\E|-i-1)$.
%

Hence, in the worst case, the overall cost of \bnd   is  same as \bsc; i.e.,
$O(|\U||\C|+|\E||\T||\U|+k|\E||\T|+ k|\E||\U| - k^2|\T|-k^2|\U|)$.
Finally, the space complexity   is $O(|\E||\T|+|\T|)$.
}
{Similar to \bsc,  in \bnd the cost in the first loop (\textit{line}~2)
 is $O(|\E| |\T||\U|)$  and the second loop (\textit{line}~6) performs $k$ iterations.
 The overall cost for the $\topAss$ operation (\textit{line}~7) is $O(\sum_{i=0}^{k-1}|\T|)$.
 Further, the loop in  \textit{line}~11 performs $|\T|$ iterations, 
 whereas in the worst case, in $|\T|-1$ of these iterations, 
\bnd performs a $\topAss$  operation which costs $O(|\E|-(i+1))$, with $0 \leq i \leq k-1$.
Thus, the overall cost for this $\topAss$ operation is  $O(\sum_{i=0}^{k-1}(|\T|-1)(|\E|-i-1))$.
{Particularly, the aforementioned worst case occurs when all the assignments included 
in $\M$ refer to the same event.}
Next (\textit{loop in line}~17), in the worst case, each time, 
all the not updated assignments are updated (same as \bsc). 
{The aforementioned worst case occurs, when the bound $\Phi$ is always smaller than 
all the not updated assignments.}
Therefore, in the worst case, the overall cost of \bnd   is  same as \bsc; i.e.,
$O(|\U||\C|+|\E||\T||\U|+k|\E||\T|+ k|\E||\U| - k^2|\T|-k^2|\U|)$.

{Even though, in the worst case,  \bnd has the same complexity as \bsc.
However, in the \textit{best case}, \bnd does not perform any computations for assignment updates,  while in \bsc, in every case, the cost for updates is  $O(|\U|\sum_{i=0}^{k-2}|\E|-i-1)$.
Further, as \textit{average case} for \bnd, we can assume the case that the number of updates (as well as the cost) is  half compared to \bsc.
\RED{Note that, in several cases in our experiments,
 \bnd is more than three times faster than \bsc.}
Finally, the space complexity (with out considering the input data) is $O(|\E||\T|+|\T|)$.}
}

 \begin{figure}[t]
    \vspace*{-8pt}
\center
\tiny
\begin{tabular}[t]{c ccc ccc cc}
& $t_1$  & $t_2$ & &$t_1$ & $t_2$ & & $t_1$ & $t_2$\\ \cmidrule(lr){2-2} \cmidrule(lr){3-3} \cmidrule(lr){5-5}\cmidrule(lr){6-6} \cmidrule(lr){8-8} \cmidrule(lr){9-9}
													
 $e_1$ & \cellcolor{gray!30} 0.59  & \cellcolor{gray!30} 0.53 & \hspace{0pt} $\rightarrow$ \hspace{0pt}& \RED{{\textbf{0.59}}} &  \cancel{0.53} &\hspace{0pt} $\rightarrow$ \hspace{0pt}& \cellcolor{gray!30}   -- &\cellcolor{gray!30} -- \\ 	

$e_2$ & \cellcolor{gray!30} 0.52 &  \cellcolor{gray!30} 0.57 & & {0.52} &  0.57 && \cellcolor{gray!30} \xcancel{0.52}  &  \cellcolor{gray!30} \RED{{\textbf{0.16}}}\\

$e_3$ & \cellcolor{gray!30}  0.10 &  \cellcolor{gray!30}  0.09  &&{0.10}  & 0.09 && \cellcolor{gray!30} {0.05} & \cellcolor{gray!30}{0.03} \\ 																				

$e_4$ &  \cellcolor{gray!30}  \cancel{{0.64}} &  \cellcolor{gray!30}  \RED{\textbf{0.66}} & & -- & --& &\cellcolor{gray!30}  --  & \cellcolor{gray!30}-- \\
\specialrule{0.5 pt}{2pt}{3pt}

\rowcolor{gray!30}																				

 \scriptsize{\hspace{-4pt}\textbf{Select:}}\hspace{4pt}&\multicolumn{2}{c}{ {\small\ding{172}} \hspace{1pt} $\alpha_{e_4}^{t_2}$} & \hspace{1pt} $\rightarrow$ \hspace{1pt}&\multicolumn{2}{c} {{\small\ding{173}} \hspace{1pt} $\alpha_{e_1}^{t_1}$} & \hspace{1pt} $\rightarrow$ \hspace{1pt}&\multicolumn{2}{c} {{\small\ding{174}} \hspace{1pt} $\alpha_{e_2}^{t_2}$}
\vspace{3pt}\\
%

\scriptsize{\textbf{Update:}}& & &--  &\multicolumn{2}{c} {}&\hspace*{-19pt} $\alpha_{e_3}^{t_1}$  $\alpha_{e_2}^{t_2}$ $\alpha_{e_3}^{t_2}$ \hspace*{-19pt}& &\\

\end{tabular}
\vspace{-5pt}
\caption{\hor algorithm example} 
 \label{fig:horEx}
 \vspace{5pt}
 \end{figure}

 \subsection{Horizontal Assignment Algorithm  \textnormal{{({\hor})}}}
 \label{sec:hor}
%
 
In this section we propose the  \textit{Horizontal Assignment algorithm} 
(\hor), which in general, is more efficient than the \bsc and \bnd algorithms and in
most cases in practice provides same solutions.
The goal of \hor is twofold. 
First, to reduce the number of updates by  performing only a part of the required updates; 
and, at the same time,  minimize the impact of  not regular updates,  in the solution quality. 
In \hor both of these issues are realized by the  policy that are employed to select the assignments. 
In our experiments, in several cases \hor \textit{is around 5 and 3 times faster} than \bsc and \bnd, respectively. Also, in more than   $70 \%$ of our experiments, \hor reports the same solution as  \bsc, while in the rest cases the difference is marginal.


\vspace*{1pt}
\stitle{Horizontal Selection Policy.}
The key idea of \hor is that it adopts a  selection policy, named 
\textit{horizontal selection policy}, that selects assignments in a ``horizontal'' fashion. 
In this policy, in each  {iteration} the algorithm selects a set of assignments consisting of one assignment from each interval. 
Particularly, the top assignment from each interval is selected.  
This way, essentially, a layer  of assignments is generated in each iteration. 
For example, consider the scenario where $k> |\T|$
(and the assignments are feasible in all cases). 
In the first iteration, \hor will assign one event in each interval;
equally, in the $n^{th}$  {iteration}, $n$  events will have been  assigned in each interval.%
\footnote{With an exception in the last iteration $l$,  in which if  $k \Mod |\T| \neq 0$,
then, ${|\T|-(k \Mod |\T|)}$ intervals will have $l-1$ events.}

This policy allows \hor to  avoid performing updates after each assignment selection.
This holds, since, in each iteration at most one assignment per interval is selected.
Thus, during an iteration, when an assignment is selected for an interval, the algorithm
stops examining the selection of further assignments for this interval.
As a result, there is no need to perform any updates until the end of each iteration, 
where the scores for all the assignments have to  recomputed.
%



 In what follows we outline the  intuition behind the horizontal selection policy.
Considering that the users' attendance is shared between the events 
that take place during the same or overlapping time intervals. 
The horizontal  policy assigns the same number of events to each
interval, ignoring the possibility that it may be preferable to assign a 
different number of events to some intervals.


    \begin{myExample} [\myFontP{\textit{\hor Algorithm}}]
\label{ex:hor}
 Figure~\ref{fig:horEx} outlines the execution of  the  \hor  algorithm, 
presenting assignments following an interval-based organization.
Initially, \hor selects the assignment with the largest score.  
Since the first selected assignment refers to $t_2$, 
in the next selection, \hor will select the top assignment from $t_1$. 
After selecting assignments from both intervals, 
\hor has to update all the available assignments in order to perform the third selection.
 Therefore, \hor performs three updates, whereas it finds the same
  schedule as \mbox{\bsc  /\ \bnd}.
  \end{myExample}

 \vspace{-5pt}


%
%
%
%
%
%
%
%
%
%


\stitle{Algorithm Description.}
Algorithm~\ref{algo:hor} presents the pseudocode of \hor.
Note that, since the horizontal selection policy performs selections at the interval-level, 
we implement interval-based assignment organization.
Finally, similarly to \bnd, \hor  uses the $|\T|$   lists $\L_i$ and the  list $\M$.
%
 %
At the beginning of each iteration the algorithm calculates the scores  for all possible assignments (\textit{loop in line}~4) and initializes $\M$ (\textit{line}~8). 
%
In the next phase (\textit{line}~9), 
the algorithm selects the assignments  based on $\M$.
Particularly, in each step, the top valid assignment from $\M$ is selected. 
%



\begin{algorithm2e}[t]
\scriptsize
\setstretch{0}
\SetInd{0.4em}{1em}
\caption{\hor ($k$, $\T$, $\E$, $\C$, $\U$)}
\label{algo:hor}
\KwIn{
$k$: number of scheduled events;  \: 
$\T$:  time intervals;  \linebreak
$\E$: candidate events;  \:  
$\C$: competing events;  \:  
$\U$: users;
}
\KwOut{$\S$: feasible schedule containing $k$ assignments}
 \KwVar{$\L_i$: assignment lists for interval $i$;  \: \:
$\M$:  top assignments list 
}
\vspace{2pt}
 
 $\S \gets \varnothing;$\:\: 
$\M \gets \varnothing;$\:\:  

\While{$|\S|<k$ }{

$\L_i \gets \varnothing;$ \: \:  $1 \leq i \leq |\T|$

 \ForEach(\Comment*[f]{\mycomment{{generate assignments}}})
 {$(e, t) \in \{\E \backslash \E(\S)\} \times \T$}{

 	\If{$\alpha_{e}^{t}$  \text{is  \myvalid} }{ 
 	
 	compute $\alpha_e^t.S$;
  \Comment*[r]{\mycomment{{{by Eq.\ \ref{eq:gain}}}}}

  insert  $\alpha_e^t$ into $\L_t$
 \Comment*[r]{\mycomment{{{initialize assignment  lists}}}}
 
\scalebox{0.9}[1]{ $\M_{[t]} \gets  \maxAss(\M_{[t]},  \alpha_e^t)$} \hspace{0pt}
\scalebox{0.78}[1]{\Comment*[r]{\mycomment{{{insert into  $\M$ the top assignment from each interval}}}}}
}

}




\While(\Comment*[f]{\mycomment{{select assignments from $\M$}}})
{$\M \neq \varnothing$ \normalfont{\textbf{and}} $|\S|<k$}{

$\alpha_{e_p}^{t_p}$ $\gets$ $\popAss(\M)$  


\uIf{$e_p \notin \S$ }{

		insert $\alpha_{e_p}^{t_p}$ into $\S$
		  \Comment*[r]{\mycomment{{{insert into schedule}}}}

}

\Else(\Comment*[f]{\mycomment{{assignment is invalid; select the top valid from  $\L_{t_p}$   and insert it into $\M$}}}){




insert the  top  assignment $\alpha_{e_i}^{t_p}$ from $\L_{t_p}$ into $\M$,  s.t.\ $e_i \notin \S$


}
	}
}
 \Return $\S$\;
\end{algorithm2e}


\vspace{0pt}

\subsubsection{\hor Algorithm Analysis}
\label{sec:horAnal}
\hfill \break

\vspace{-11pt}
 
 \stitle{\bsc  vs.\ \hor  Score Computations Analysis.}
Here we study the number of score computations, comparing the \bsc and the \hor algorithms. 
The following proposition specifies the cases where \hor performs less score computations than \bsc.


\begin{myproposition}
\label{prop:scoresComp}
\hor performs less score computations than \bsc when  
$k\leq |\T|$  or $|\E|<\frac{k}{2}(3|\T|+1)$.
\end{myproposition}

 \vspace*{-2pt}

\begin{myproofSk}
 In case that $k\leq |\T|$, \hor computes only the scores for the initial assignments (i.e., $|\T||\E|$) without performing any updates.
In case that $k > |\T|$, \hor computes the same initial assignments,  as well as
the scores for  $\sum_{i=0}^{(k/|\T|)-1}  |\T| (|\E| - i|\T|-|\T|)$ updates.
On the other hand, in the \bsc algorithm, in any case, we have to compute the same number of initial assignments as in \hor, as well as 
$k|\E|+ \frac{k}{2}-\frac{k^2}{2}  -|\E|$  updates.~~~~~~~\qed
 \end{myproofSk}

From Proposition~\ref{prop:scoresComp}  we can infer that in ``rational/typical'' (i.e.,~real-world) scenarios, \hor perform fewer computations than \bsc. 
Particularly,  even in cases that $k>|\T|$, it should also hold that 
\linebreak 
 ${|\E|\geq \frac{k}{2}(3|\T|+1)}$ in order for \hor to perform more computations. 
Considering the setting of our problem,  
the second relation   is difficult to hold in   practice. 
For example, consider the scenario where $|\T|=10$ and $k =20$. 
Then, in order for \hor to perform more computations, it should hold
 that $|\E|\geq 301$, 
which   seems unrealistic due to the noticeable difference between  the number of scheduled ($k=10$) and candidate events ($|\E|\geq 301$).



\stitle{Worst Case w.r.t.\ $k$ and $|\T|$.}
Considering the horizontal selection policy, beyond the size of the input (e.g., $|\E|$, $|\U|$, $k$), the number of computations in \hor
is also influenced  by the ratio between parameters $k$ and $|\T|$.
During the execution, \hor performs $\lceil k / |\T|\rceil$ iterations. 
At the beginning of each iteration,   
it computes the scores according to which  $|\T|$  assignments are selected.
In the last iteration, if  $k \Mod |\T| \neq 0$, then only 
$k \Mod |\T|$ assignments  need to be selected, while the algorithm
has already performed the computations that are required to select $|\T|$ assignments.
Thus, in this case,   \hor has performed  more 
computations than the ones required for selecting its assignments.
For example, assume that we have $|\T| = 10$ and $k=11$. 
In this case, the score computations performed by  \hor are the same as in the case that we have
$k=20$. 

The case in which the difference between the number of 
computed assignment selections 
 and the number of the selections  that need to 
be performed is maximized,  is referred as  \textit{worst case w.r.t.}\  $k$ \textit{and} $T$.
 Note that, even in the worst case, in our experiments, \hor outperforms \bnd in several cases.



%

\vspace{2pt}
\begin{myproposition}
\label{prop:worstcase}
In \hor, the  worst case w.r.t.\ $k$ and $|\T|$  occurs when
 $k >|T|$ and   $k \Mod |\T| = 1$.
\end{myproposition}

%
%
%

\vspace{3pt}
 \stitle{Complexity Analysis.}
%
In the first while loop (\textit{line}~2), \hor performs $\lceil k / |\T|\rceil$  
iterations.
In each iteration, in the worst case, there are
\linebreak \scalebox{1}[1]{${|\E| - i|\T|}$} available events, where 
$0\leq  i\leq  \frac{k}{|\T|}$. 
Thus, in each iteration, it computes $|\T| (|\E| - i|\T|)$ assignments (\textit{line}~6).
The overall cost for computing the assignments is  
\scalebox{1}[1]{$O(|\U|\sum_{i=0}^{k/|\T|}  |\T| (|\E| -i|\T|))$}.%
\alt{}
{Note that this cost is defined when $k>|\T|$. 
In case that ${k\leq |\T|}$, there are not assignment updates; 
so, the overall cost computing the assignments is  $O(|\U||\T||\E|)$.}
Further,  in each iteration of the first  loop (\textit{line}~2),  
 the nested loop (\textit{line}~9)  performs $|\T|$ iterations (referred as nested iterations).
In each nested iteration, in the worst case, 
  \hor performs  two $\popAss$ operations (\textit{lines}~10~\&~14).
The cost for the first and the second $\popAss$ operations 
is $O(|\T|)$ and ${O(|\E|-i|\T|)}$, respectively. 
Hence, in sum, the cost for $\popAss$ operations is \linebreak
$O( \sum_{i=0}^{k/|\T|} |\T| ( |\T| + |\E| - i |\T|))$.
The worst case can occur when all the assignments contained in $\M$ refer to the same event.

Thus, the overall cost of \hor is 

\alt{$O( |\U||\C|+|\E| |\T||\U| +   k|\E||\U|+  |\T|^2 - k|\T||\U|-k^2|\U|)$.}
{\\ $O(|\U||\C|) + $
$O(|\U|\sum_{i=0}^{k/|\T|}  |\T| (|\E| - i|\T|))+$ \\
${O( \sum_{i=0}^{k/|\T|} |\T| ( |\T| + |\E| - i |\T|)) }= $ \\
$O( |\U||\C|+|\E| |\T||\U| +   k|\E||\U|+  |\T|^2 - k|\T||\U|-k^2|\U|)$.}
Finally, the space complexity   is $O(|\E||\T|+|\T|)$.

\alt{ 
}
{Note that, since the \hor algorithm does not perform updates in the scores of the assignments.
We can use heaps instead of lists in the $\L$ and $\M$ structures,  
and replace the  {$insert$} operation (\textit{lines} 7~\&~14) with $push$. 
In this case, assuming a Fibonacci heap, the $push$ and $pop$ (\textit{line}~10~\&~14) operations,  can be performed in  $O(1)$ amortized time.
Hence, in a heap-based implementation, the overall (amortized) cost of \hor  is 
$O(|\U||\C|) +O(|\U|\sum_{i=0}^{k/|\T|}  |\T| (|\E| - i|\T|))+$ 
${O( \sum_{i=0}^{k/|\T|} |\T|)}=$\\
$O{( |\U||\C| +|\E| |\T||\U| + k|\E||\U|- k|\T||\U|-k^2|\U|)}$.
Note that to ensure that \hor is not favored compared to the other methods, 
in our experiments we use a list-based \hor implementation. 
}

\vspace{-1pt}

 \subsection{Horizontal Assignment with Incremental Updating Algorithm  \textnormal{{(\horb)}}}
\label{sec:horb}
\vspace{-0pt}

 This section introduces the \textit{Horizontal Assignment with Incremental Updating algorithm}  (\horb).
\horb combines the basic concepts from the \bnd and \hor algorithms, in order to further reduce the computations performed by \hor. 
Particularly,  \horb adopts {an incremental updating scheme}, similar to \bnd (Sect.~\ref{sec:bscheme}), 
as well as the horizontal selection policy employed by \hor (Sect.~\ref{sec:hor}). 
Note that, in several cases, in our  experiment \horb \textit{performs about   half computations and is up to two times faster compared to} \hor. 


Recall that,  at the beginning of each iteration, \hor calculates the scores for all available assignments. 
Particularly, in the first iteration, the algorithm generates the assignments and calculates their (initial) scores, 
while in each of the following iterations the scores for all the   assignments are updated.
On the other hand, after the first iteration, \horb  instead of updating all the assignments, 
 uses   an incremental updating scheme.
This way, in each iteration, a reduced number of updates are performed.
%


Note that  since the  updates are performed after the first iteration, 
it is obvious that \horb is  identical to \hor in cases where only one iteration is required (i.e., $k \leq|\T|$).
\alt{}{In general, more than one iteration is performed when $k>|\T|$, 
assuming that there is at least one valid assignment for each interval.
If the latter does not hold,  in all cases, more than one iteration is performed.}

 \begin{algorithm2e}[t]
\setstretch{0}
\scriptsize
\SetInd{0.8em}{1em}
\caption{\horb ($k$, $\T$, $\E$, $\C$, $\U$)}
\label{algo:horb}
\KwIn{
$k$: number of scheduled events;  \: 
$\T$:  time intervals;  \linebreak
$\E$: candidate events;  \:  
$\C$: competing events;  \:  
$\U$: users;
}
\KwOut{$\S$: feasible schedule containing $k$ assignments}
\KwVar{$\L_i$: assignment lists for interval $i$;  \: \:  $\Phi$:    bound; 
$\M$:  top assignments list }

\vspace{1mm}
 
 $\S \gets \varnothing;$\:\:
$\M \gets \varnothing;$\:\:  
$\L_i \gets \varnothing$ \:  $1 \leq i \leq |\T|$

\While{$|\S|<k$ }{

\uIf(\Comment*[f]{\mycomment{{first iteration}}}){$\S= \varnothing$}{

 \ForEach(\Comment*[f]{\mycomment{{generate assignments}}})
 {$(e, t) \in \{\E \backslash \E(\S)\} \times \T$}{

 compute $\alpha_e^t.S$;  \:  \:  $\alpha_e^t.U$ $\gets$  \myupdated;
 \Comment*[r]{\mycomment{{{by Eq.\ \ref{eq:gain}}}}}

  insert  $\alpha_e^t$ into $\L_t$
 \Comment*[r]{\mycomment{{{initialize assignment  lists}}}}
 
 $\M_{[t]} \gets  \maxAss(\M_{[t]},  \alpha_e^t)$ \hspace{-0pt}

}
}\Else(\Comment*[f]{\mycomment{{incremental assignments updating}}}){

\For{$i=1$ \emph{\KwTo} $|\T|$}{

	$\Phi \gets 0$   \Comment*[r]{\mycomment{{{initialize   bound}}}}
  
 \ForEach{$\alpha_{e}^{t} \in \L_i$}{


 \uIf{$\alpha_{e}^{t}$  is \myvalid}{

 \uIf{$\alpha_{e}^{t}.S \geq \Phi$ }{

	compute new $\alpha_{e}^{t}.S;$ \:  \: $\alpha_{e}^{t}.U \gets \myupdated$; 

	$\Phi \gets \maxAss(\Phi, \alpha_{e}^{t}.S)$ 
	\Comment*[r]{\mycomment{{{update   bound}}}}

}\Else{
	$\alpha_{e}^{t}.U \gets \mynotupdated$
\Comment*[r]{\mycomment{{{partially updated}}}}
}

}\Else{
remove $\alpha_{e}^{t}$ from $\L_i$
}

}
$\M_{[i]} \gets \Phi$ \Comment*[r]{\mycomment{{{update top assignment}}}}

}
  

}

\While(\Comment*[f]{\mycomment{{select assignments from $\M$}}}){$\M \neq \varnothing$ \normalfont{\textbf{and}} $|\S|<k$ }{

$\alpha_{e_p}^{t_p}$ $\gets$ $\popAss(\M)$

 	\uIf{$e_p \notin  \S$}{

		insert $\alpha_{e_p}^{t_p}$ into $\S$
		  \Comment*[r]{\mycomment{{{insert into schedule}}}}

}

\Else(\Comment*[f]{\mycomment{{select the top, valid \& updated from  $\L_{t_p}$ and insert it into $\M$}}}){
 	 $\alpha_{e_p}^{t_p} \gets$ top \& updated assignment from  $\L_{t_p}$, s.t.\ $e_p \notin \S$
		 
\If{$\alpha_{e_p}^{t_p} = \varnothing$ \Logand $\exists \alpha_{e}^{t} \in \L_{t_p}$ s.t.\ $\alpha_{e}^{t}$  is \myvalid}{
\vspace{2pt}
\hspace{6pt}	 {$\cdot$}			  \Comment*[r]{\mycomment{{{incremental updates in interval $p$}}}} 
\hspace{6pt}	$\cdot$ \:  {\myFontP{\textit{Same as lines 10 to 20, with}}} $i=p$\vspace{0pt} \;
\hspace{6pt}	$\cdot$\; 

	}

}

}
}
 \Return $\S$\;
  \end{algorithm2e}


    \begin{myExample} [\myFontP{\textit{\horb Algorithm}}]
\label{ex:horb}
The difference between \horb and \hor example (Example~\ref{ex:hor}), 
  appears at the third selection,  where from $t_2$  only the $\alpha_{e_2}^{t_2}$ is updated. 
This happens  because after updating $\alpha_{e_2}^{t_2}$, its score (0.16) 
is the current bound for this interval. 
Then, when checking $\alpha_{e_3}^{t_2}$  for update, its score (0.09) 
is lower than the bound, so there is no need to update it. 
Hence, \horb performs   two of the three updates performed by \hor. 
\end{myExample}

  \vspace{-8pt}
\stitle{Algorithm Description.}
Algorithm~\ref{algo:horb} presents \horb; \horb 
uses the same structures as \hor, as well as a   bound $\Phi$.
 %
At the first iteration (\textit{loop in line}~4), as is the case with \hor, \horb generates the assignments and initializes $\M$. 
In the next iterations (\textit{loop~in~line}~9), it performs  incremental updates for each interval, 
determining a different bound $\Phi$ for each interval. 
Then, similarly to \hor, \horb  performs the assignment selection based on $\M$ (\textit{loop in line}~21).  
In contrast to \hor, \horb has also to examine the update status of the assignments.
In case that there is not a valid and updated assignment left 
on an interval (\textit{lines}~27-30),
\horb has to perform incremental  updates in  this interval.

 
\subsubsection{\horb Algorithm Analysis}  \label{sec:horbanal}
\hfill \break
 \vspace*{-9pt}

  \stitle{\horb Solution \& Worst Case w.r.t.\ $k$ and $|\T|$.}
The following propositions state that both  \horb and \hor
return the same solutions and also have the same worst case w.r.t.\ $k$ and $|\T|$.

\begin{myproposition}
\label{prop:horb}
\horb and \hor always return the same solution.
\end{myproposition}

 \begin{myproposition}
\label{prop:worstcasehorb}
In \horb, the  worst case w.r.t.\ $k$ and $|\T|$  occurs when
 $k >|T|$ and   $k \Mod |\T| = 1$.
\end{myproposition}

\vspace{-3pt}

%
%
%
%


\stitle{Complexity Analysis.}
In the worst case, the computation cost in \linebreak  \horb, 
is the same as   \hor.
Particularity, in the worst case,  the bound employed by \horb 
cannot prevent any of the assignment updates (\textit{line}~9). 
This case arises, when  the assignments in each interval list $\L_i$ 
are sorted in ascending order by its score, and there are no assignments having the same score.
Thus, the computation cost for \horb is 
\linebreak
$O( |\U||\C| +|\E| |\T||\U| + k|\E||\U|+ |\T|^2 - k|\T||\U|-k^2|\U|)$.
Note that, in the \textit{Best case} \horb does not perform any computations for updates,  while in \hor, in any case (where $k>|\T|$), the cost for updates is 
 $O(\sum_{i=0}^{(k/|\T|)-1}  |\U||\T| (|\E| - i|\T|-|\T|))$.%
\alt{}
{As an \textit{average case} for \horb, we can assume the case that cost of updates is  half compared to \hor.}
Finally, the space complexity   is  $O(|\E||\T|+|\T|)$.



  \section{Experimental Analysis}

 \label{sec:eval}


  \subsection{Setup}
 \label{sec:setup}


\stitle{Datasets.}
\alt{In our experimental evaluation we present the results from four datasets, \textit{two real} and \textit{two synthetic}.
The \textit{first real} is the  \textit{Meetup dataset} (\meetupc) from \cite{Pham2015},  which contains data from {California}, and is the dataset used in \cite{ses}.
We follow the same approach as in \cite{ses,She2015,She2015a,She2016,Tong2015},  
in order to define the interest of a user to an event. 
 %
After preprocessing, we have the  \meetupc dataset  
containing   $42$,$444$ users and about $16$K events.
}
{In our experimental analysis we present the results from four, two real and two synthetic.
The first real is the largest Meetup dataset (\meetupc) from \cite{Pham2015},  which contain data from  {California}. 
Note that, in Meetup, the users and the groups are associated with a set of tags. 
On the other hand, the events are not associated with tags; however, each event is organized by a group. 
Thus,  similarly to \cite{She2015,She2015a,Tong2015},  in order to define the interest of a user to an event,
we associate the events with the tags of the group who organize it.
Then, we compute the interest value using Jaccard similarity over the user-event tags. 
 Note that, the tags in adopted datasets are represented using code numbers.

During  the data preprocessing phase, we consider only users that are associated with tags, and  in cases where a user appears more than once in the data, we construct a ``merged'' version of this user, containing the union of the user's tags. 
After preprocessing, we have the  \meetupc dataset  
containing   $42$,$444$ users and about $16$K events. 
Note that, similar results are also reported in the second
\textit{Meetup} dataset from \cite{Pham2015} (referring  to New York).}

The next \textit{real dataset} (\music)  which is the largest, is related to music  and provided from Yahoo! (``Music user ratings of musical tracks, albums, artists and genres dataset'').
%
\music is used to demonstrate the scenario of music festival organization.
Particularly,  \music contains data for several music entities (i.e., tracks, albums, artists, genres), as well as ratings of users over these entities.%
\alt{}%
{
\music contains two tracks, here we consider all the sets (i.e., training, test, validation) from   Track1, which is the larger one.}
In this dataset, we consider albums to represent the events (i.e., music concerts).  
We select the albums that are associated with at least one genre, which results to $89$K albums.
Further, as users we select the users that have rated at least $10$ genres, 
which result to $379$,$391$ users.

In order to compute   user interest over the albums, 
we consider the users ratings over the music genres,  
as well as the genres that are associated with the music albums.
Let a user $u$,  $R_u$ denote the set of ratings $r_i$ over genres, 
where $r_i \in [0,1]$ is the  rating over the $i$ genre. 
Also, let $G_a$ be a set of genres associated with a music album $a$. 
Here, we define the interest  of a user $u$ over the album $a$ as 
$(\sum_{\forall g \in   G_a }   r_g )   /  |G_a|$, 
where  $r_g=1$ if the genre $g$ is not  specified in $R_u$. 
Note that, similar results are reported  using alternative methods, such as setting  
${r_g=0}$  for genres not  specified in $R_u$, or considering only the common user-album genres.

Finally, regarding \textit{synthetic datasets} (Table~\ref{tab:param}), we generate the users' \textit{interest values} for the  events, following the \textit{three distribution types} examined in the related literature 
\cite{Li2014,She2015,She2015a,She2016,Tong2015}: 
\textit{Uniform} (\uni), \textit{Normal} (\norm) and \textit{Zipfian} (\zip). 
Note that, for brevity, the results for the Normal distribution
are not presented here since they are similar to Uniform. Further for Zipfian,
we present   only the results with parameter equal to 2 which are similar
to those of 1 and 3.

\alt{
 
\begin{figure*}[t]
\centering
\vspace{-39pt}
\hspace*{-0.35cm}\mbox{
\subfloat[Utility (\meetupc) ]{\includegraphics[height=1.28in]{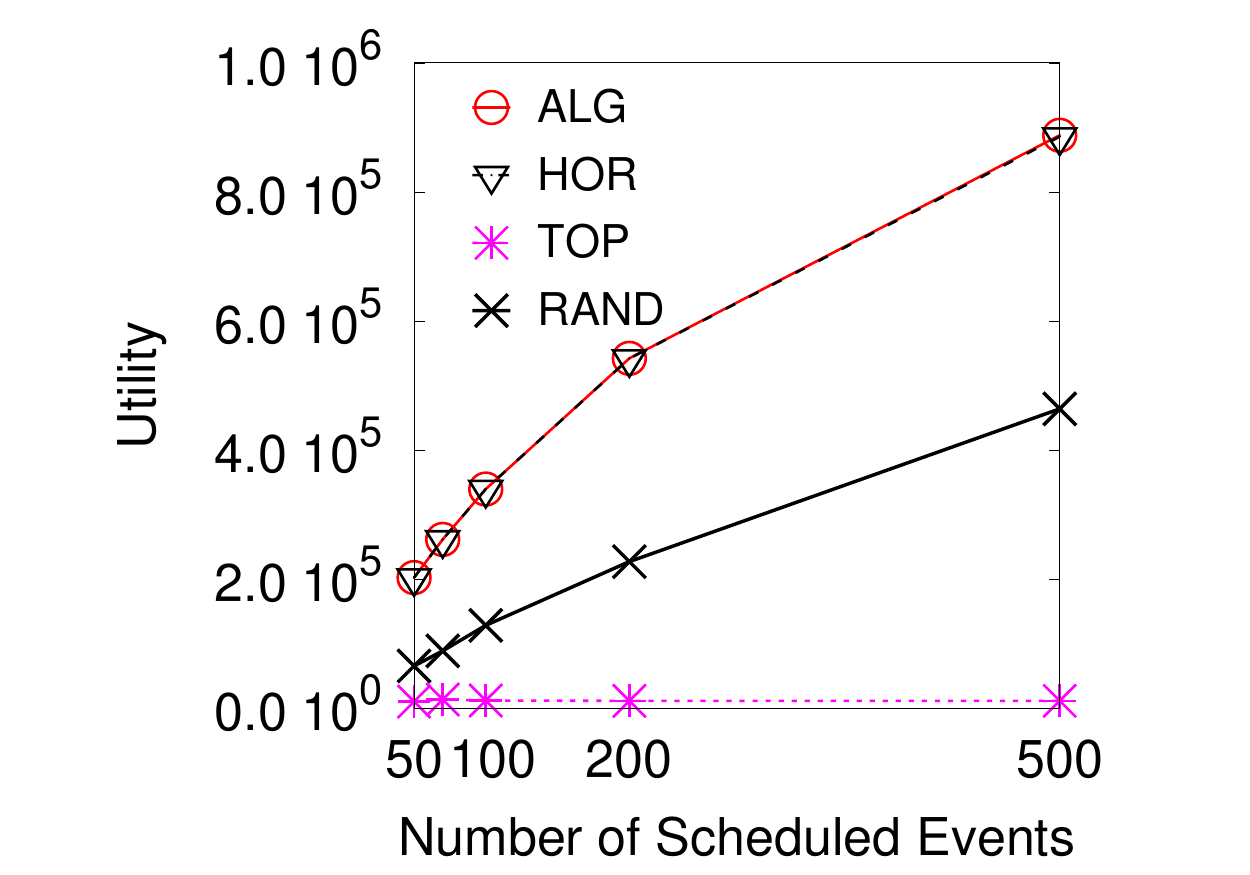}\label{fig:u_CA_k}}\hspace{-0.1cm}
\subfloat[Utility (\music) ]{\includegraphics[height=1.28in]{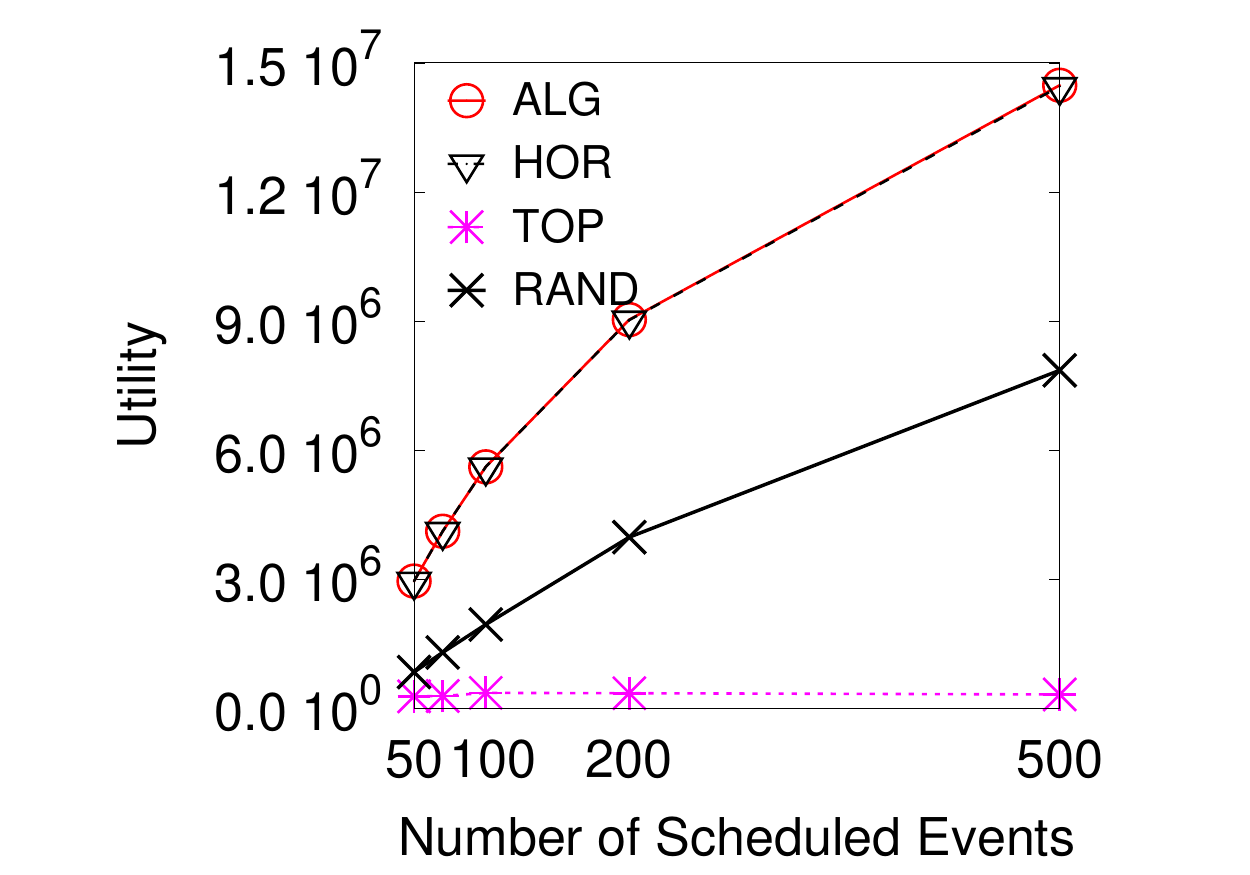}\label{fig:u_YA_k}}\hspace{-0.1cm}
\subfloat[Utility (\uni) ]{\includegraphics[height=1.28in]{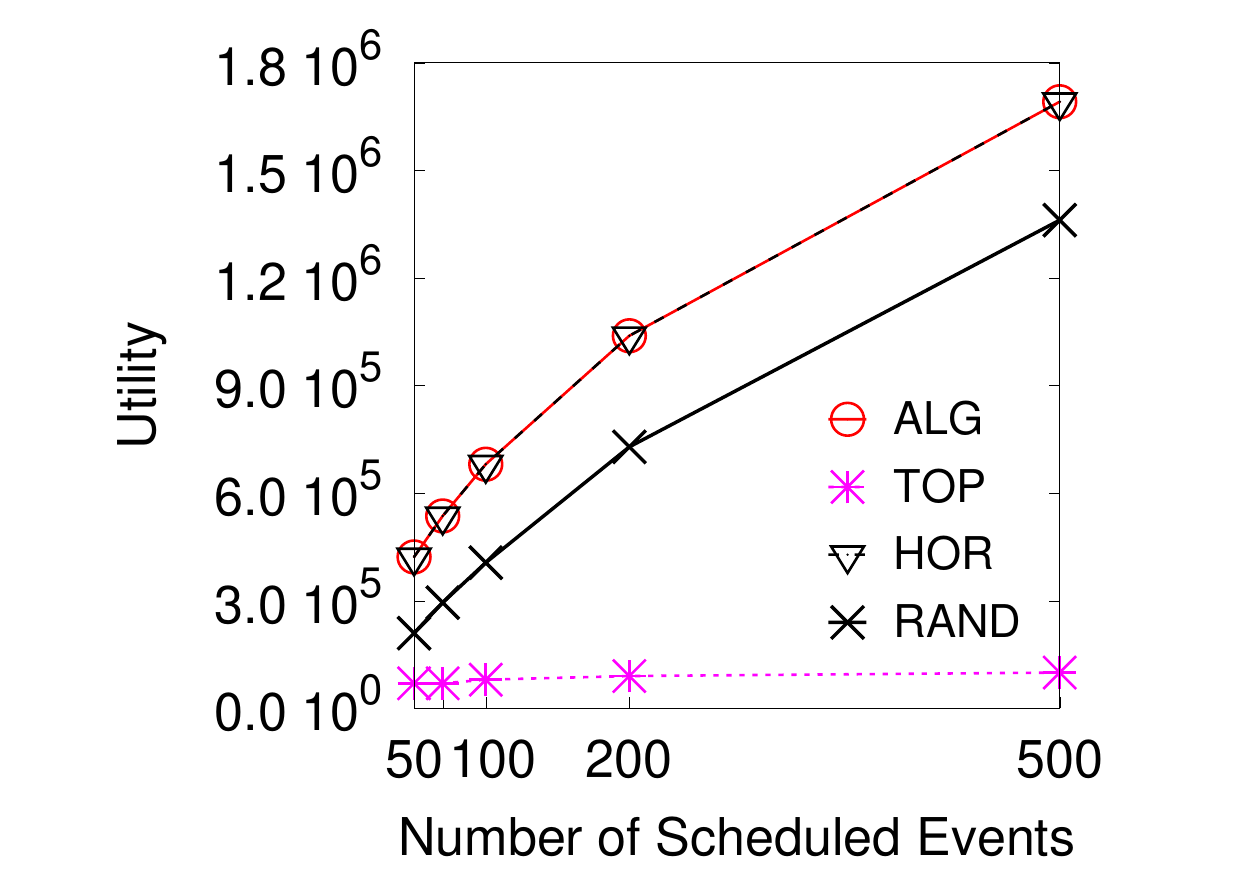}\label{fig:u_IN_k}}\hspace{-0.1cm}
\subfloat[Utility (\zip) ]{\includegraphics[height=1.28in]{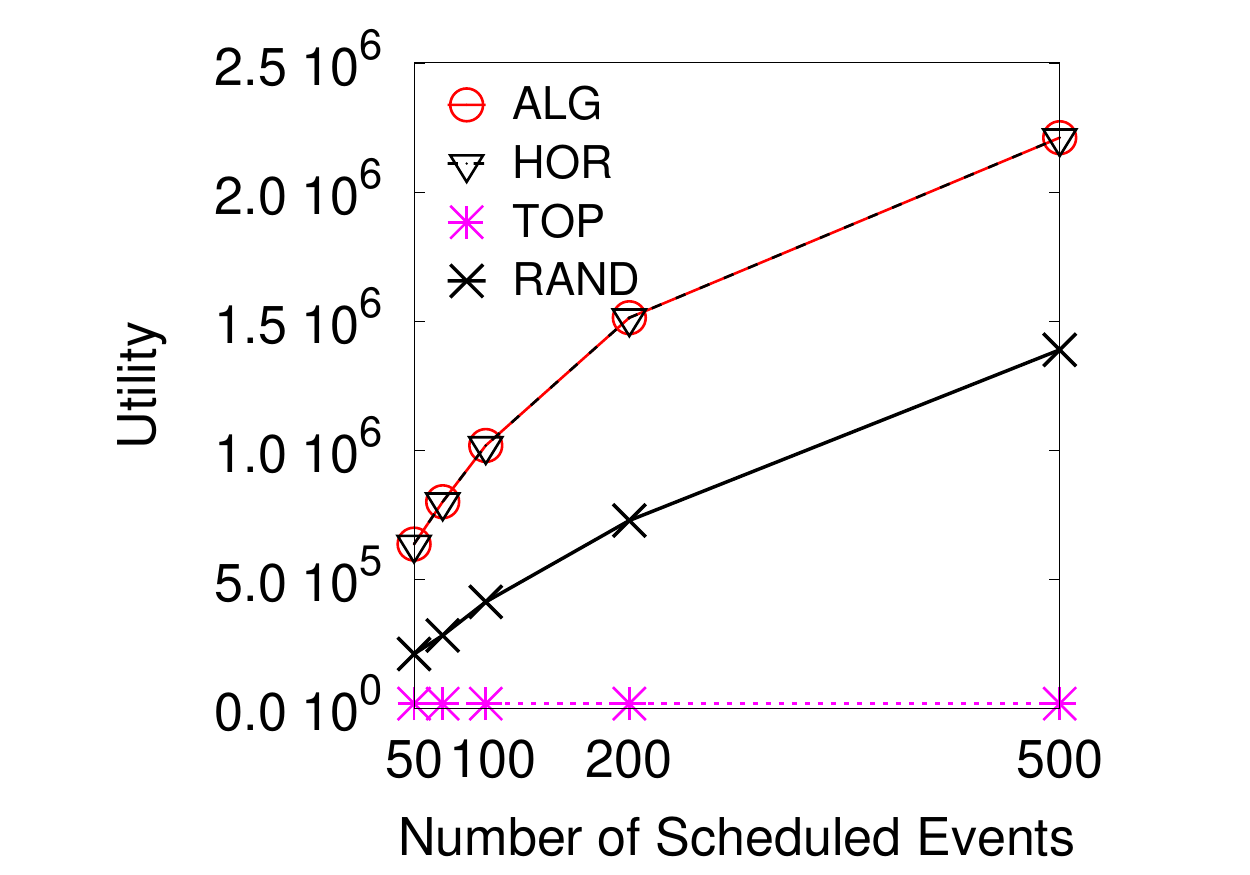}\label{fig:u_ZP_k}}\hspace{0cm}}
\\ \vspace{-10pt}
\hspace*{-0.40cm}\mbox{
\subfloat[Computations  (\meetupc) ]{\includegraphics[height=1.28in]{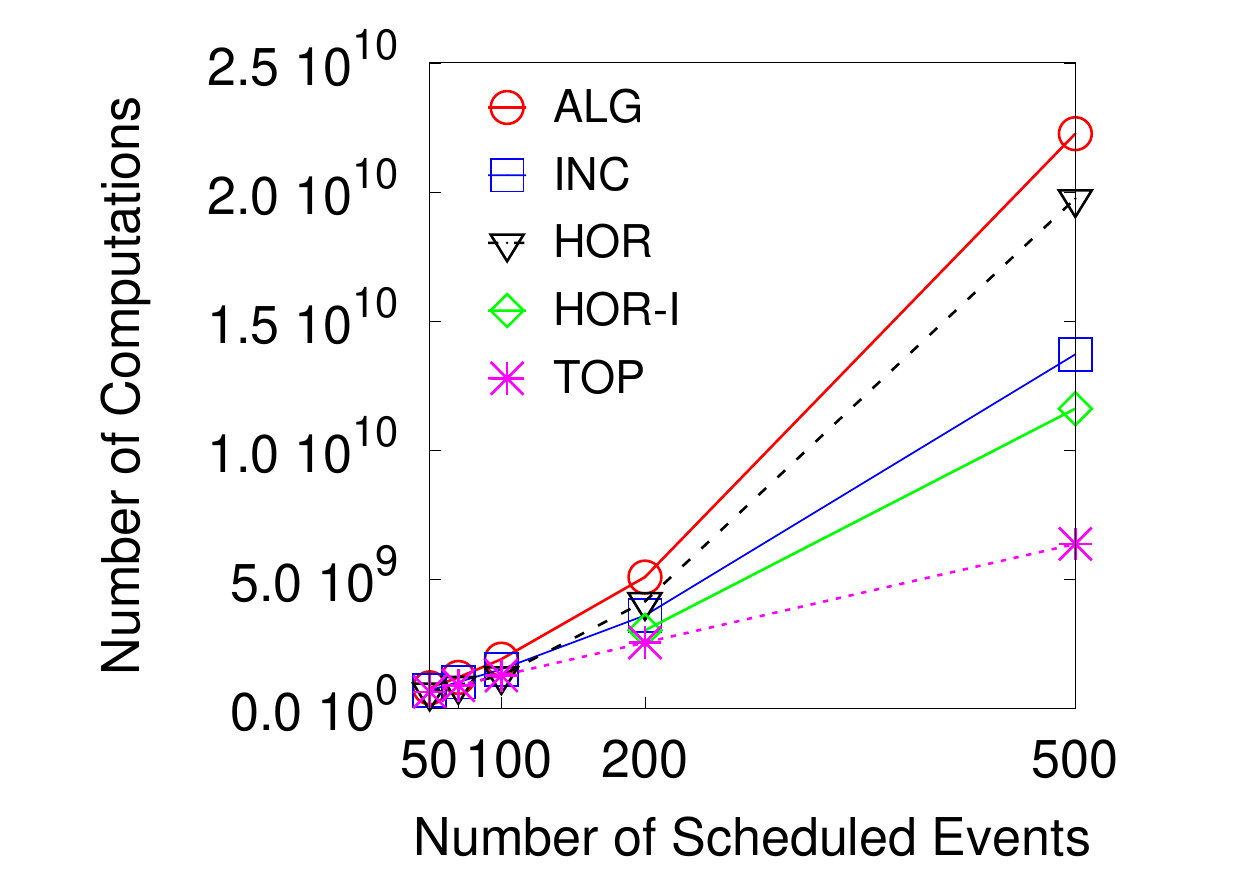}\label{fig:c_CA_k}}\hspace{-0.1cm}
\subfloat[Computations  (\music) ]{\includegraphics[height=1.28in]{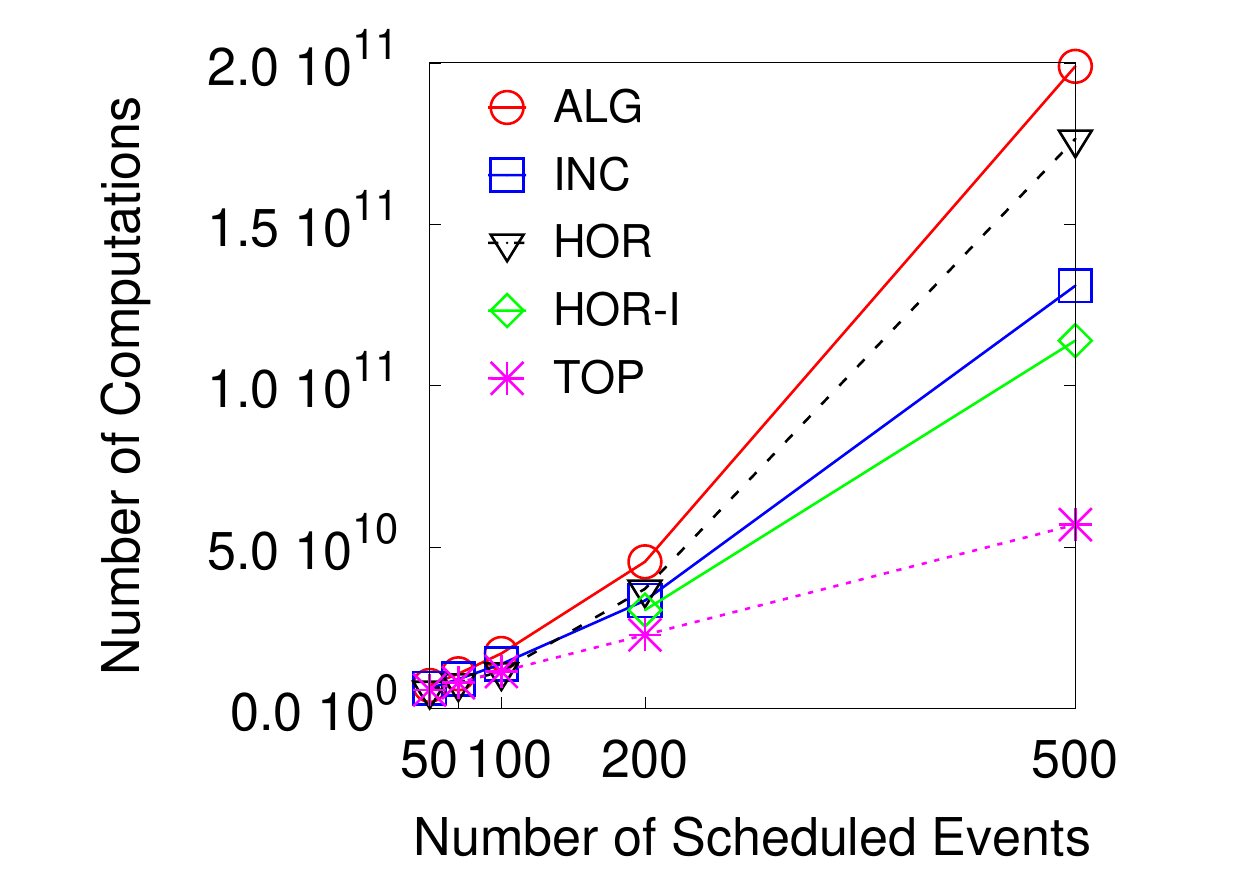}\label{fig:c_YA_k}}\hspace{-0.1cm}
\subfloat[Computations (\uni) ]{\includegraphics[height=1.28in]{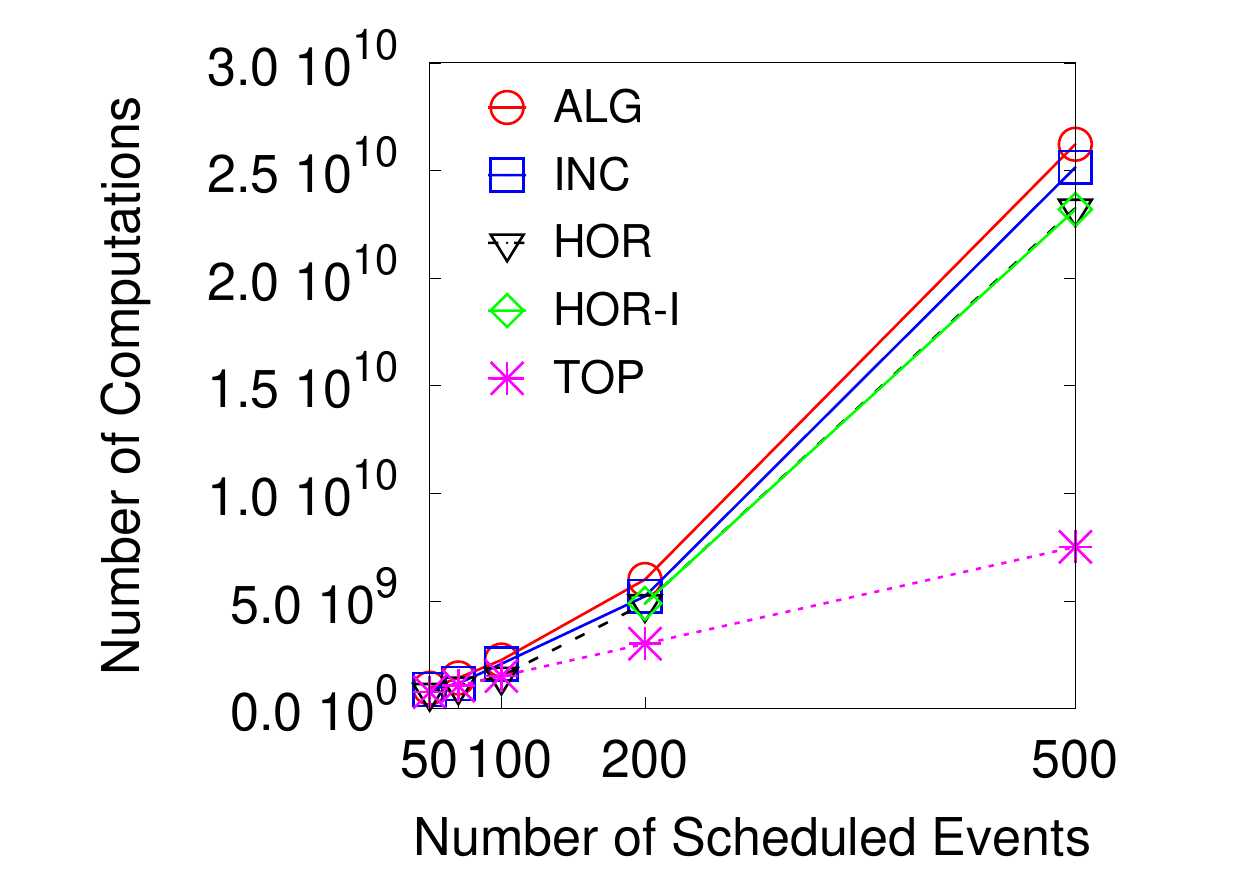}\label{fig:c_IN_k}}\hspace{-0.1cm}
\subfloat[Computations  (\zip)]{\includegraphics[height=1.28in]{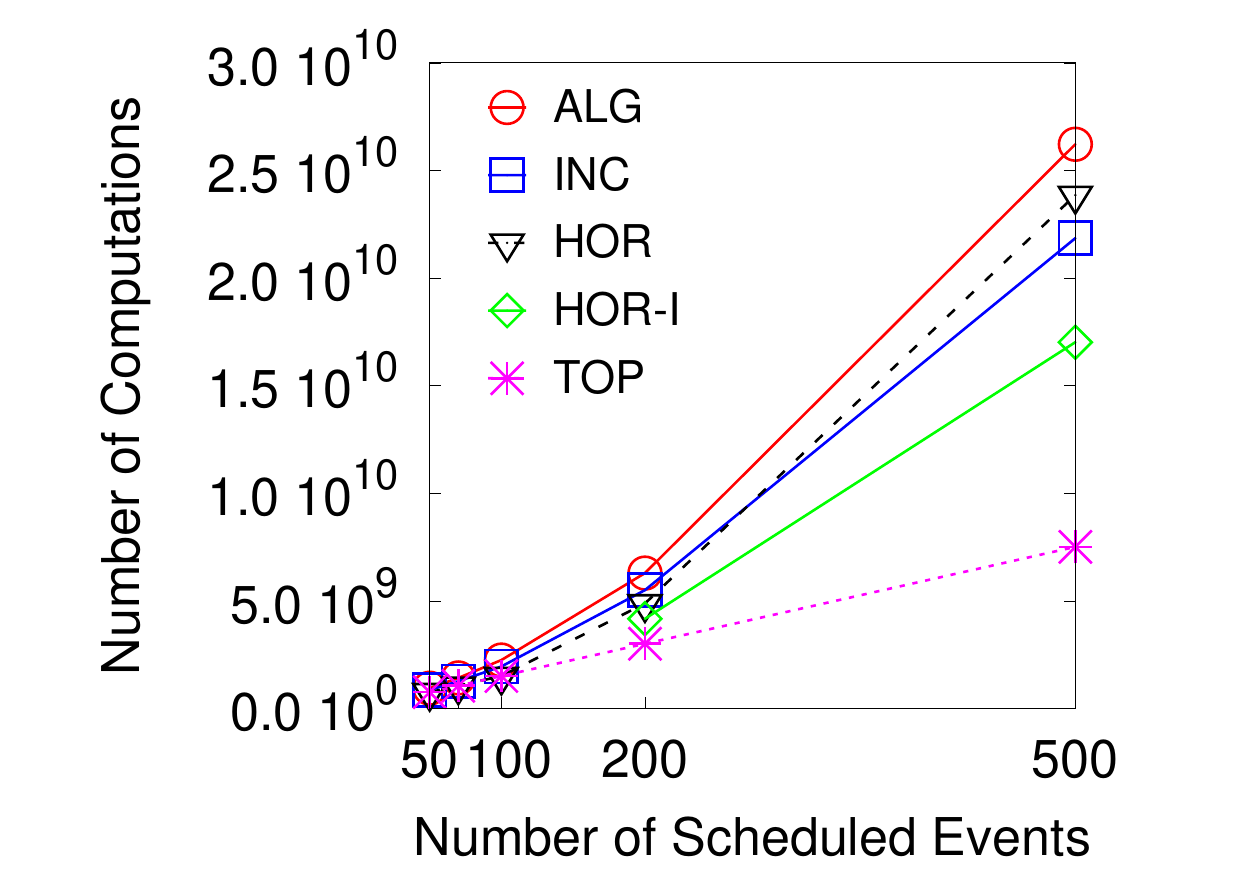}\label{fig:c_ZP_k}}\hspace{0cm}}
\\ \vspace{-10pt}
\hspace*{-0.35cm}\mbox{
\subfloat[Time (\meetupc) ]{\includegraphics[height=1.28in]{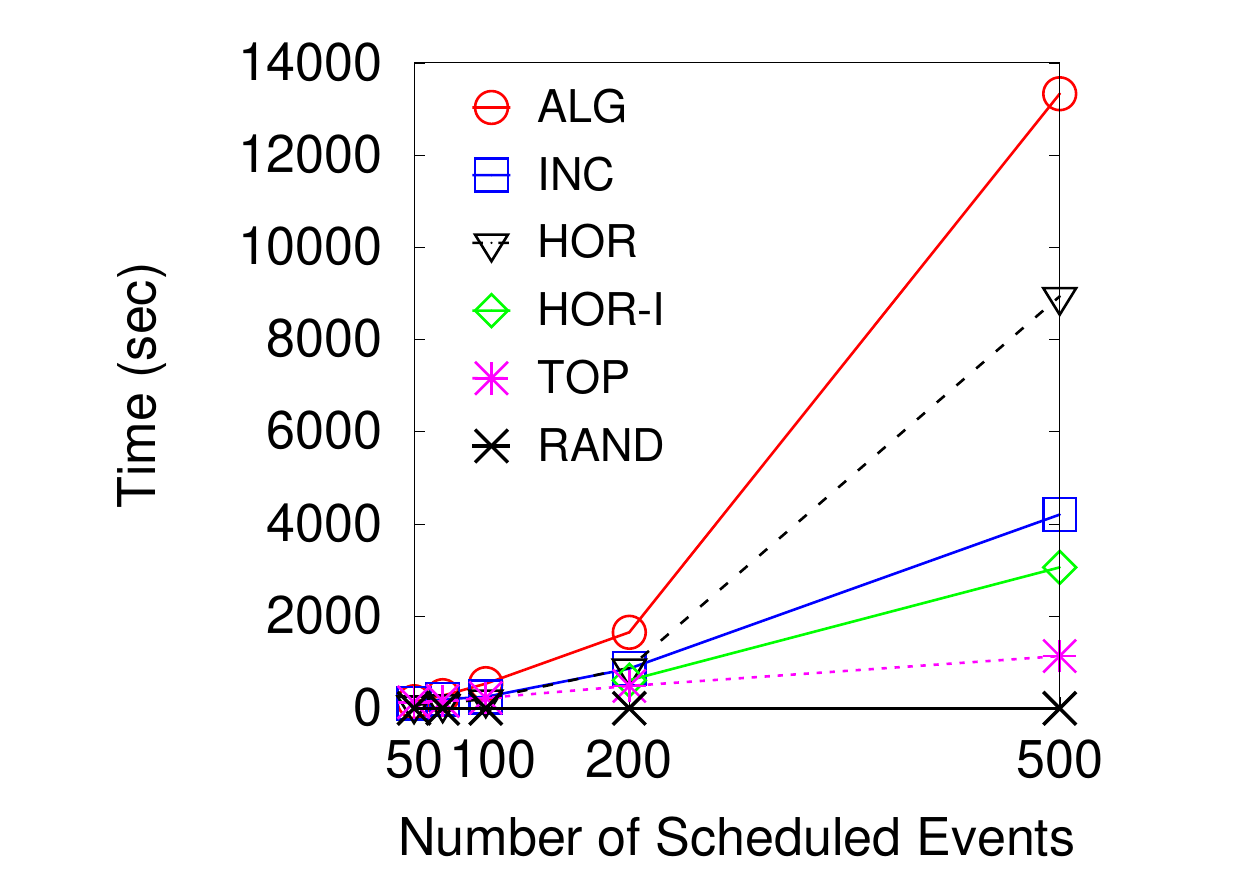}\label{fig:t_CA_k}}\hspace{-0.1cm}
\subfloat[Time (\music) ]{\includegraphics[height=1.28in]{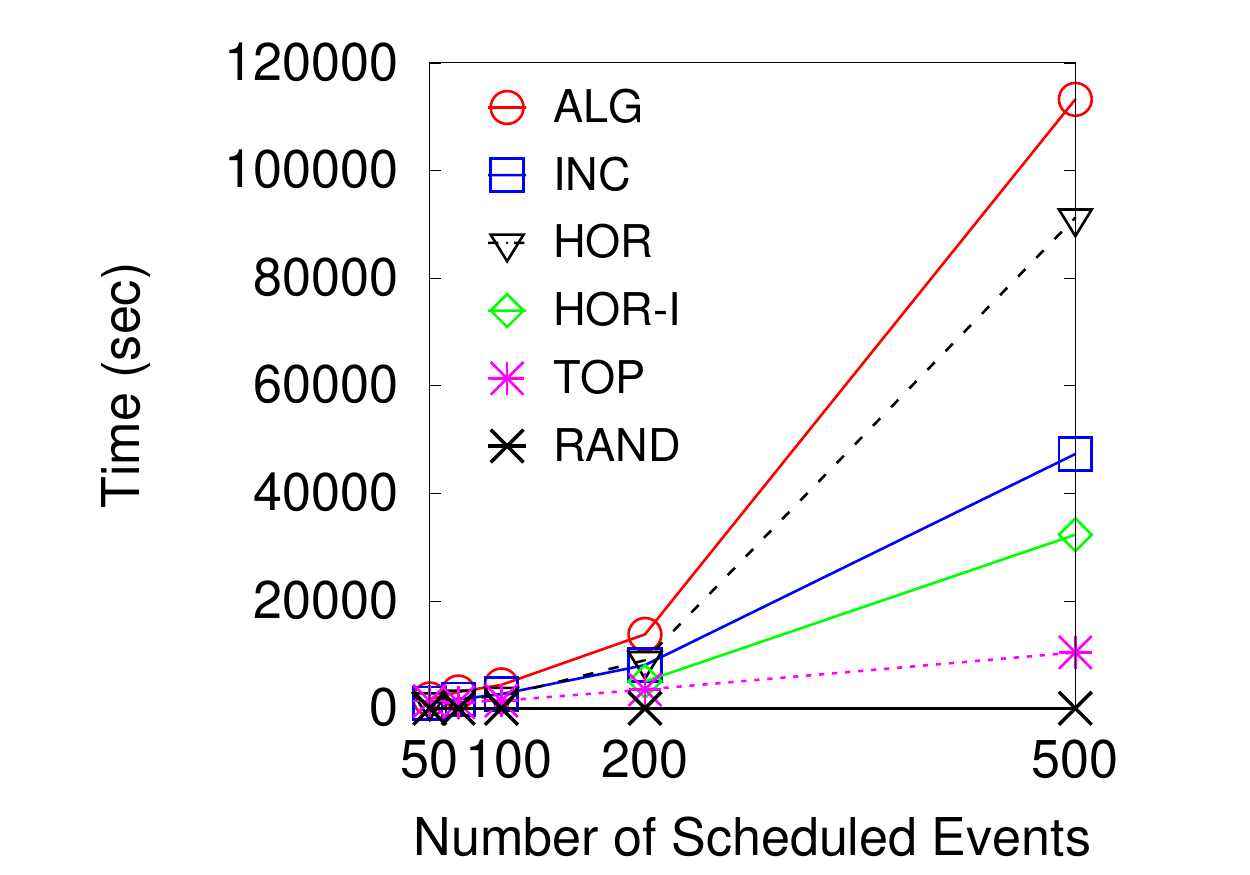}\label{fig:t_YA_k}}\hspace{-0.1cm}
\subfloat[Time (\uni) ]{\includegraphics[height=1.28in]{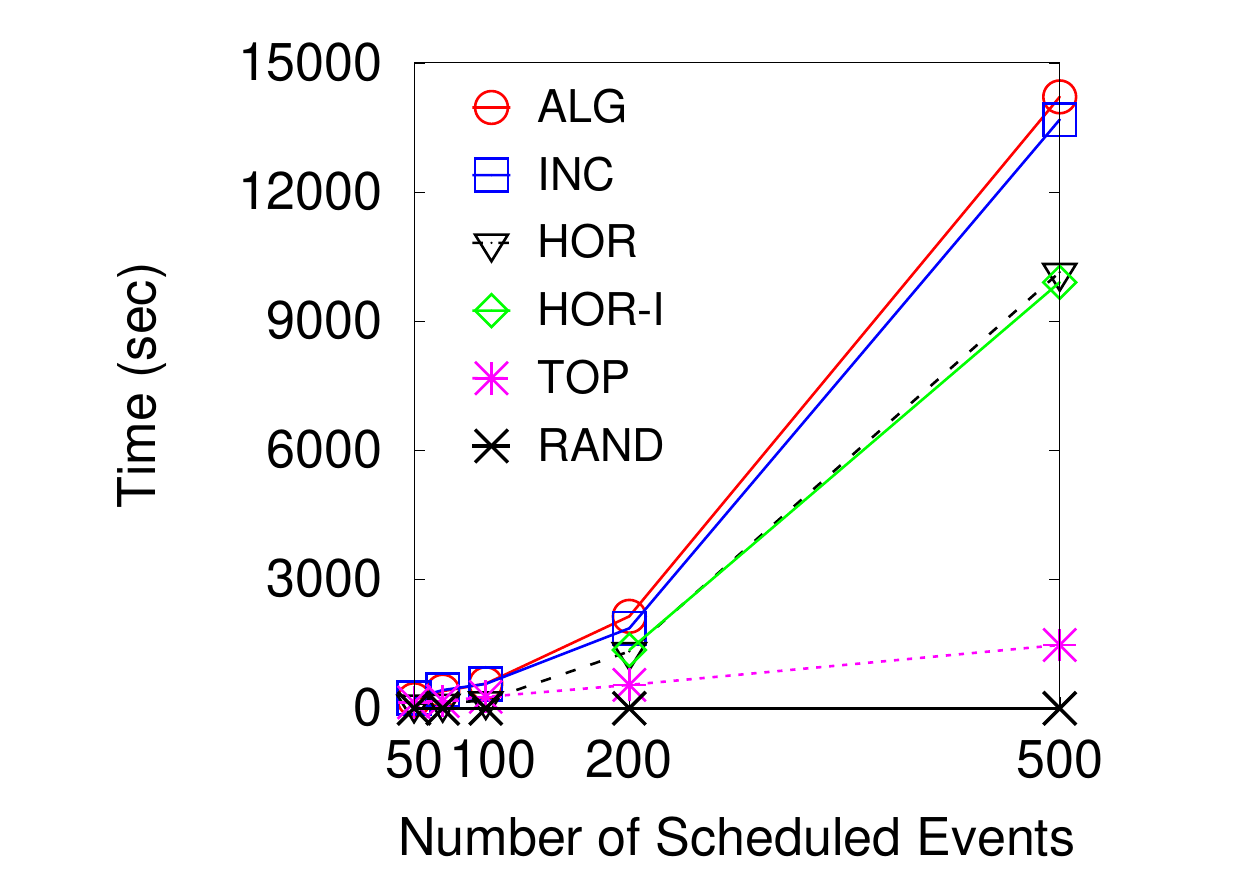}\label{fig:t_IN_k}}\hspace{-0.1cm}
\subfloat[Time (\zip) ]{\includegraphics[height=1.28in]{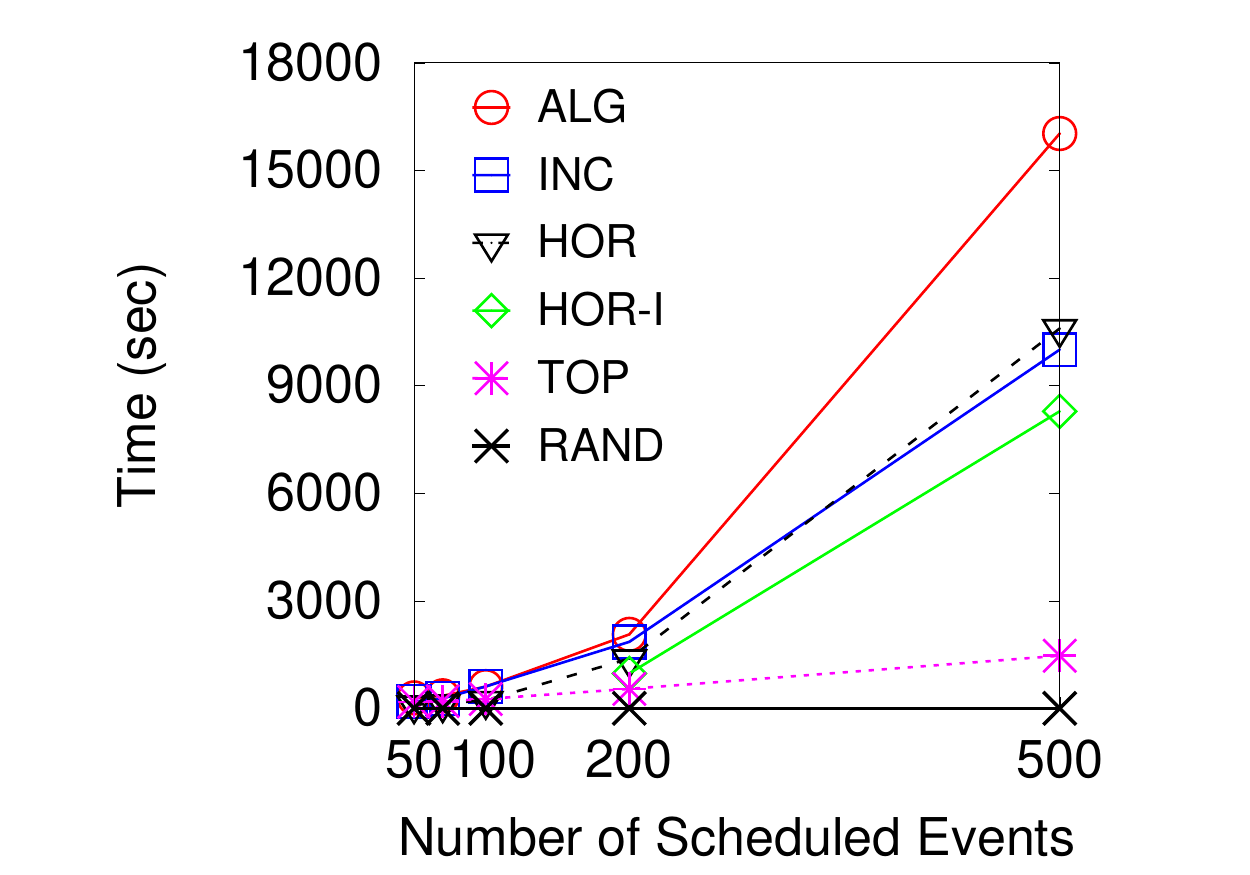}\label{fig:t_ZP_k}}\hspace{0cm}}
 \vspace{-9pt}
\caption{Varying the number of scheduled events  $k$}
\label{fig:vark}
\vspace{-8pt}
\end{figure*}
}{
\begin{figure*}[t]
\centering
\vspace{-12pt}
\hspace{-0.35cm}\mbox{
\subfloat[Utility (\meetupc) ]{\includegraphics[height=1.3in]{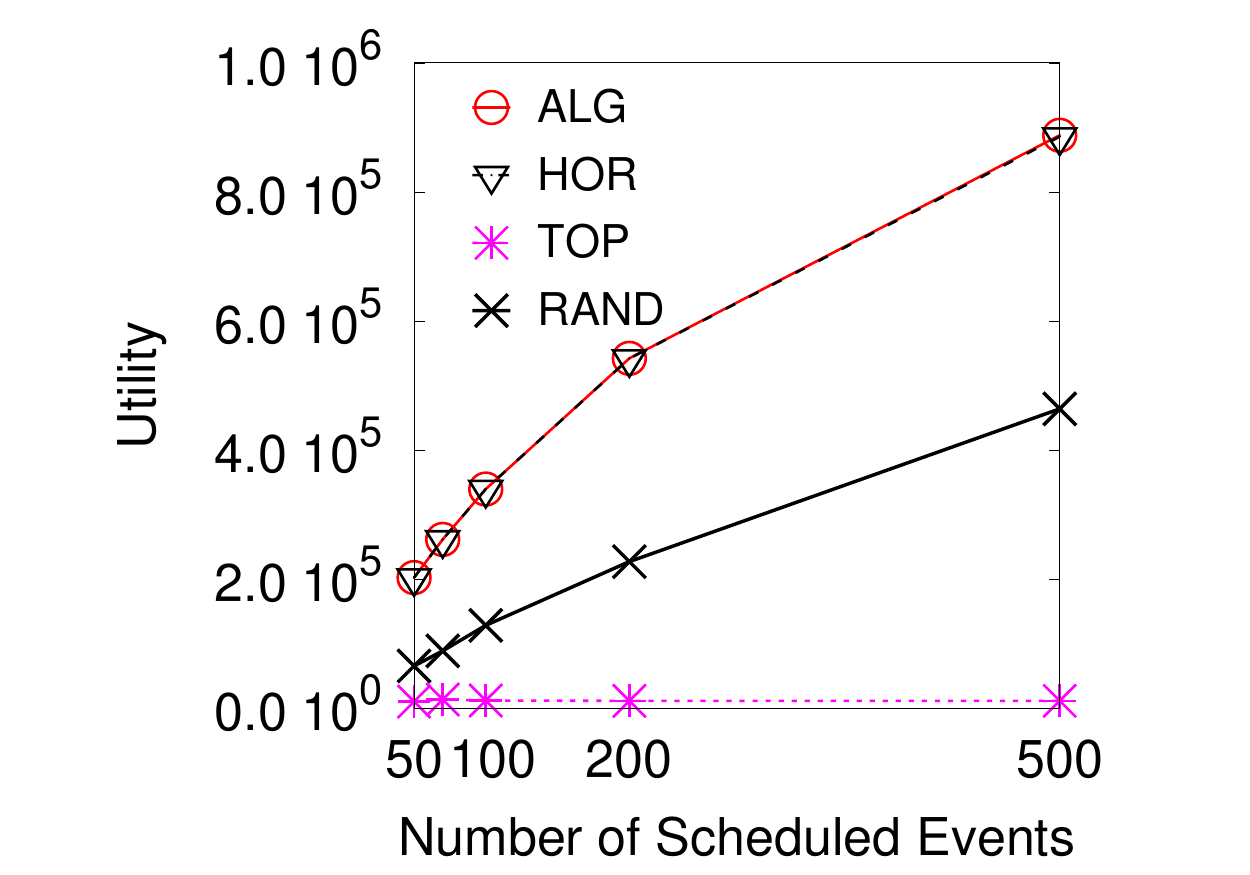}\label{fig:u_CA_k}}\hspace{-0.4cm}
\subfloat[Utility (\music) ]{\includegraphics[height=1.3in]{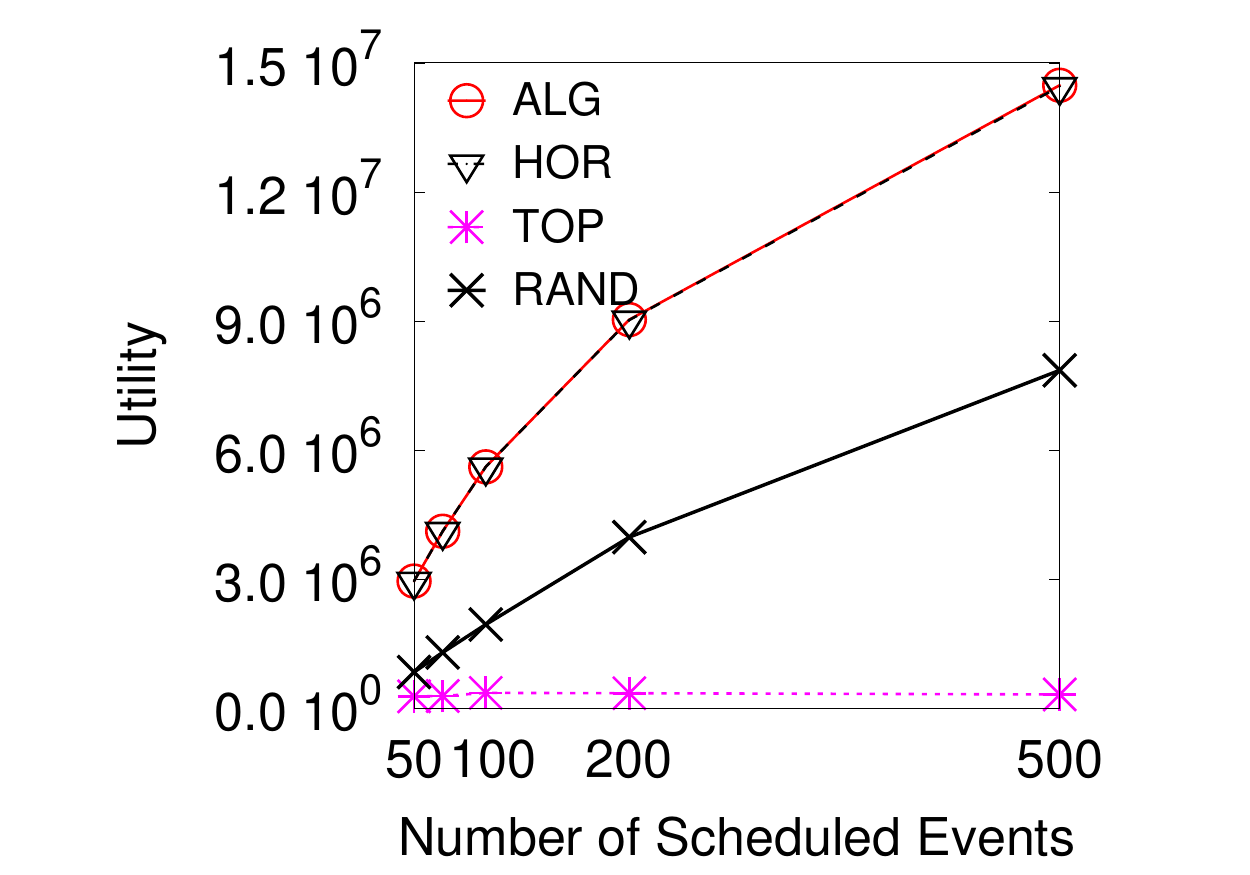}\label{fig:u_YA_k}}\hspace{-0.4cm}
\subfloat[Utility (\uni) ]{\includegraphics[height=1.3in]{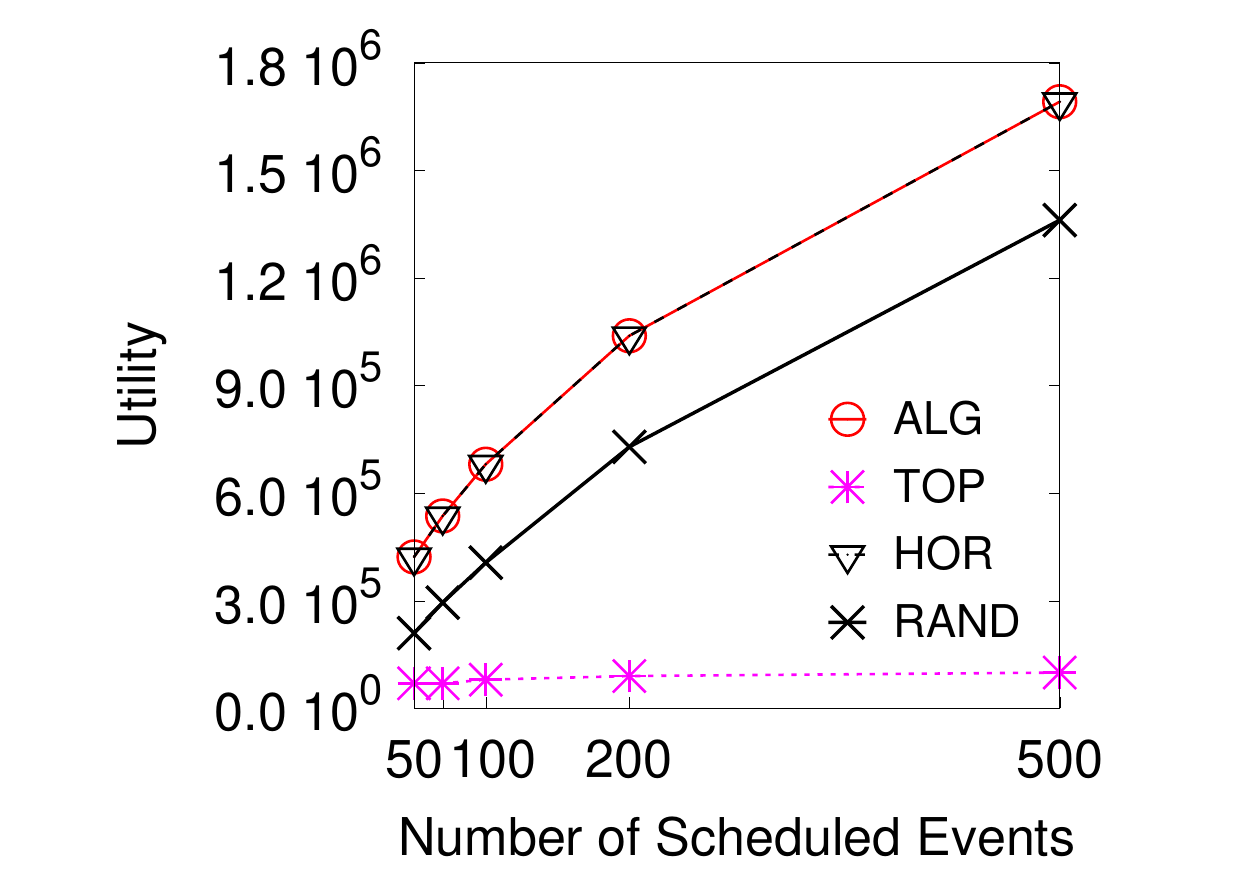}\label{fig:u_IN_k}}\hspace{-0.4cm}
\subfloat[Utility (\zip) ]{\includegraphics[height=1.3in]{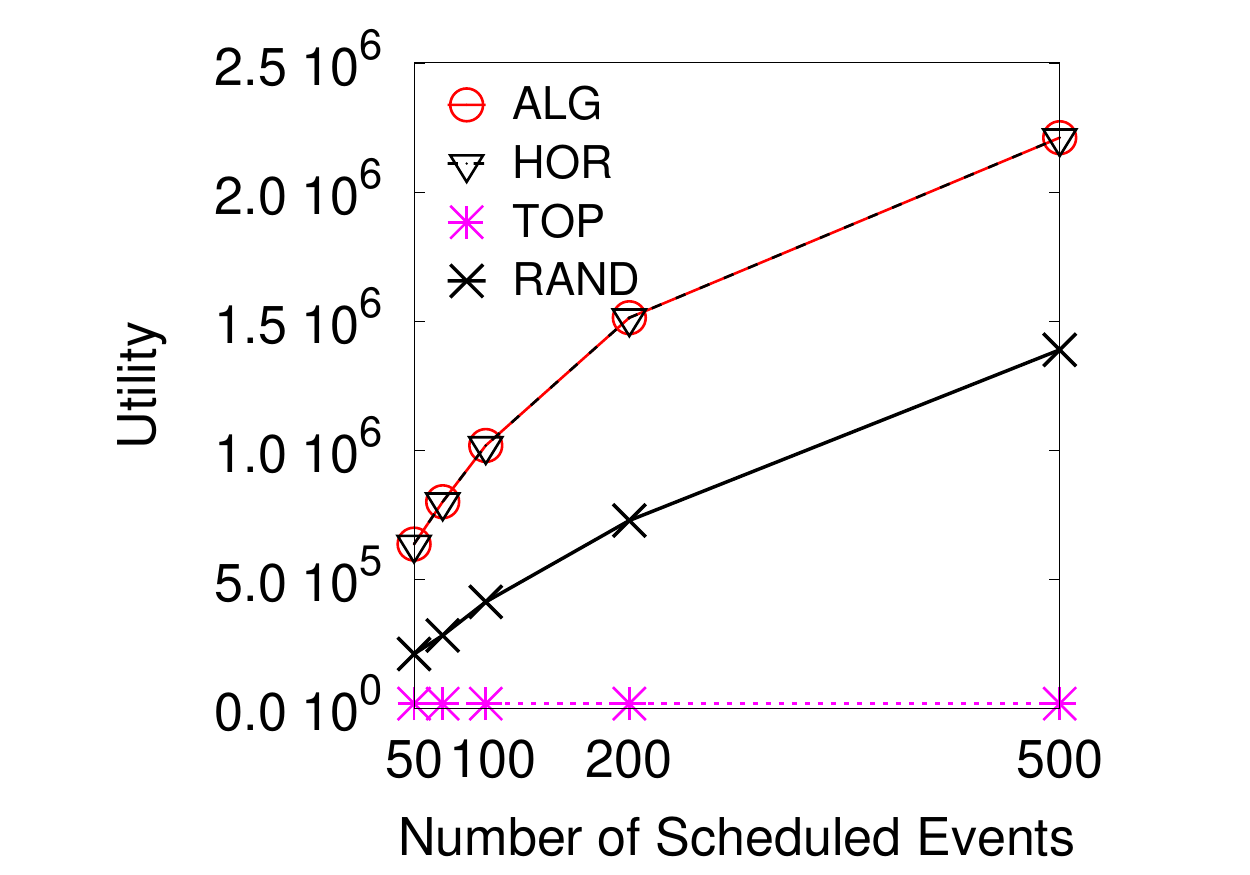}\label{fig:u_ZP_k}}\hspace{0cm}}
\\ \vspace{-5pt}
\hspace{-0.35cm}\mbox{
\subfloat[Computations  (\meetupc) ]{\includegraphics[height=1.3in]{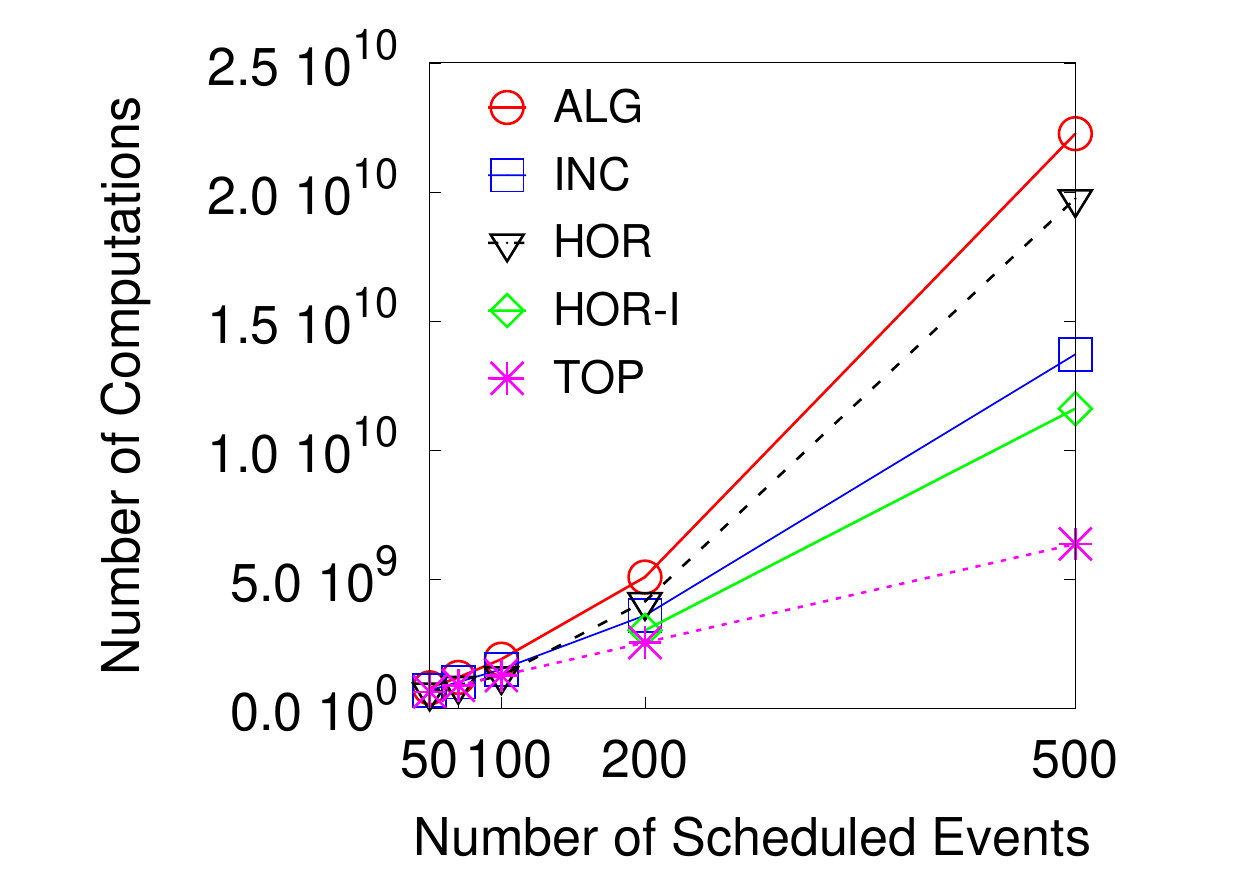}\label{fig:c_CA_k}}\hspace{-0.4cm}
\subfloat[Computations  (\music) ]{\includegraphics[height=1.3in]{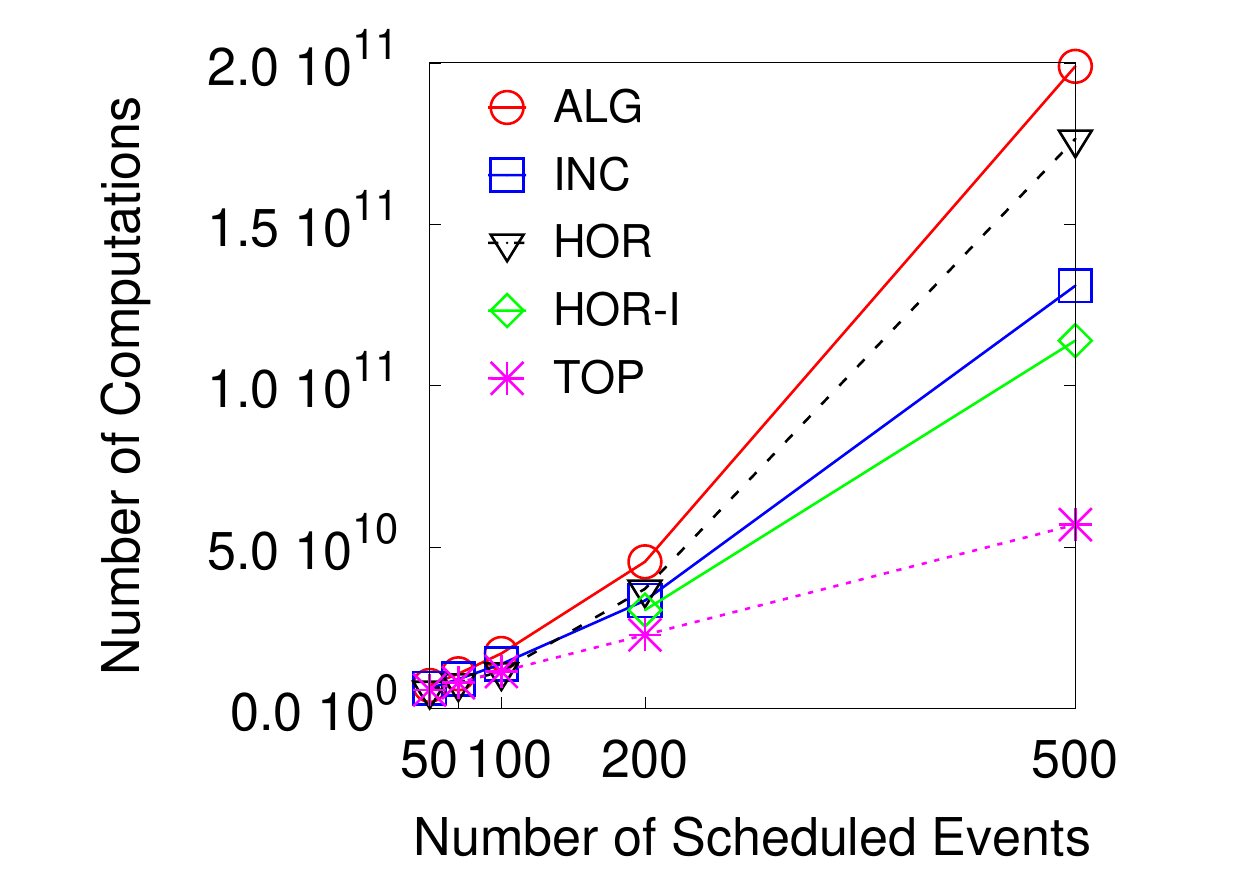}\label{fig:c_YA_k}}\hspace{-0.4cm}
\subfloat[Computations (\uni) ]{\includegraphics[height=1.3in]{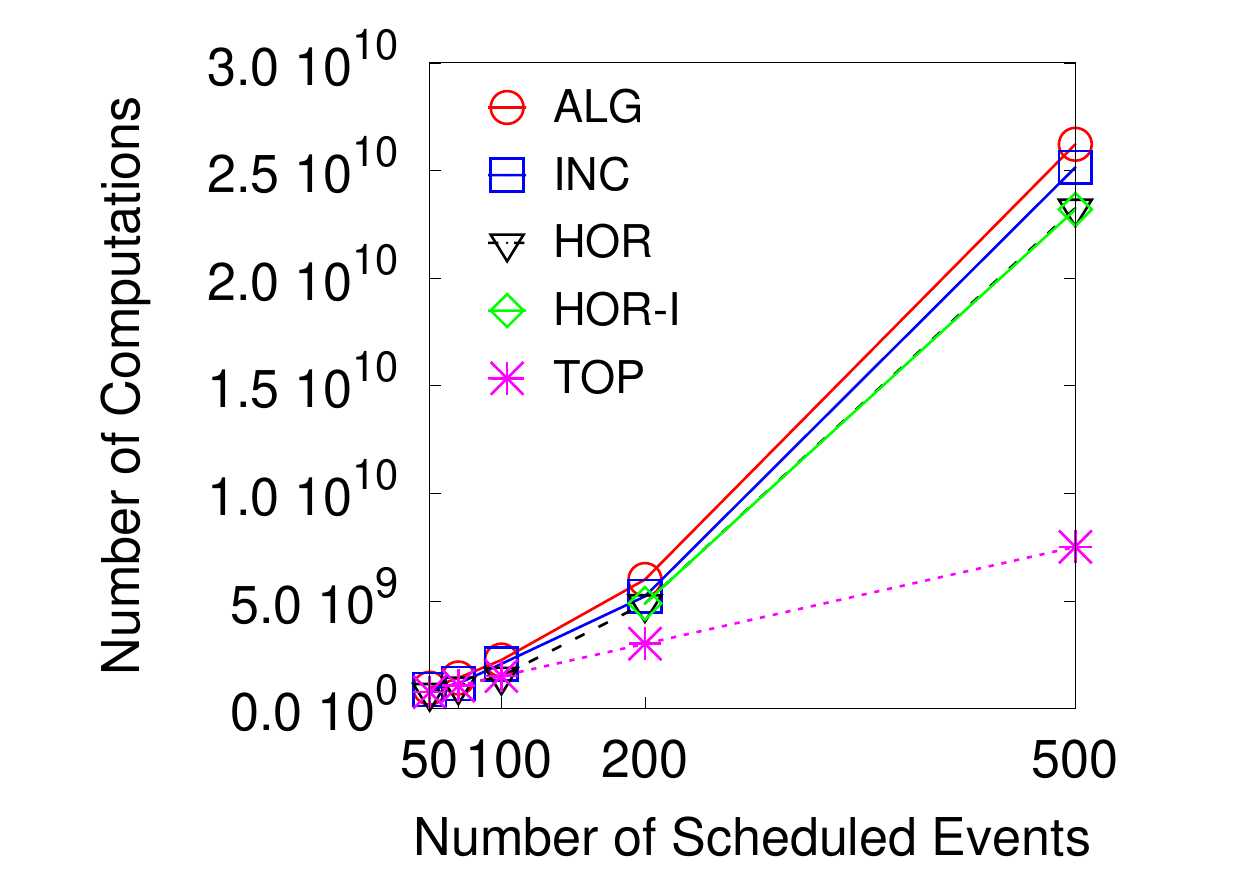}\label{fig:c_IN_k}}\hspace{-0.4cm}
\subfloat[Computations  (\zip)]{\includegraphics[height=1.3in]{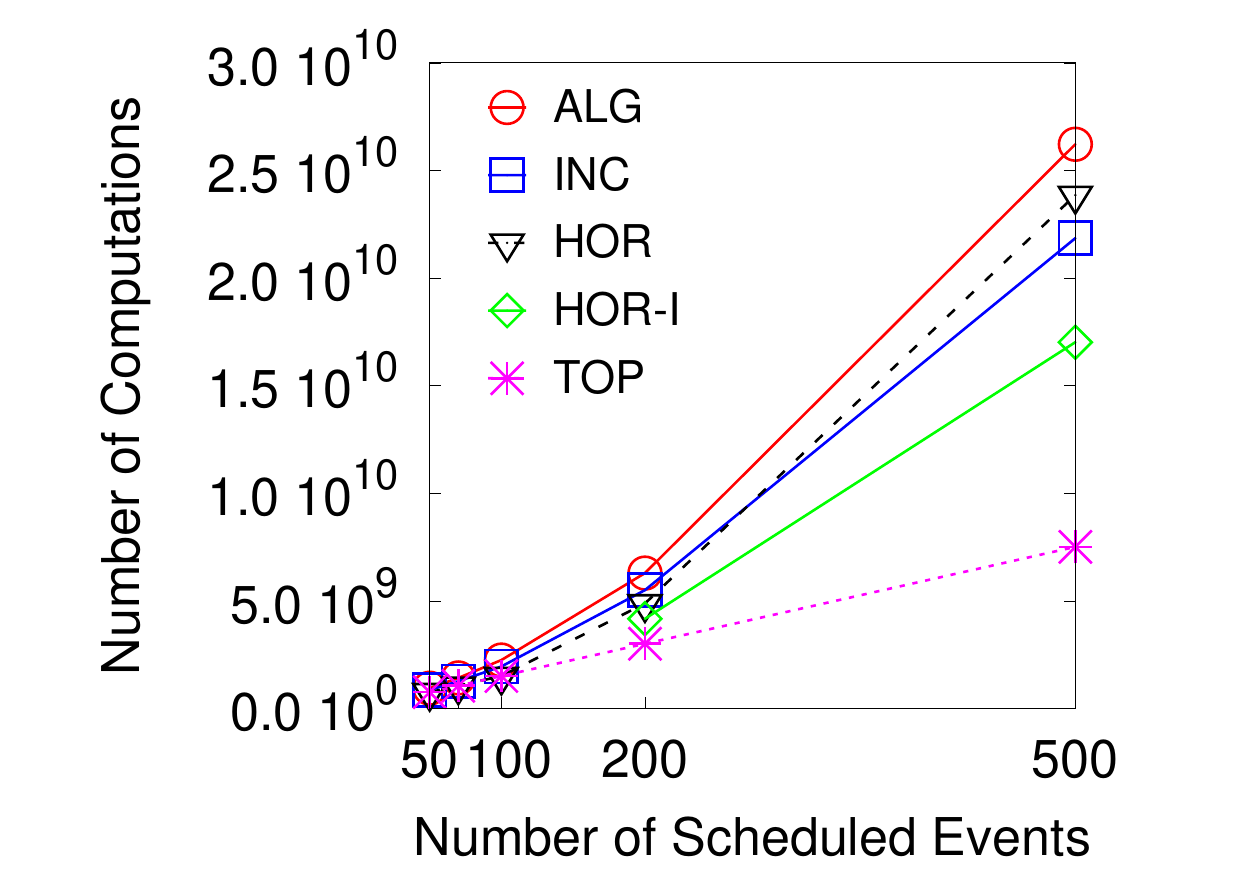}\label{fig:c_ZP_k}}\hspace{0cm}}
\\ \vspace{-5pt} 
\hspace{-0.35cm}\mbox{
\subfloat[Time (\meetupc) ]{\includegraphics[height=1.3in]{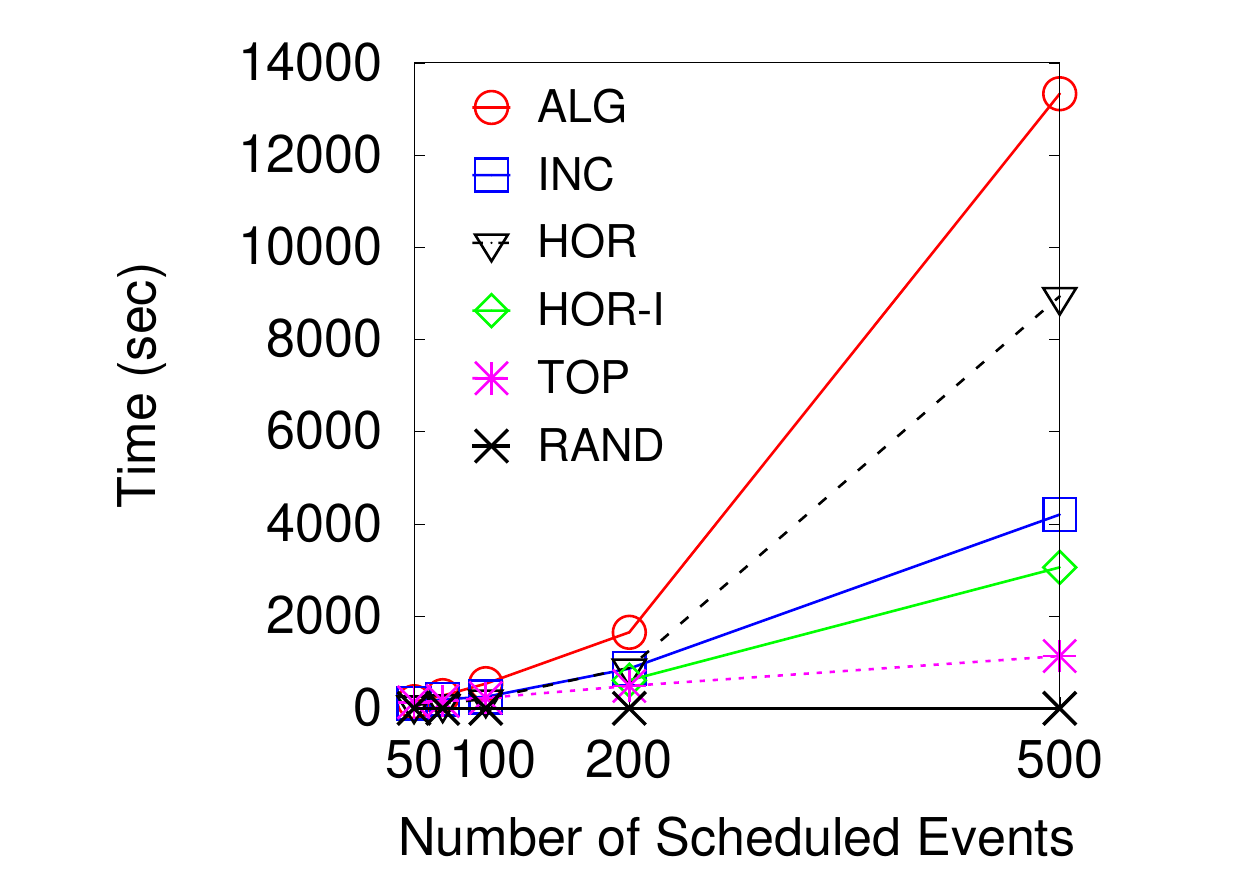}\label{fig:t_CA_k}}\hspace{-0.4cm}
\subfloat[Time (\music) ]{\includegraphics[height=1.3in]{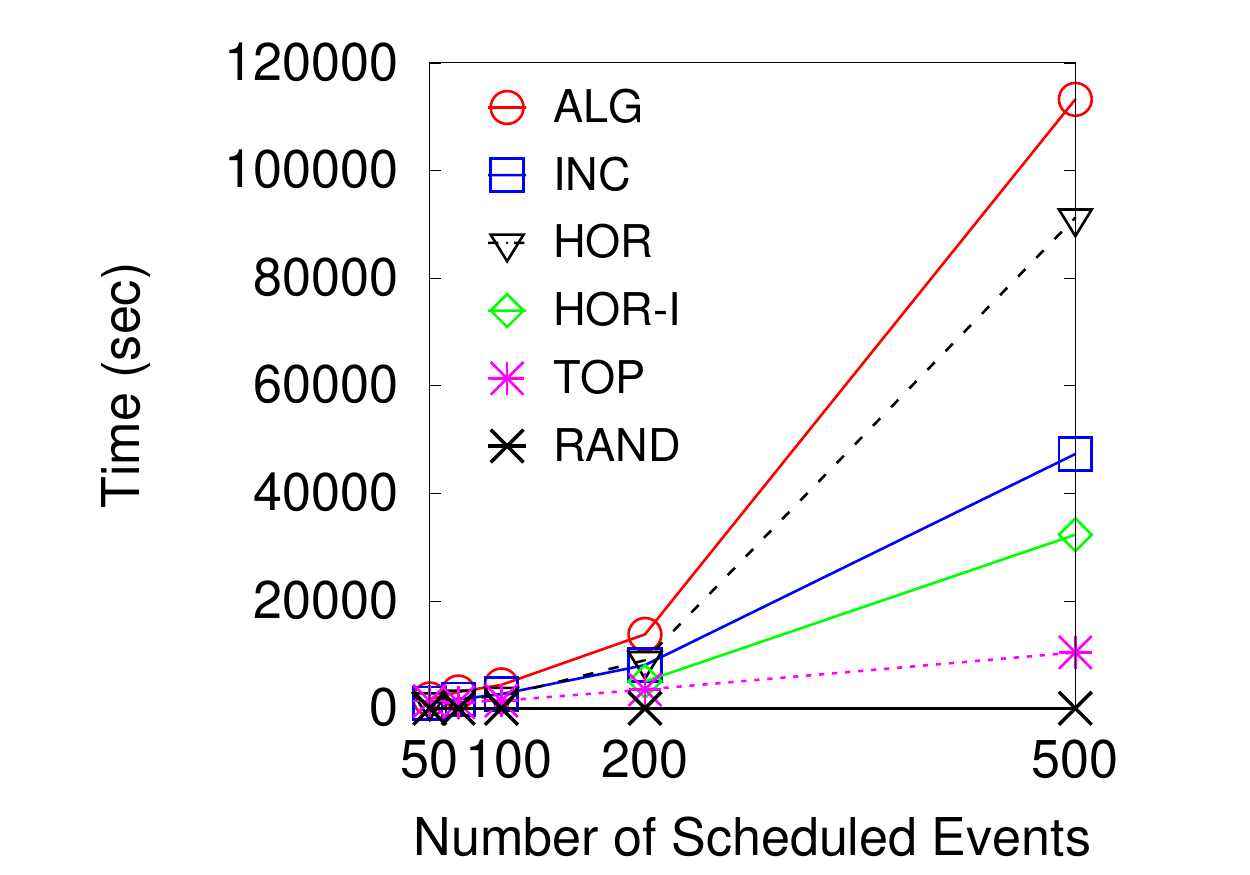}\label{fig:t_YA_k}}\hspace{-0.4cm}
\subfloat[Time (\uni) ]{\includegraphics[height=1.3in]{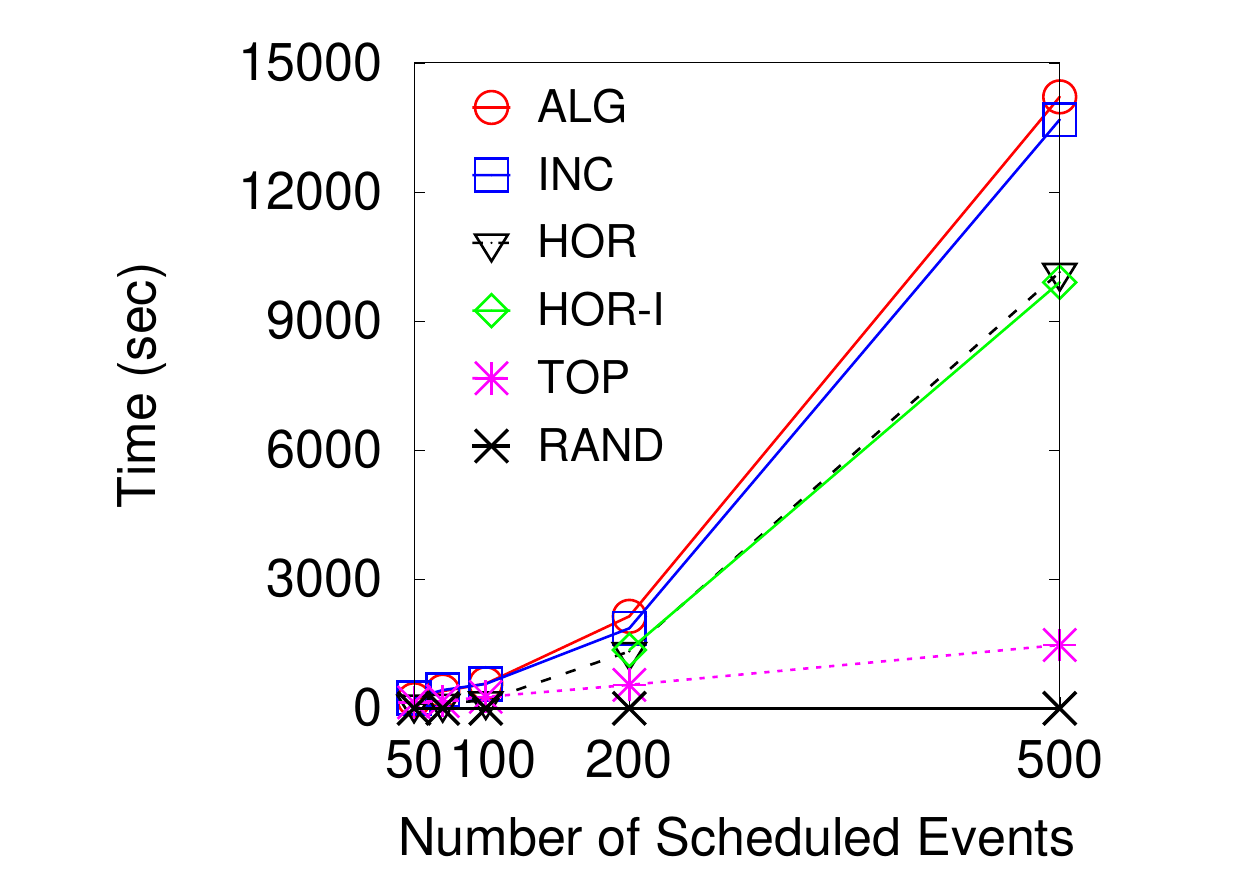}\label{fig:t_IN_k}}\hspace{-0.4cm}
\subfloat[Time (\zip) ]{\includegraphics[height=1.3in]{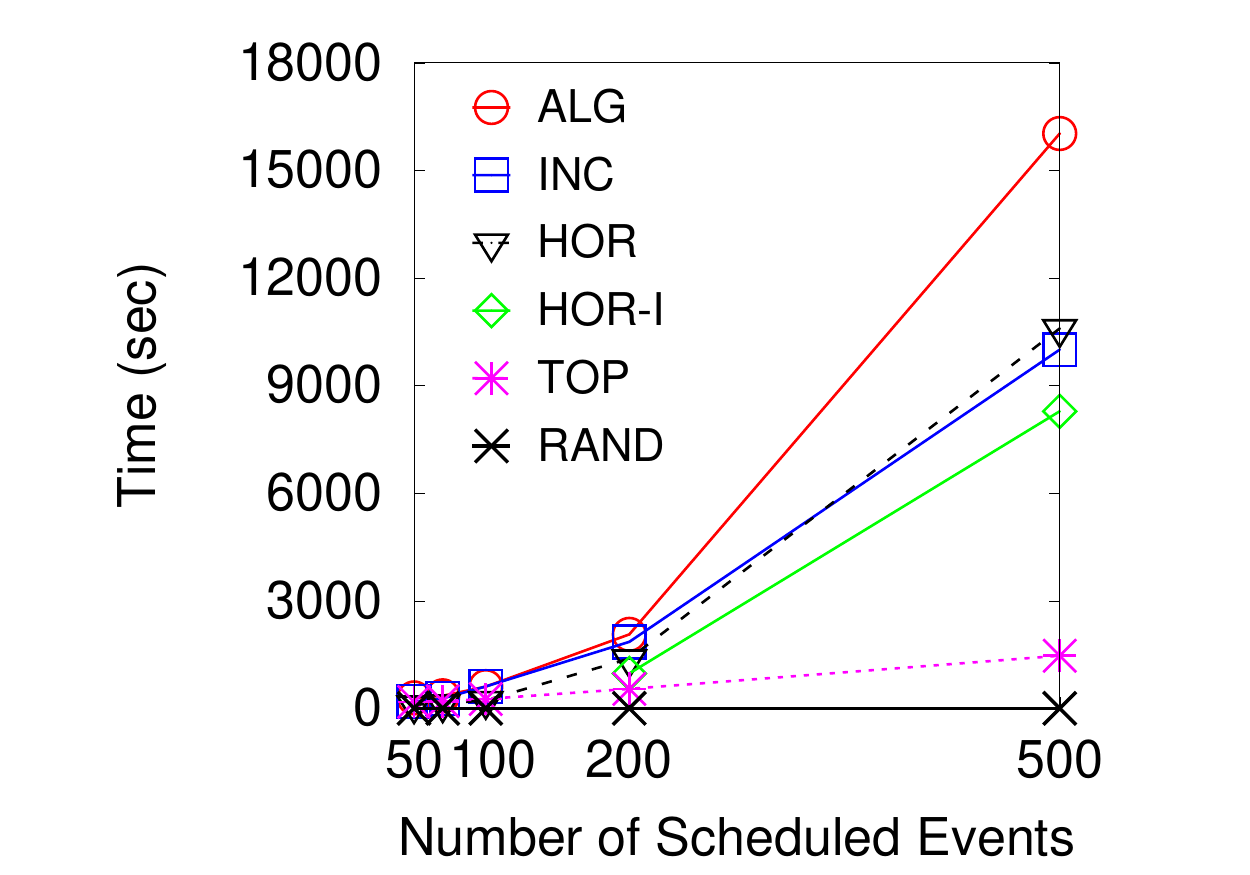}\label{fig:t_ZP_k}}\hspace{0cm}}
\caption{Varying the number of scheduled events  $k$}
\label{fig:vark}
\vspace{-0pt}
\end{figure*}
}

 \begin{table}[t]
\vspace{-1pt}
\centering
 \caption{Parameters}   
\vspace{-10pt}
\label{tab:param}
\setlength{\tabcolsep}{2pt}
\scriptsize

\begin{tabular}{lc}
\tline 
\hspace{0pt} \textbf{ Description (Parameter)} & \hspace{-0pt}\textbf{Values} \\   \dline

\rowcolor{gray!30}
  \multicolumn{2}{l}{\textbf{\hspace{0pt} Synthetic \& Real Datasets}}  \hspace{-2pt} \vspace{2pt} \\

\hspace{0pt} Num of scheduled events ($k$) &\hspace{0pt} 50, {70}, \textbf{100}, 200, 500 \\

\hspace{0pt} Num of candidate events ($|\E|$) &\hspace{0pt}   $k$,  $\boldsymbol{2k}$,  {$3k$}, $5k$, $10k$\\

\hspace{0pt} Num of time intervals ($|\T|$) &\hspace{0pt} 
$\frac{k}{5}$, $\frac{k}{2}$, $k$,  $\boldsymbol{\frac{3k}{2}}$, $2k$, $3k$  \vspace{0.2mm}\\



\hspace{0pt} Competing events per interval   &\hspace{-4pt}  
{ Uniform: $[1, 4]$, $[1, 8]$,  $\boldsymbol{[1, 16]}$, $[1, 32]$, $[1, 64]$}  \vspace{0.2mm}\\
%


\hspace{0pt} Num of available locations &\hspace{-0pt} 
$5, 10, \textbf{25}, 50, 70$\\

\hspace{0pt} Num of available resources ($\theta$) &  \hspace*{-0pt} 10, \textbf{20},  30, 50, 100 \\

\hspace{0pt} Num of required resources per event  ($\xi_e$) &\hspace{-4pt}  {Uniform:} 
$[1,  \frac{\theta}{4}]$,  $\boldsymbol{[1,  \frac{\theta}{3}}]$, ${[1,  \frac{\theta}{2}]}$, $[1, \frac{3 \theta}{4}]$, $[1,  \theta]$ \vspace{1pt}\\

\hspace{0pt} Distribution of  social activity probability ($\sigma_u^t$) &\hspace{-3pt} \textbf{Uniform}, {Normal (0.5, 0.25)}  
\vspace{4pt}\\

\rowcolor{gray!30}
 \multicolumn{2}{l}{\textbf{\hspace{0pt} Synthetic   Datasets}} \vspace{2pt}\\
 
\hspace{0pt}  Num of users  ($|\U|$) & \hspace{-0pt} 10K, \textbf{50K}, 100K, 500K, 1M  \\

\hspace{0pt} Distribution of  interest ($\mu_{u, e}$) & \hspace{-7pt} Uniform, {Normal (0.5, 0.25)},  Zipfian:  {1}, 2, 3  \hspace{2mm}\\
\bline
\end{tabular}
 \end{table}

\vspace{-3pt}
\stitle{Parameters.}
Table~\ref{tab:param} summarizes the parameters 
that we vary and the range of values examined;
default values are presented in bold. 

\alt{}
{The first rows list the parameters  used in experiments for real and synthetic datasets, 
while the last two present the parameters used only in synthetic datasets.}
Adopting the same setting as in the related works \cite{ses,Li2014,She2015,She2015a,She2016,Tong2015}, 
we set the the default  and maximum value of the of \textit{scheduled events} $k$,  to $100$ and $500$, respectively.
 %
%
In order to select the values for  the number of \textit{competing events per interval}, 
we analyze the two Meetup datasets used in our evaluation \cite{Pham2015}.   
Particularly, we are interested in the number of events taking place during overlapped time intervals. 
As event   interval  we consider the period spanning  
from one hour before to two hours after the  event's scheduled time. 
From the analysis, we found that, on average, $8.1$ events are taking place during overlapping  intervals.
Therefore, in the default setting the number of competing events per interval 
is selected by a uniform distribution having $8.1$ as mean value.
Further, we vary the mean value from $2$ to $32$ (Table~\ref{tab:param}). 
In our experiments, the reported results are similar to the default setting, 
with the utility score being slightly lower for larger numbers of competing events, as expected (results are omitted due to lack of space).

\alt{In order to select  the default and the examined values 
for the \textit{number of available events' locations},
we consider the percentage of pairs of  events that are spatio-temporally conflicting, as specified in  \cite{She2015}. 
}
{
 In analogy to \cite{She2015}, that specifies 
the percentage of pairs of  events that are spatio-temporally conflicting, 
we select the values for the event location parameter $\ell$. 
Particularly, we assume a set of available locations; 
then,  the location for each event is randomly selected from these locations.
In default setting, we considered  25 available locations; that is, 
$0.25$  spatio-temporal conflict ratio for the default number of scheduled events.


Regarding the social activity probability $\sigma_u^t$, we use  Uniform and distribution. Note that, the same results are also reported for Normal distribution.
Finally, we should note that, the minimum \textit{number of available locations}, 
as well as  the default value for the resources per event (i.e, $[1, \theta/3]$),
are selected so there are at least $k$ feasible assignments in each experiment. 
}
\alt{Also, we vary the \textit{number of required resources} for each event,
 as well as the number of \textit{available resources} (Table~\ref{tab:param}).
Here, as resources we consider agents (i.e., organizer's staff). 
In the aforementioned experiment,  the  methods are marginally  affected by the examined parameters. 
Thus, due to lack of space, the results are omitted.
Finally, regarding the \textit{social activity probability},
we use Uniform and Normal distribution.
Note that, the results for Normal distribution are not presented here, 
since they are the same as in Uniform.
}
{Moreover, the difference in terms of utility between the baselines and the other methods  slightly increases with the number of competition events. 
The reason is that, increasing the max value of competing events
increases the variance in the number of competition events per interval.
The results for this experiment are omitted for brevity.
}

%
%
%
 

 


%
%
\stitle{Methods.} 
In our evaluation we study the three proposed algorithms (\bnd, \hor, \horb),
as well as the \bsc algorithm proposed  in \cite{ses}.
%
Further, we  include the baselines used in \cite{ses}.
The first, denoted as 
 \stat,  computes the assignment scores for all the events and selects the events with
top-k score values. 
Since \stat  computes the scores only once, \stat is always performing the minimum number of computations.
The second, \rand assigns events to intervals, randomly. 
Note that, since the objective,  the solution and  the setting of our problem 
are substantially different (see Sect.~\ref{sec:intro}) from the related works \cite{Li2014,She2015a,She2016,She2015,Tong2015,She2017,ChengYCGW17}, 
the existing methods cannot be used to solve the \ses problem. 

\stitle{Metrics \& Implementation.}
In each experiment, we measure: 
$(1)$~the \textit{total utility score}; 
$(2)$~the \textit{execution time}; and 
$(3)$~the \textit{number of  computations} for assignment scores  ($|\U|$ per assignment score).
All algorithms were written in C++ and the experiments were 
performed on an 2.67GHz Intel Xeon  E5640 with 32GB of RAM. 

%



%
%


\vspace{-4pt}
\subsection{Results}
\label{sec:results}
\vspace{-2pt}

Recall that, the \hor and \horb algorithms return the same solutions (i.e., equal utilities); 
the same also holds for the \bsc and \bnd algorithms. 
Hence, only the former utility plots are presented.
{Further,  in cases where $k<|\T|$, the \horb algorithm is identical to \hor (Sect.~\ref{sec:horb});
thus, in these cases only  the \hor results are included in the plots.
}

\alt{
\begin{figure*}[t]
\centering
\vspace{-39pt}
\hspace*{-0.35cm}\mbox{
\subfloat[Utility (\meetupc) ]{\includegraphics[height=1.28in]{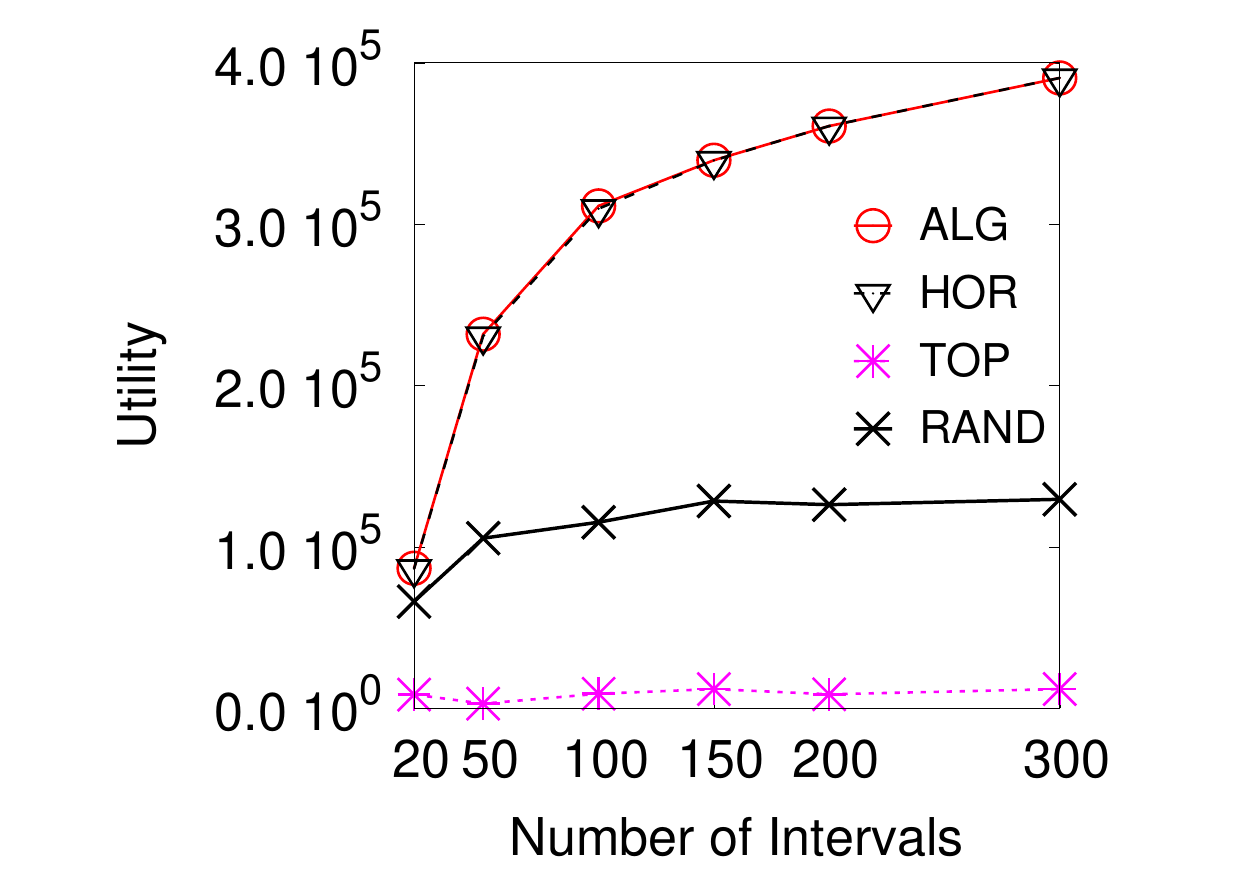}\label{fig:u_CA_i}}\hspace{-0.1cm}
\subfloat[Utility (\music) ]{\includegraphics[height=1.28in]{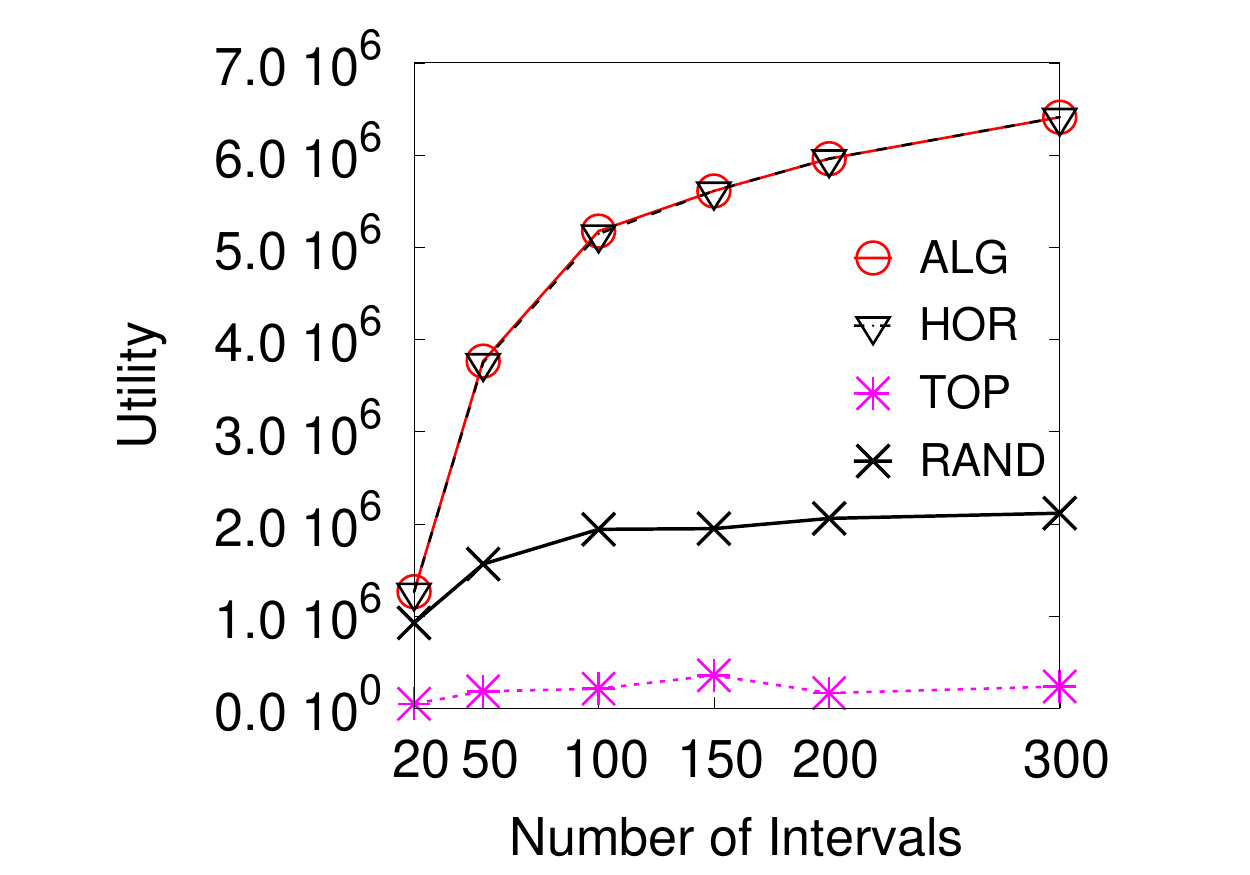}\label{fig:u_YA_i}}\hspace{-0.1cm}
\subfloat[Utility (\uni) ]{\includegraphics[height=1.28in]{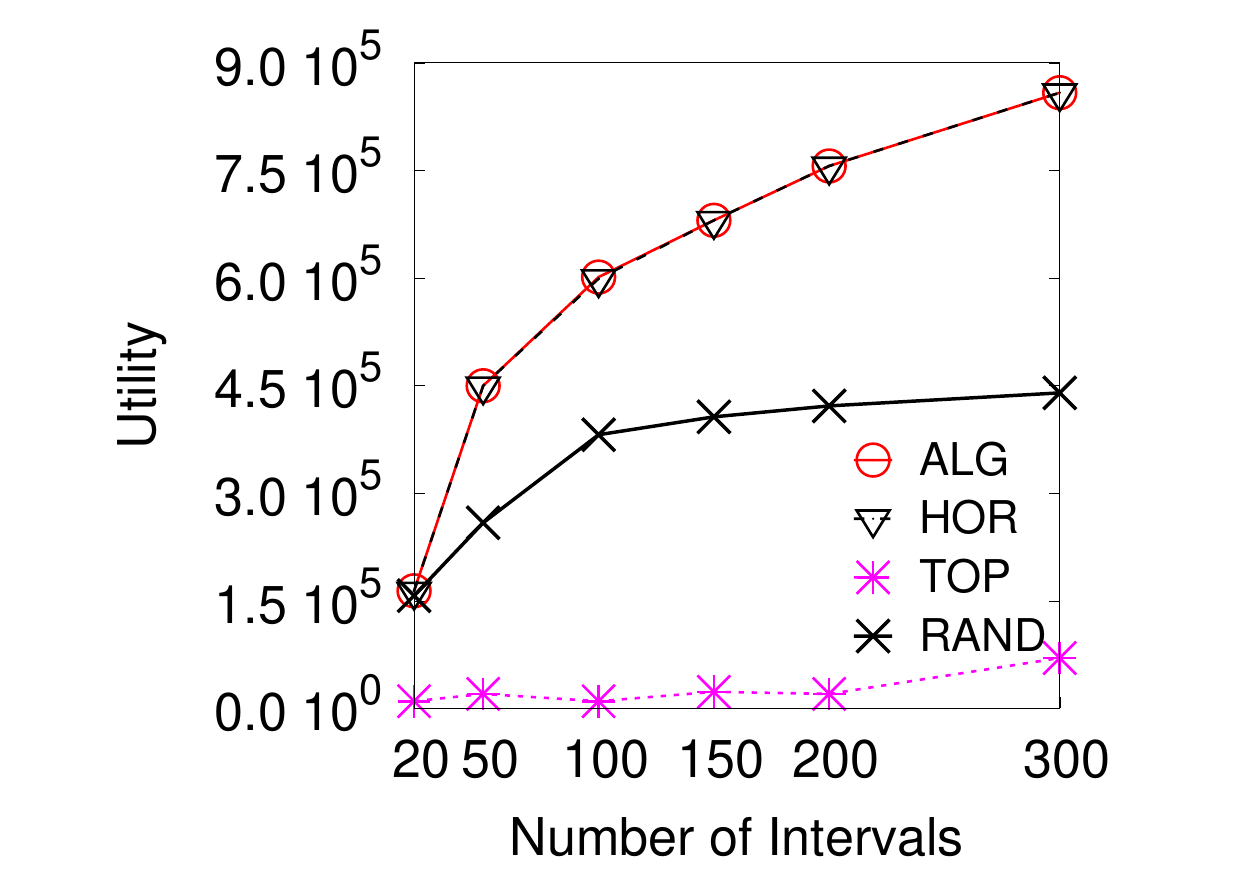}\label{fig:u_IN_i}}\hspace{-0.1cm}
\subfloat[Utility (\zip) ]{\includegraphics[height=1.28in]{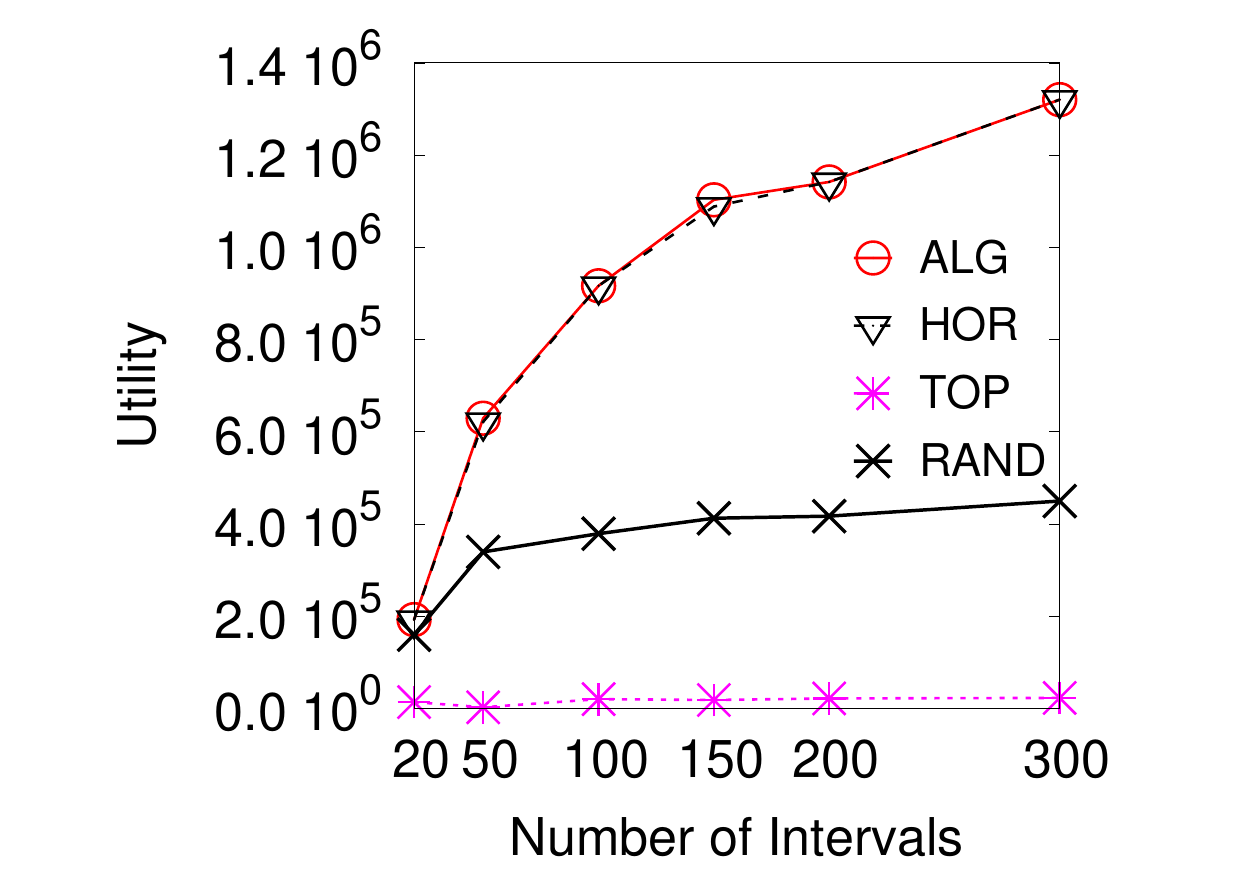}\label{fig:u_ZP_i}}\hspace{0cm}
}
\\ \vspace{-10pt}
 \hspace*{-0.27cm}\mbox{
\subfloat[Time (\meetupc) ]{\includegraphics[height=1.28in]{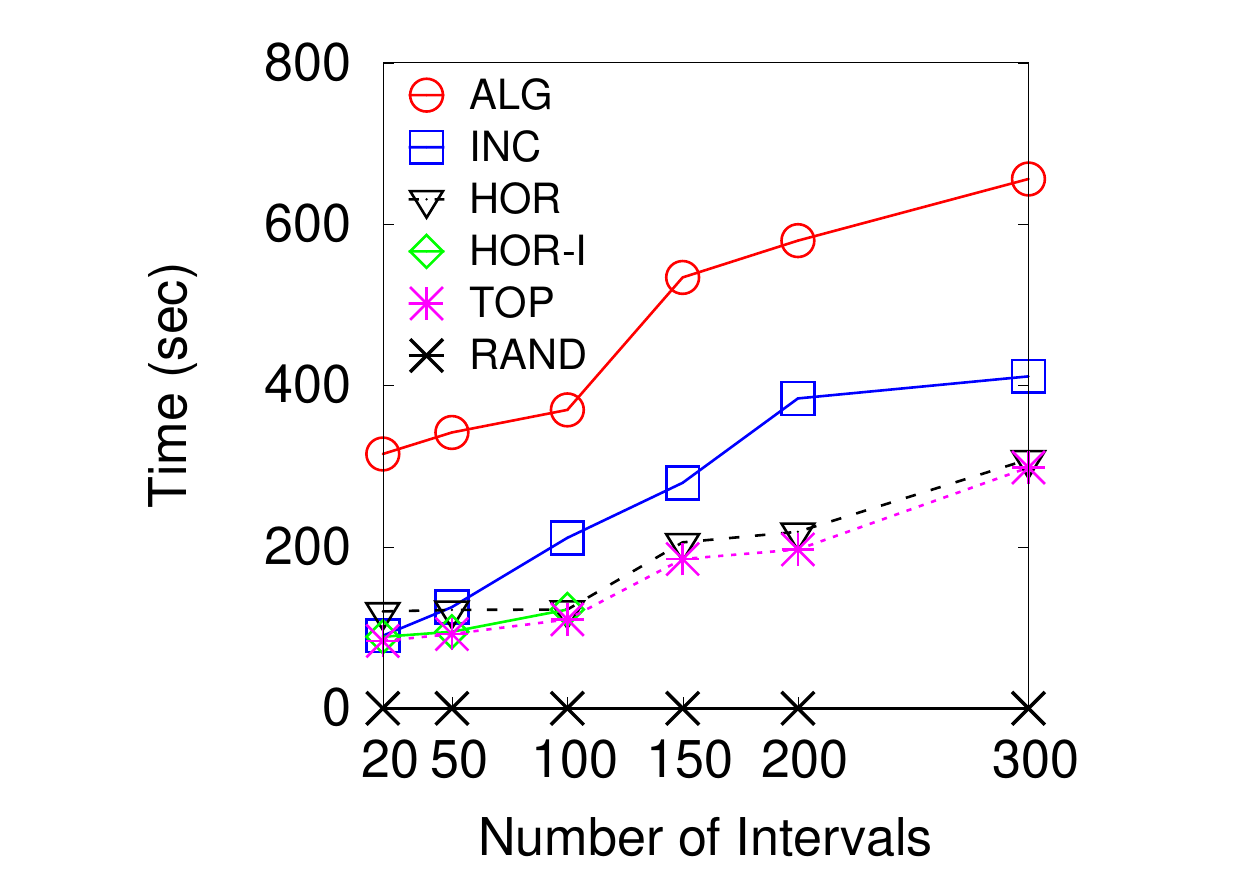}\label{fig:t_CA_i}}\hspace{-0.2cm}
\subfloat[Time (\music) ]{\includegraphics[height=1.28in]{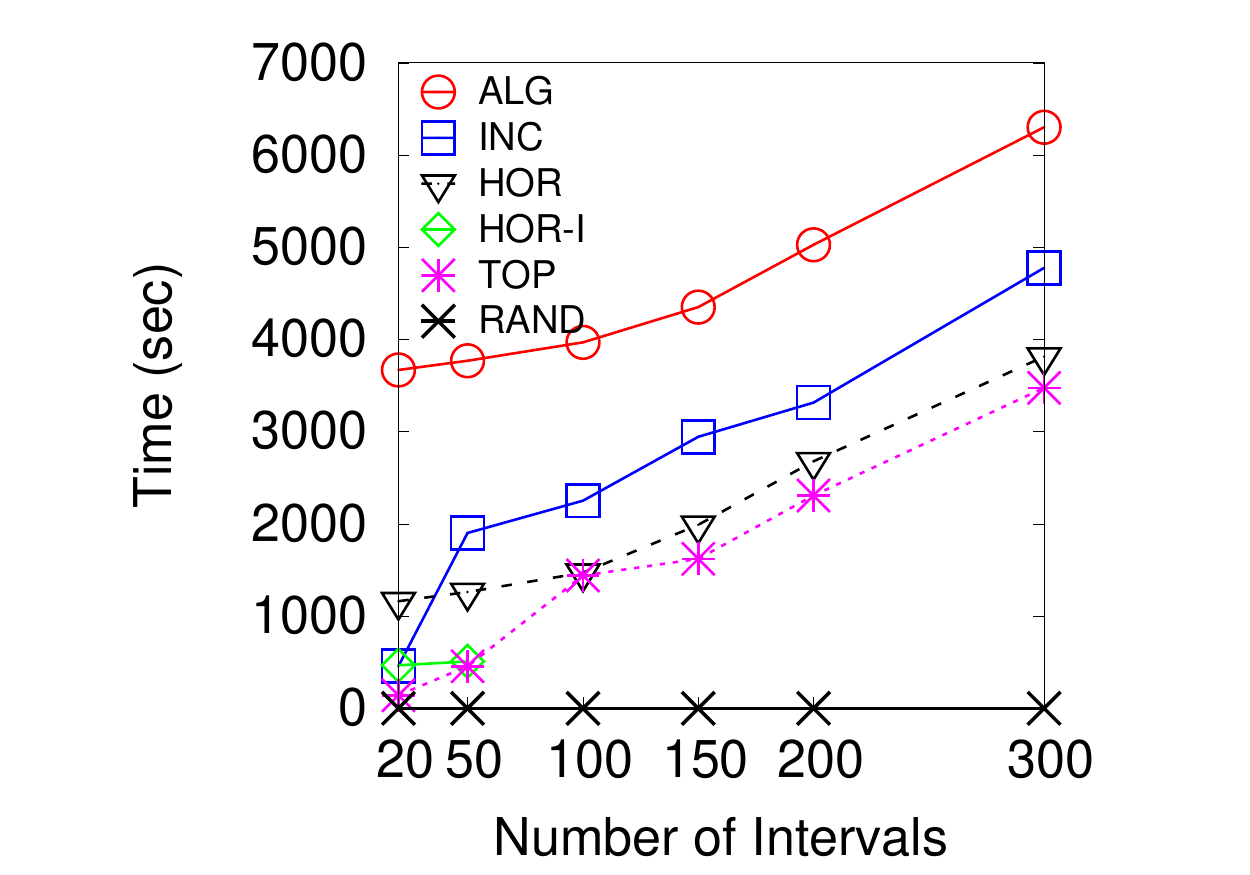}\label{fig:t_YA_i}}\hspace{-0.1cm}
\subfloat[Time (\uni) ]{\includegraphics[height=1.28in]{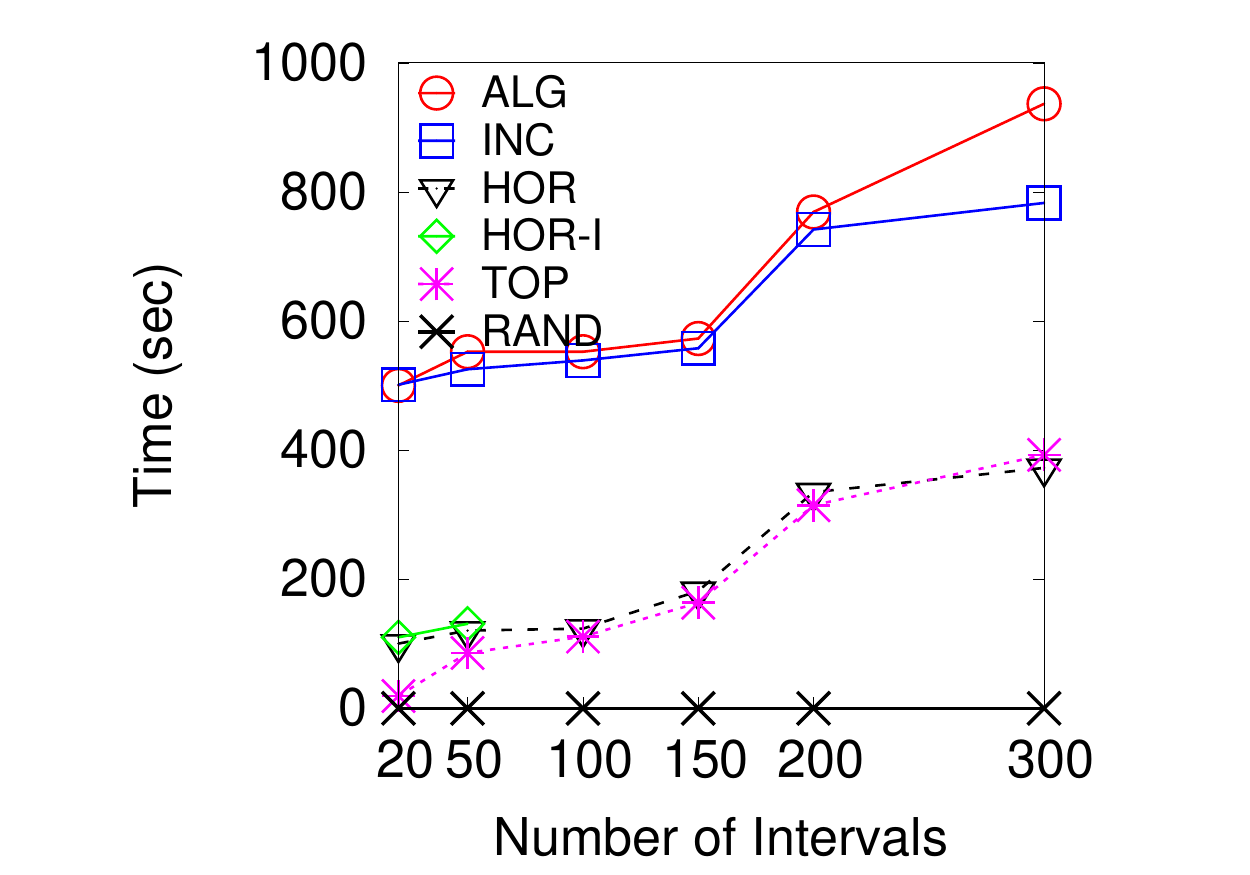}\label{fig:t_IN_i}}\hspace{-0.1cm}
\subfloat[Time (\zip) ]{\includegraphics[height=1.28in]{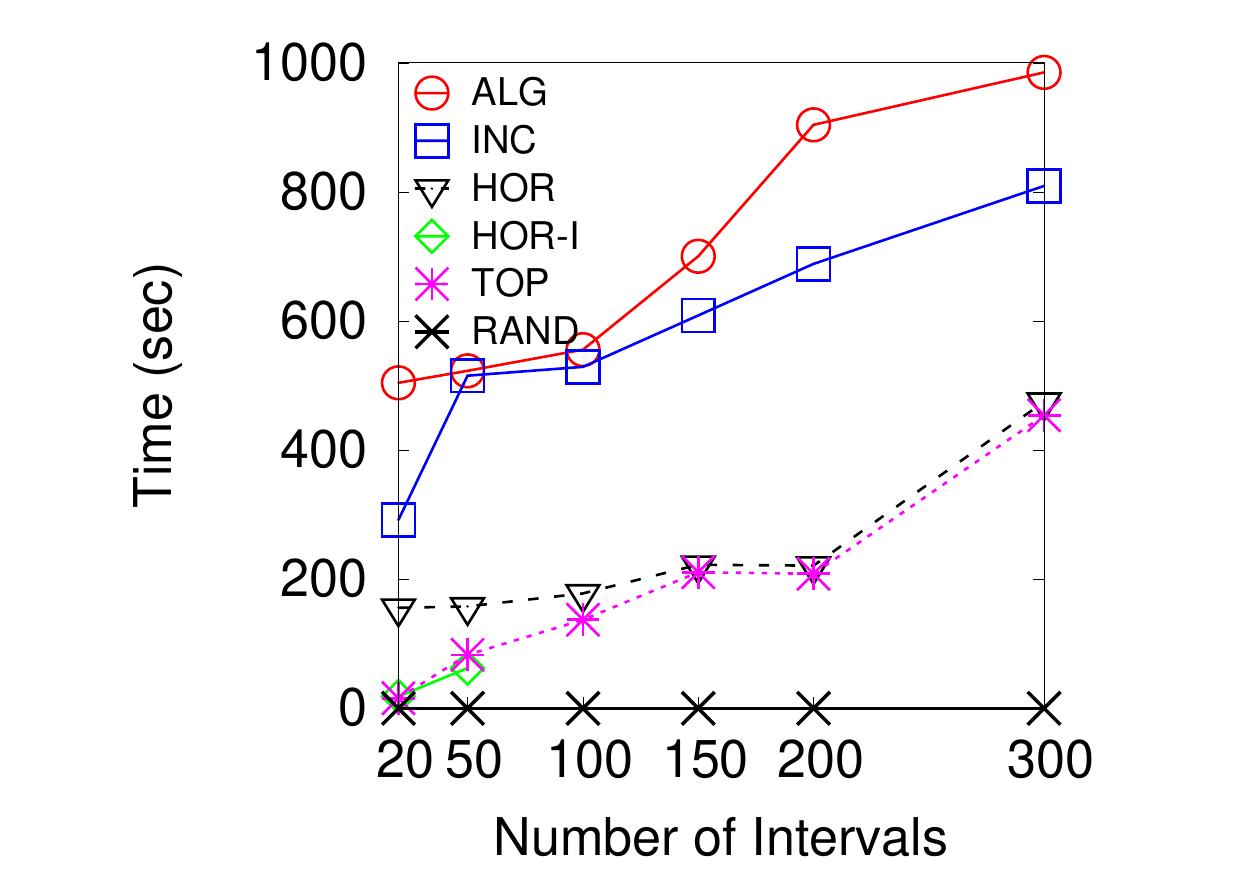}\label{fig:t_ZP_i}}\hspace{0cm}
}
\vspace{-6pt}
\caption{Varying the number of time intervals $|\T|$}
\label{fig:vart}
\end{figure*}
}
{
\begin{figure*}[t]
\centering
\vspace{-12pt}
\hspace{-0.35cm}\mbox{
\subfloat[Utility (\meetupc) ]{\includegraphics[height=1.3in]{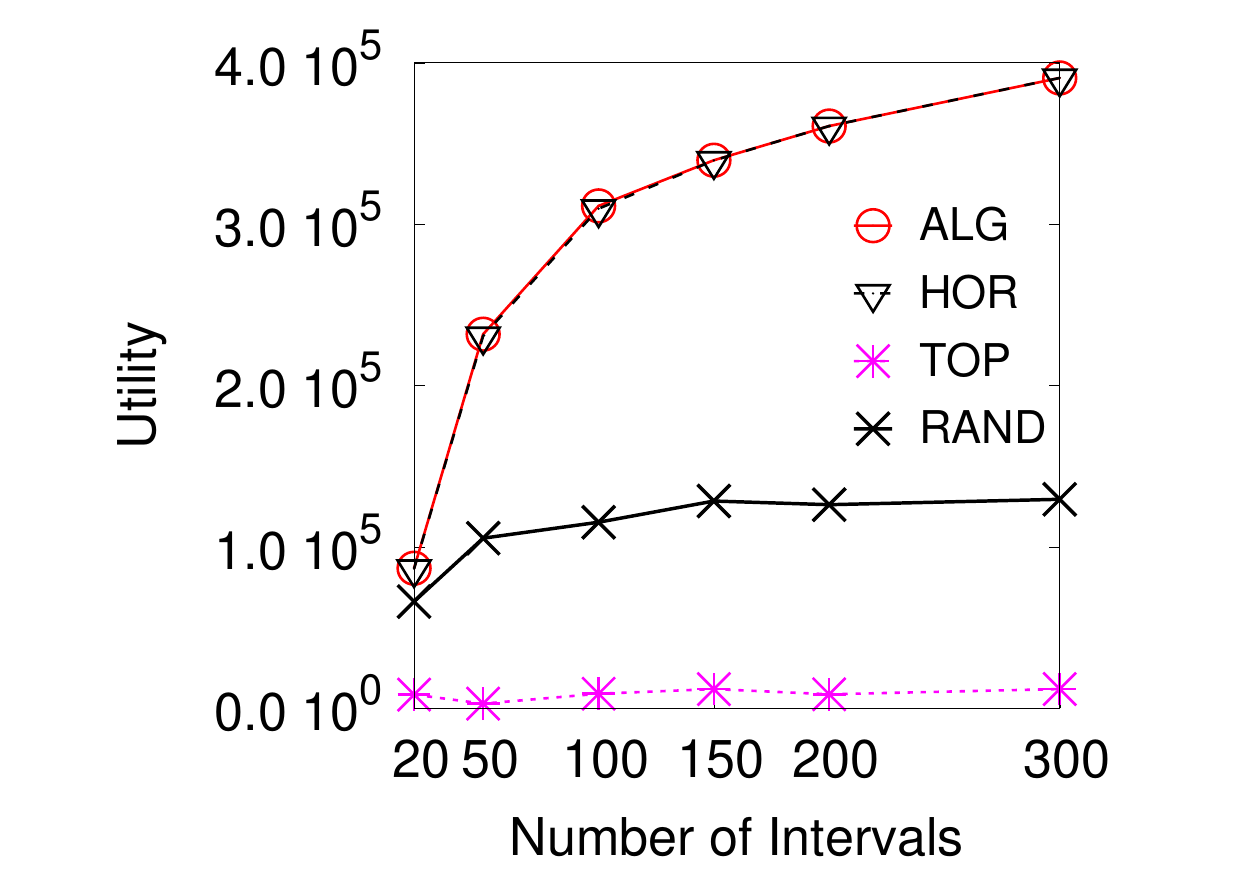}\label{fig:u_CA_i}}\hspace{-0.4cm}
\subfloat[Utility (\music) ]{\includegraphics[height=1.3in]{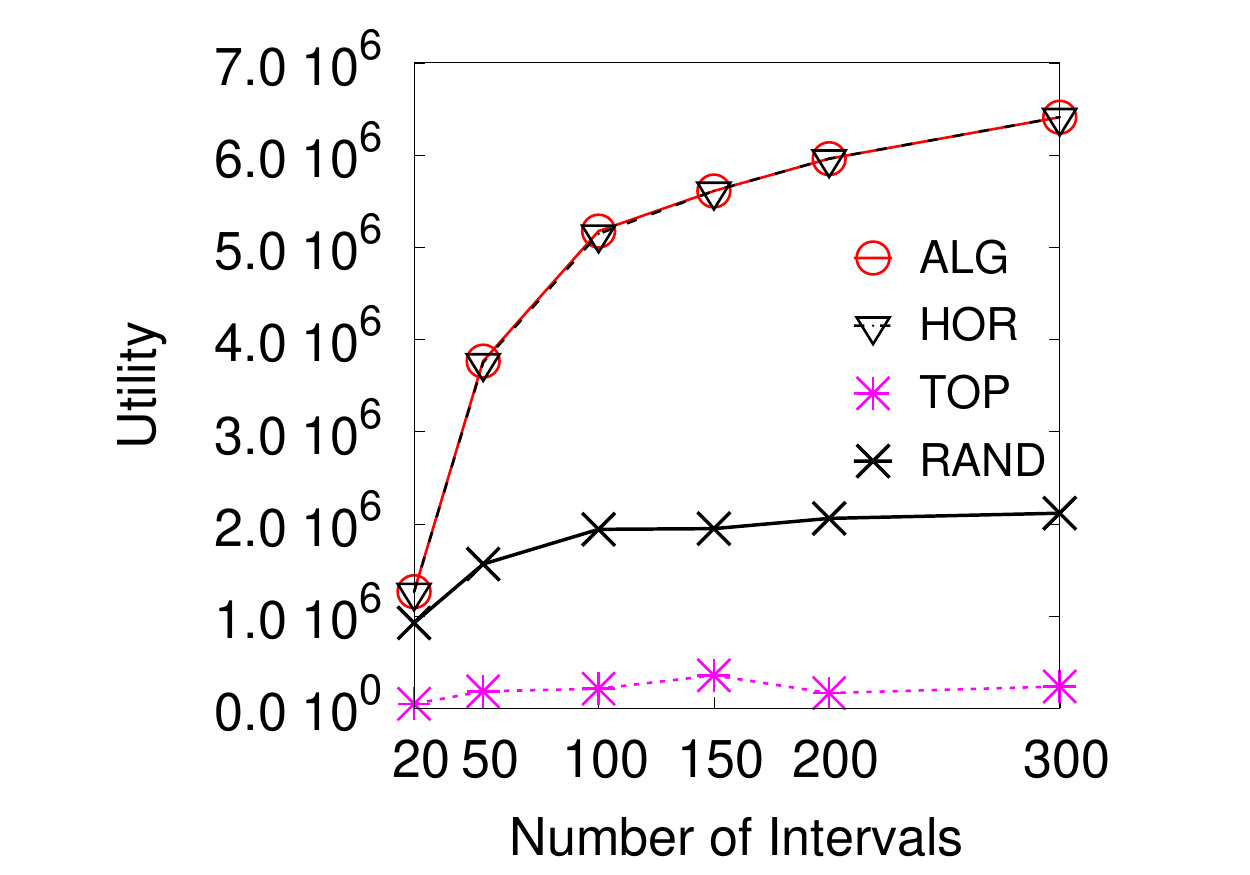}\label{fig:u_YA_i}}\hspace{-0.4cm}
\subfloat[Utility (\uni) ]{\includegraphics[height=1.3in]{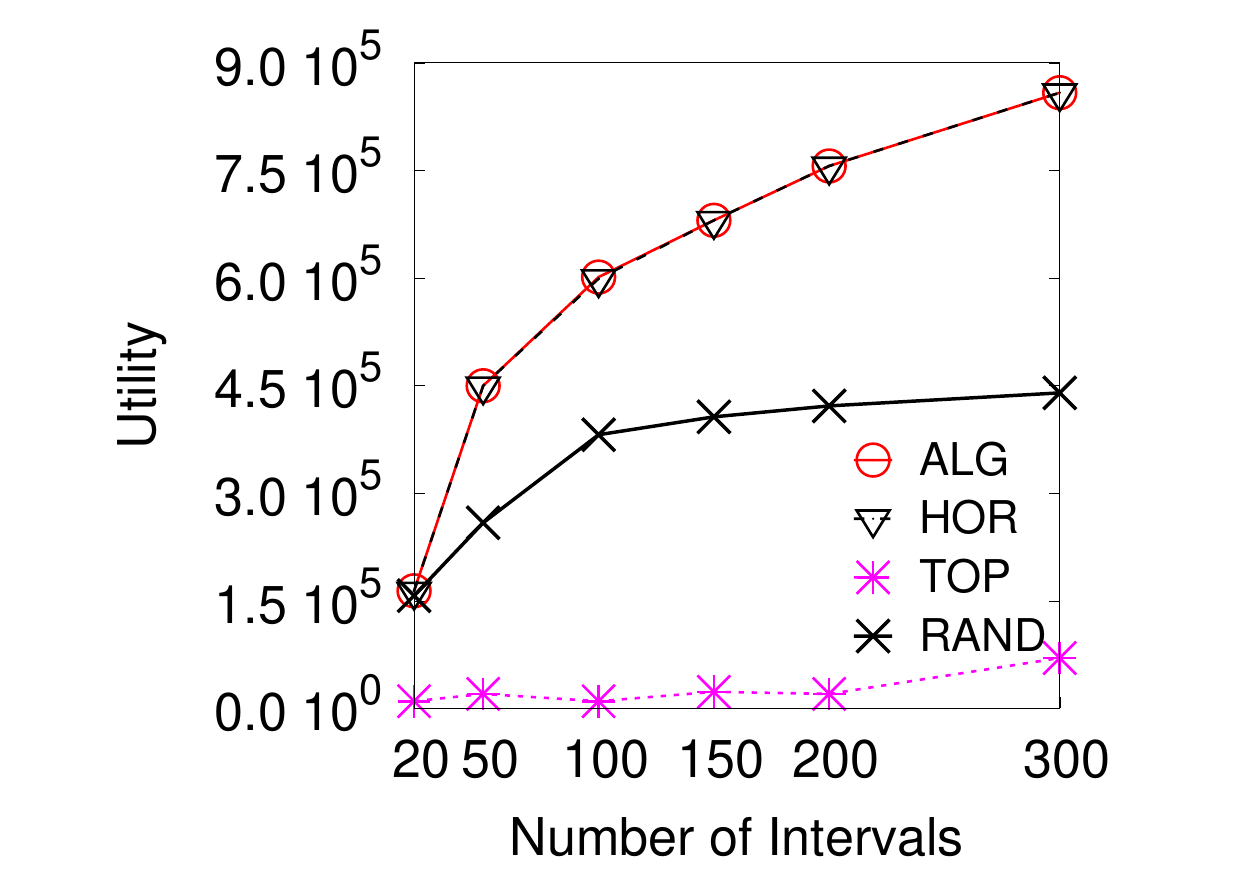}\label{fig:u_IN_i}}\hspace{-0.4cm}
\subfloat[Utility (\zip) ]{\includegraphics[height=1.3in]{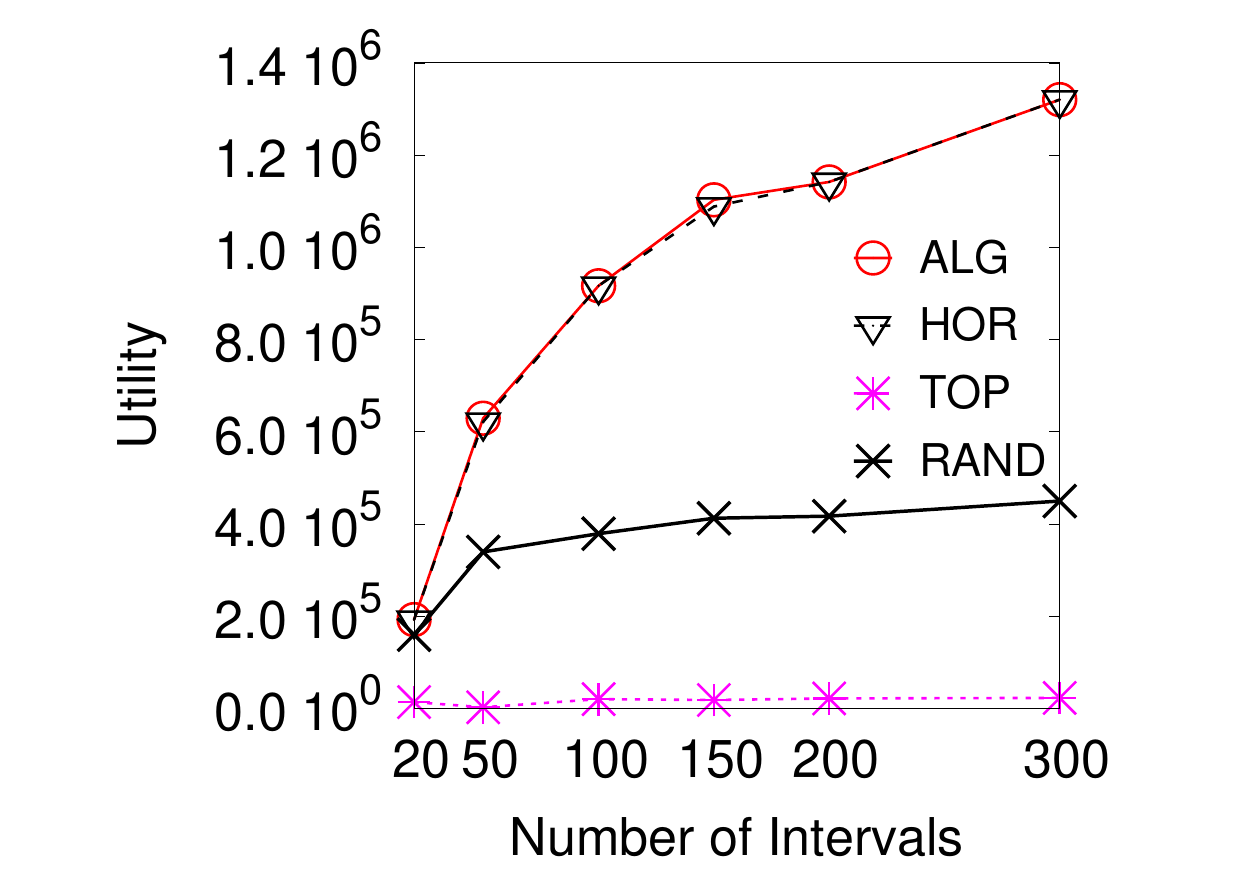}\label{fig:u_ZP_i}}\hspace{0cm}}
\\ \vspace{-5pt}
\hspace{-0.35cm}\mbox{
\subfloat[Computations  (\meetupc) ]{\includegraphics[height=1.3in]{c_CA_i-eps-converted-to}\label{fig:c_CA_i}}\hspace{-0.4cm}
\subfloat[Computations  (\music) ]{\includegraphics[height=1.3in]{c_YA_i-eps-converted-to}\label{fig:c_YA_i}}\hspace{-0.4cm}
\subfloat[Computations (\uni) ]{\includegraphics[height=1.3in]{c_IN_i-eps-converted-to}\label{fig:c_IN_i}}\hspace{-0.4cm}
\subfloat[Computations  (\zip)]{\includegraphics[height=1.3in]{c_ZP_i-eps-converted-to}\label{fig:c_ZP_i}}\hspace{0cm}}
\\ \vspace{-5pt}
\hspace{-0.35cm}\mbox{
\subfloat[Time (\meetupc) ]{\includegraphics[height=1.3in]{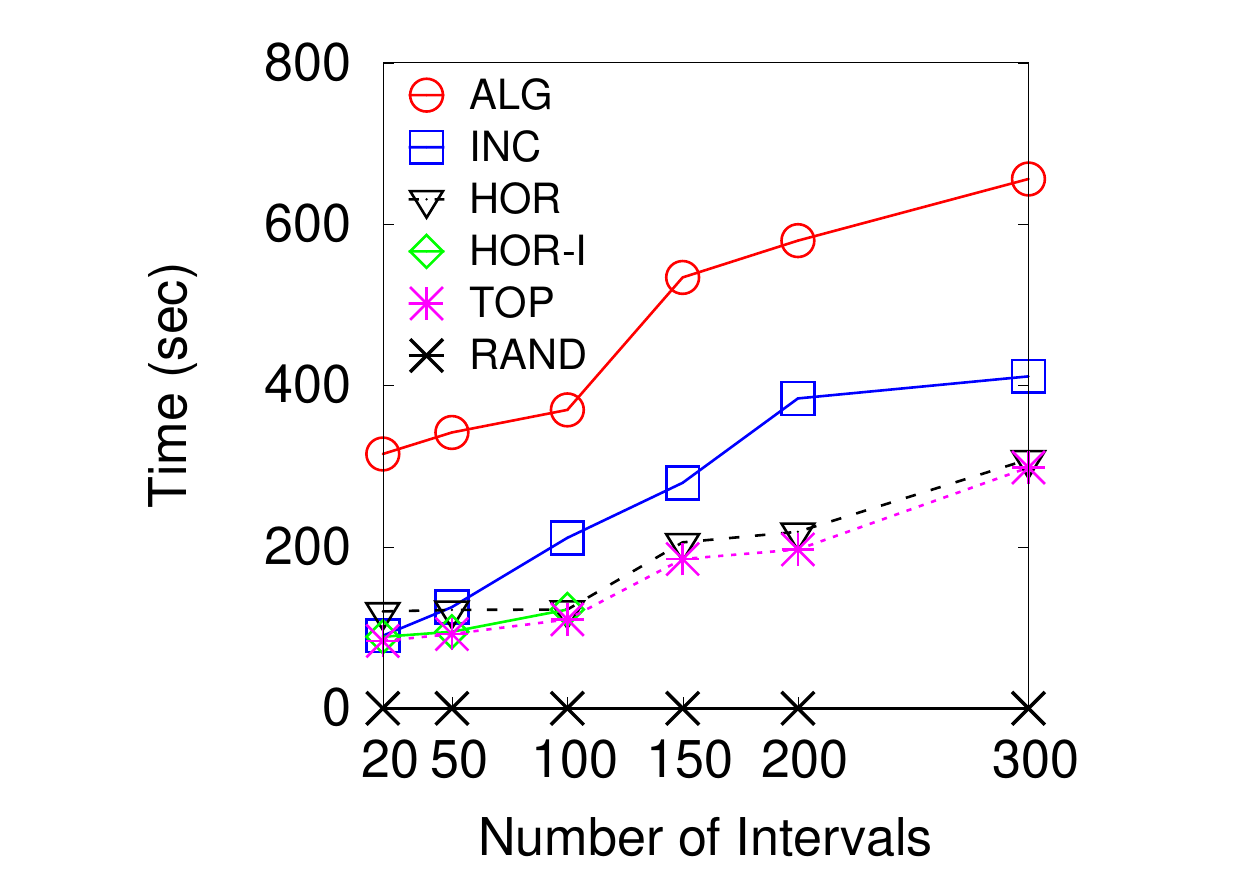}\label{fig:t_CA_i}}\hspace{-0.4cm}
\subfloat[Time (\music) ]{\includegraphics[height=1.3in]{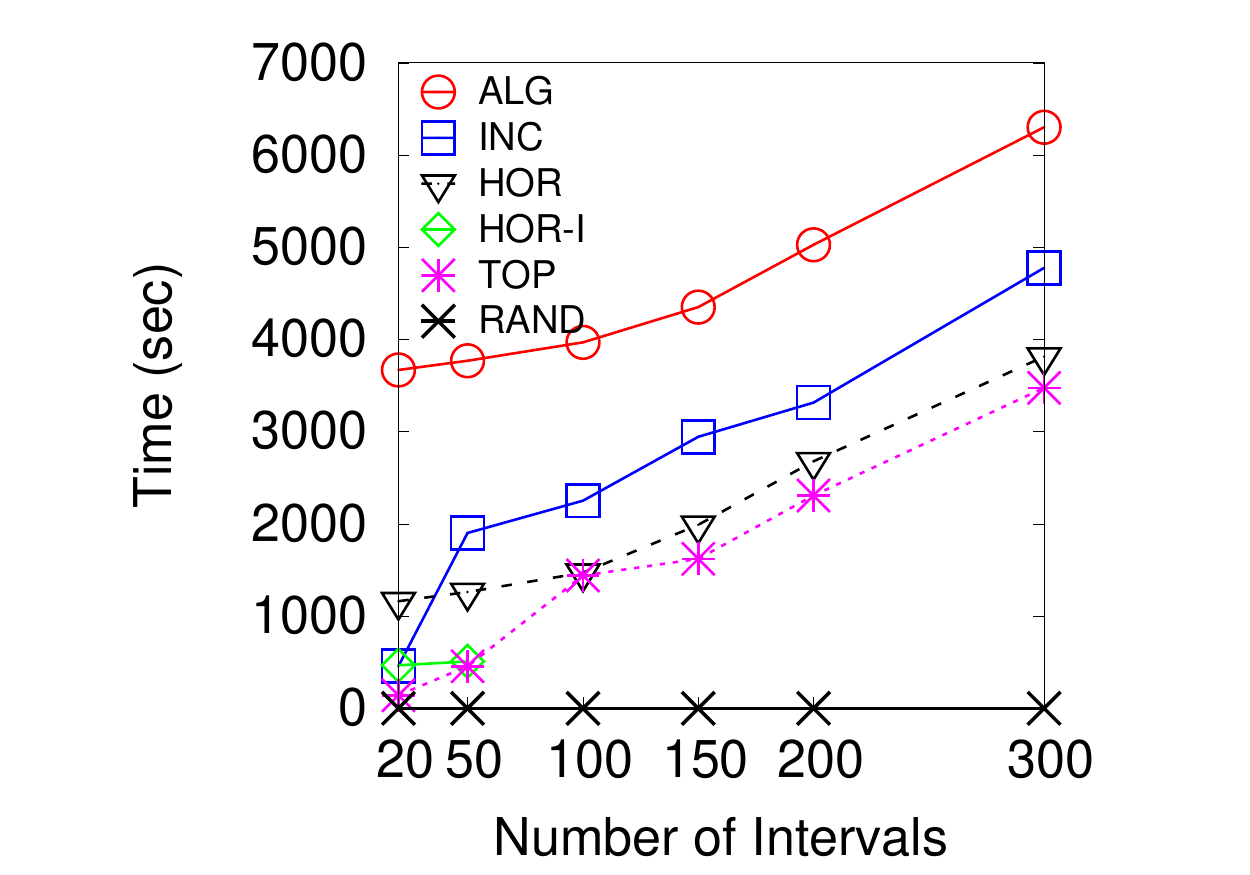}\label{fig:t_YA_i}}\hspace{-0.4cm}
\subfloat[Time (\uni) ]{\includegraphics[height=1.3in]{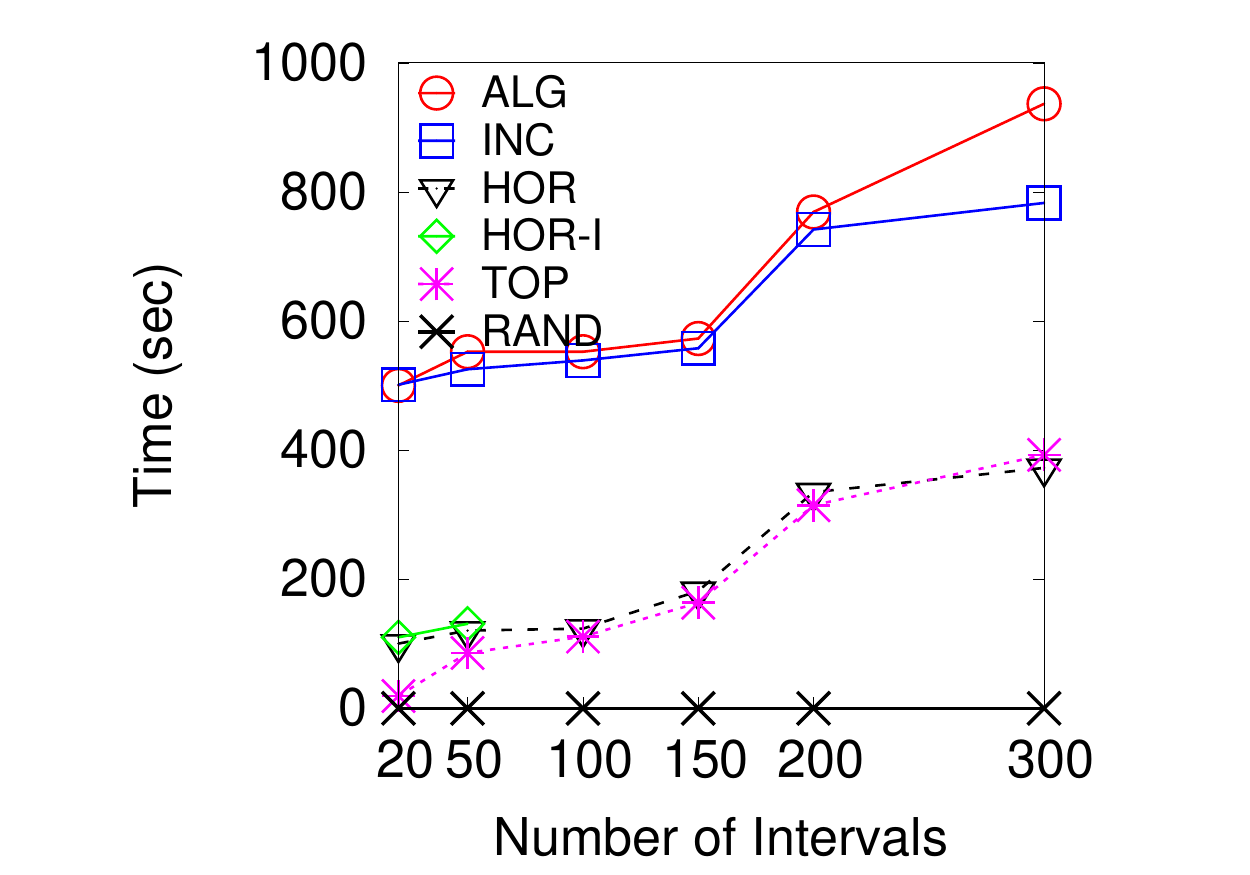}\label{fig:t_IN_i}}\hspace{-0.4cm}
\subfloat[Time (\zip) ]{\includegraphics[height=1.3in]{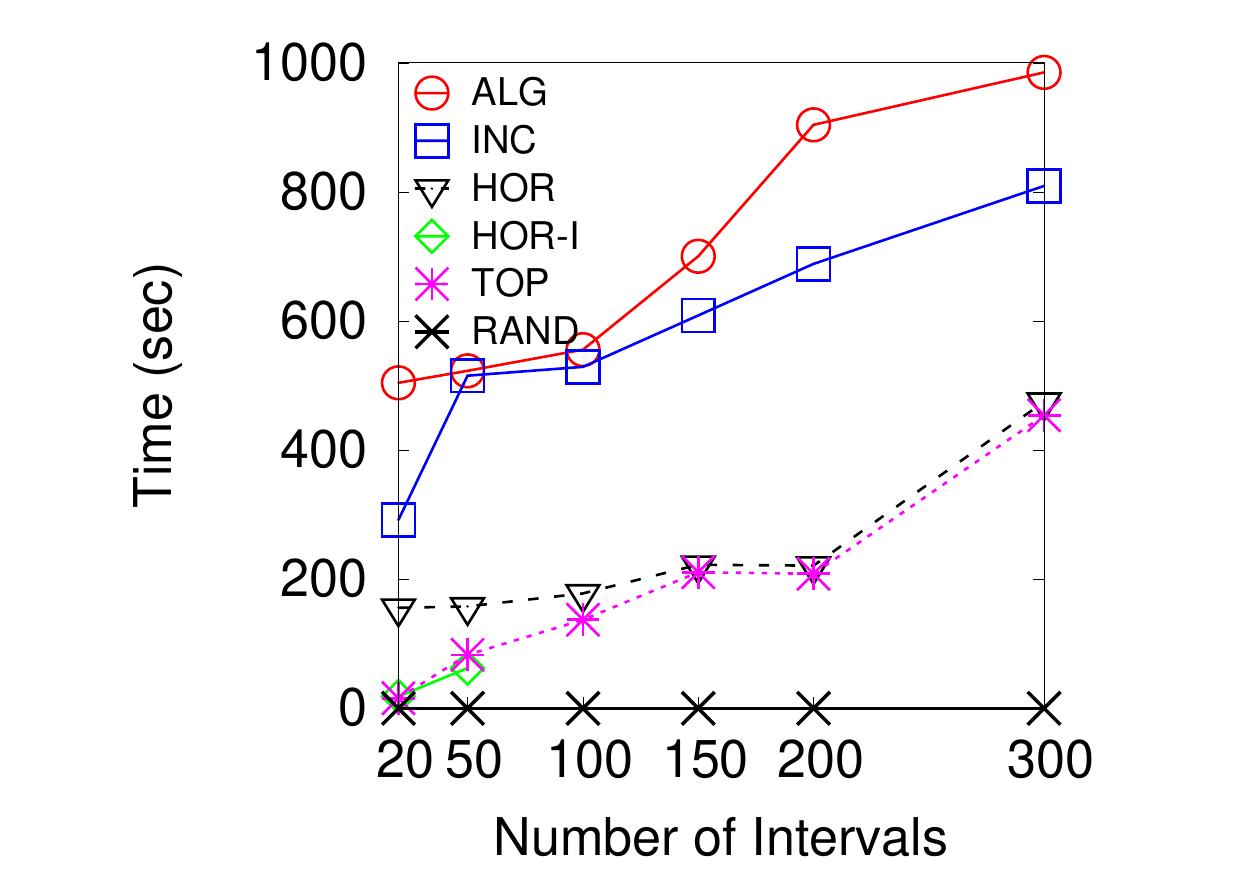}\label{fig:t_ZP_i}}\hspace{0cm}}
\caption{Varying the number of time intervals $|\T|$}
\label{fig:vart}
\vspace{-5pt}
\end{figure*}
}

\subsubsection{Effect of the Number of Scheduled Events}
\hfill \break
In the first experiment, we study the effect of varying the number of scheduled events $k$. 
%

\stitle{Utility.}
In terms of  {utility} (\mbox{Fig.~\ref{fig:u_CA_k}--\ref{fig:u_ZP_k}}),    we  observe that, in all cases, our \hor method has the same utility score as the \bsc 
(details are presented in Sect.~\ref{ex:summary}).
Further, the difference  between   \rand and the other methods increases, as $k$ increases.
This is reasonable considering the fact that the larger the $k$, the larger the number of ``better'', compared to random, selected~assignments. 

{Regarding the \uni dataset (Fig.~\ref{fig:u_IN_k}), we   observe the following. 
First, the difference between the random and the other methods is the smallest one, compared to the other datasets. 
Second, the difference between the methods is roughly the same for all $k$ values. 
The reason for the aforementioned  is that the uniform distribution results to 
utility values being very close, for all assignments. 
Thus, an effective assignment selection cannot significantly improve the overall utility. }

Finally, we can observe that \stat reports considerably low utility scores in all cases (which is also observed in the following experiments).
The reason is that \stat assigns the events to a small number of intervals. This results to a large number of parallel events which ``share'' assigned interval's utility. 

  \stitle{Computations.}
Regarding the    number of computations (\mbox{Fig.~\ref{fig:c_CA_k}--\ref{fig:c_ZP_k}}), we should mention that,  the  
computations that are performed due to updates increases with $k$,
 while the  number of {initially} computed scores is the same for all $k$.
Thus, the difference between the \bsc and our methods
increases with $k$.
Overall, we can observe that,  in all cases,  \bsc reports the 
larger number of computations, while \horb the lower (excluding the \stat baseline).
%

Regarding our methods, comparing \hor with \horb, we can observe that the difference between our \hor versions increases with~$k$,  with \horb  performing   noticeably less computations compared to \hor for large $k$.
An exception is reported in \uni dataset (Fig.~\ref{fig:c_IN_k}),
in which all bound-based methods (\bnd, \horb) report poor performance (as explained later).
  
Further, comparing the \hor with \bnd, for $k< 200$, \hor performs 
less computations  than \bnd.
However, in  the remaining cases where ${k \geq 200}$, 
\bnd outperforms \hor (with an exception in \uni).
The reason why \bnd performs better than \hor for ${k \geq 200}$ is that, 
in these cases \hor performs update computations; while, for the cases where $k \leq |\T|$  only the  {initial} computations are performed.

In the \uni dataset (Fig.~\ref{fig:c_IN_k}), we can observe that the bound-based 
 methods (i.e., \bnd, \horb) demonstrate poor performance, with \horb performing same as \hor, and \bnd performing worse than both of them. 
The reason lies to the uniform  distribution, where, as previously stated, 
the  scores are very close for all assignments. 
As a result, the values of the  bounds are larger than a small 
number of assignments' scores.
Hence, a small number of score updates can be avoided by exploiting bounds. 
%


   \stitle{Time.}
In terms of execution time (\mbox{Fig.~\ref{fig:t_CA_k}--\ref{fig:t_ZP_k}}),  we can observe
that time is determined by the number of computations performed.
\horb outperforms the other methods in all cases with \horb
being around 4 times faster than \bsc (for large $k$ in real datasets).

\vspace{2pt} 
\subsubsection{Effect of the Number of Time Intervals}
\hfill \break
In this experiment (Fig.~\ref{fig:vart}), we vary the number of time intervals~$|\T|$. 
Due to lack of space, for this and the following experiments,
the plots presenting the number of computations are omitted.


\alt{
}
{}

 \stitle{Utility.}
Regarding  {utility} (\mbox{Fig.~\ref{fig:u_CA_i} --\ref{fig:u_ZP_i}}), 
similarly to the previous experiment, our \hor algorithm performs 
the same as the \bsc. 
We   observe that, as the number of intervals increases, the utility of all methods increases too. 
This happens since the increase of available intervals  results to a smaller number of events
  assigned in the same interval, as well as to a larger number of candidate assignments. 
The former results to  the assignment scores (in general) being larger in cases where fewer parallel events take place.
The latter offers more options, which   possibly result to better assignments. 
\alt{
}{
Similarly to the previous experiment, the \hor and \bsc algorithms performs the same, in all cases. 
Additionally, the difference between the \rand and the other methods increases with $|\T|$. 
The reason is that, as the number of candidate assignments gets  larger,  
the  probability of selecting ``good'' assignments in random, gets lower.
Regarding \uni (Fig.~\ref{fig:c_IN_i}), we should mention that here, in contrast with the previous experiment (in which 
the difference between \rand and the other methods is stable) the difference increases with $|\T|$.
This explained as follows. 
As previously mentioned, due to uniform distribution the scores of the assignments are very close.  However, in this experiment, the increasing number of intervals increases the variance between scores, due to different number of competing events that may appear in the intervals. 
Therefore,  the ability of non-random methods  to select ``better'' assignments, increases with $k$. 
}

\alt{ \stitle{Time.}
 As for  execution time  (\mbox{Fig.~\ref{fig:t_CA_i}--\ref{fig:t_ZP_i}}), excluding \stat, the \horb is the most efficient in the  {cases 
 which differs from \hor} 
\linebreak 
 \mbox{(i.e., $|\T|< 100$)}; while in the rest cases, \hor is the most efficient.
Notice that, in general, \hor performs very close to \stat.  
Overall, \hor and \horb are about 2 to 4 times faster than the \bsc, and around 5 times faster for a    small number of intervals.
Finally, as explained in the previous experiment, we can observe that, also in this experiment, the bound-based methods (i.e., \bnd, \horb) are less effective in  \uni.

}{
 \sstitleM{Computations}: 
In terms of  {number of computations}, we should mention that,  in this experiment, 
the number computations that should be performed by \bsc due to updates are the same in all cases, while the  number of  {initially} computed scores increases with $|\T|$, for all methods.
This is the reason why the difference between the \bsc and the other methods 
is roughly the same in all cases. 

Overall, we can observe that,  in all cases,   the \hor and/or \horb outperform the \bsc and \bnd, while \bsc is the worst.  
We can see that \horb outperforms \hor.
Further, \bsc performs more computations than \bnd, in all cases, 
with their difference being smaller in the synthetic datasets. 
Finally, as explained in the previous experiment, we can observe that, also in this experiment, the bound-based methods (i.e., \bnd, \horb) are less effective in  \uni. 

 \sstitleM{Time}: 
As for  execution time, excluding baselines, the \horb is the most efficient in the  {cases 
which differs from \hor} \mbox{(i.e., $|\T|< 100$)}; while in the rest cases, \hor is the most efficient.
Notice that, in general, \hor performs very close to \stat.  
Overall, \hor and \horb are 2 to 4 times faster than the \bsc.
}

 \alt{
 \begin{figure}[t]
\centering
\vspace{-12pt}
\hspace*{-0.56cm}
\mbox{
\subfloat[Utility (\music) ]{\includegraphics[height=1.28in]{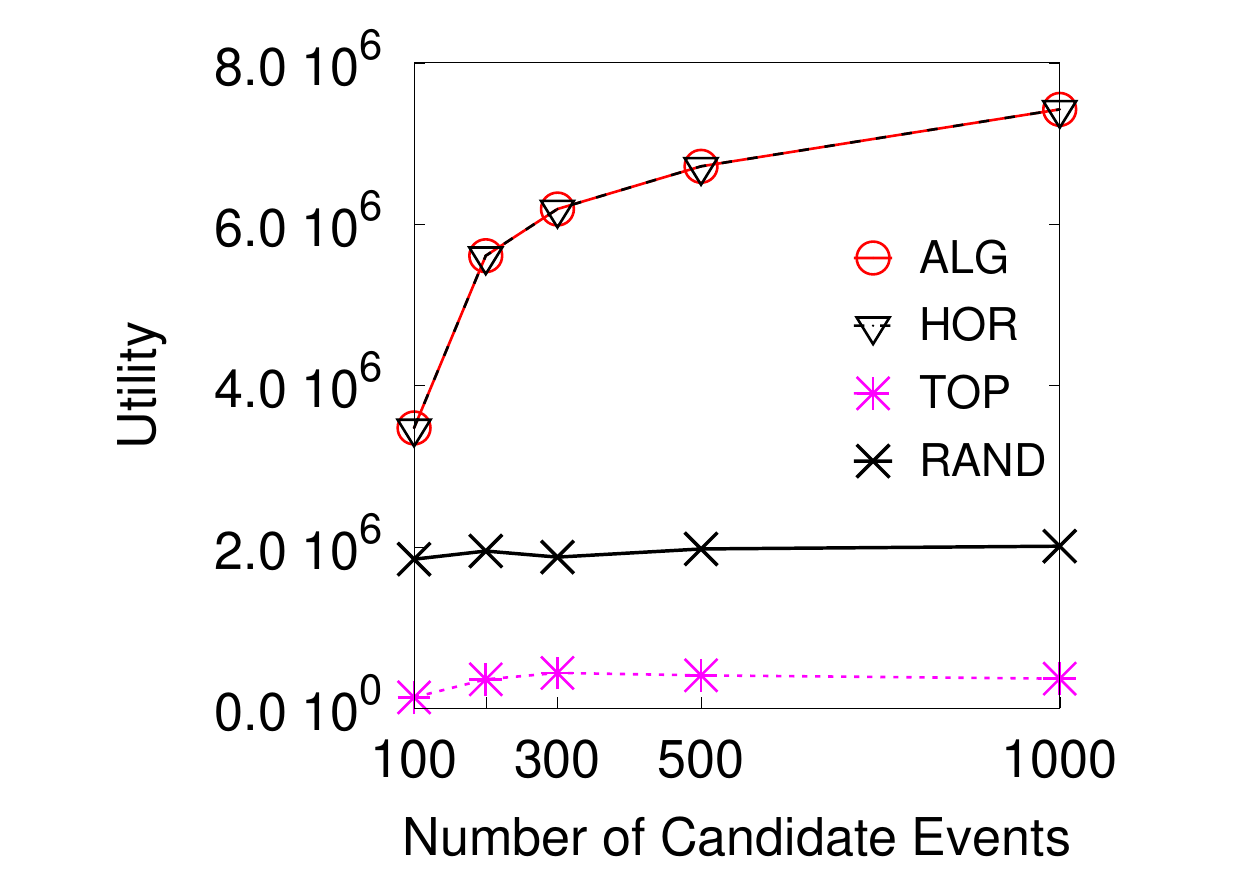}\label{fig:u_YA_e}}\hspace{-0.2cm}
\subfloat[Utility (\uni) ]{\includegraphics[height=1.28in]{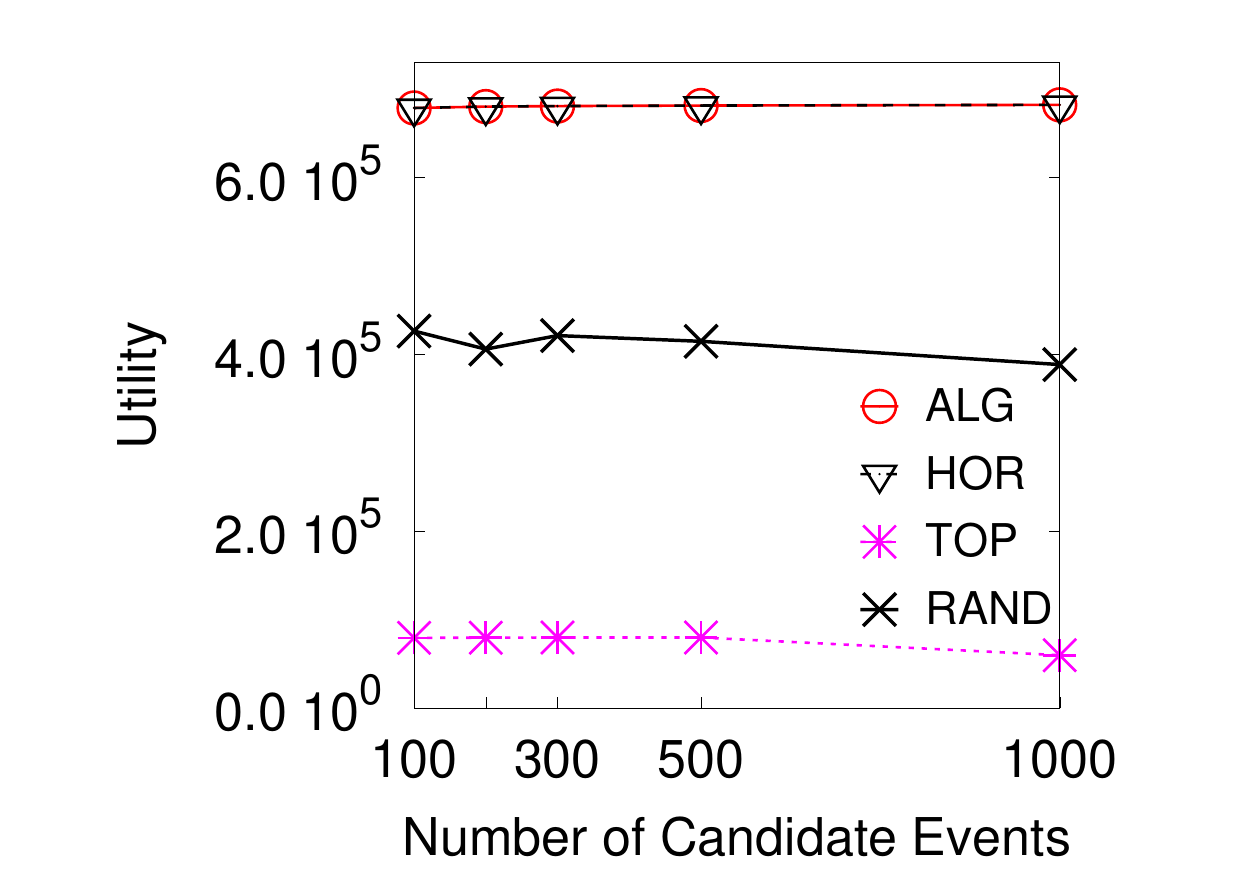}\label{fig:u_IN_e}}\hspace{-0.2cm}
}\\ \vspace{-10pt}
\hspace{-0.53cm}\mbox{
\subfloat[Time (\music) ]{\includegraphics[height=1.28in]{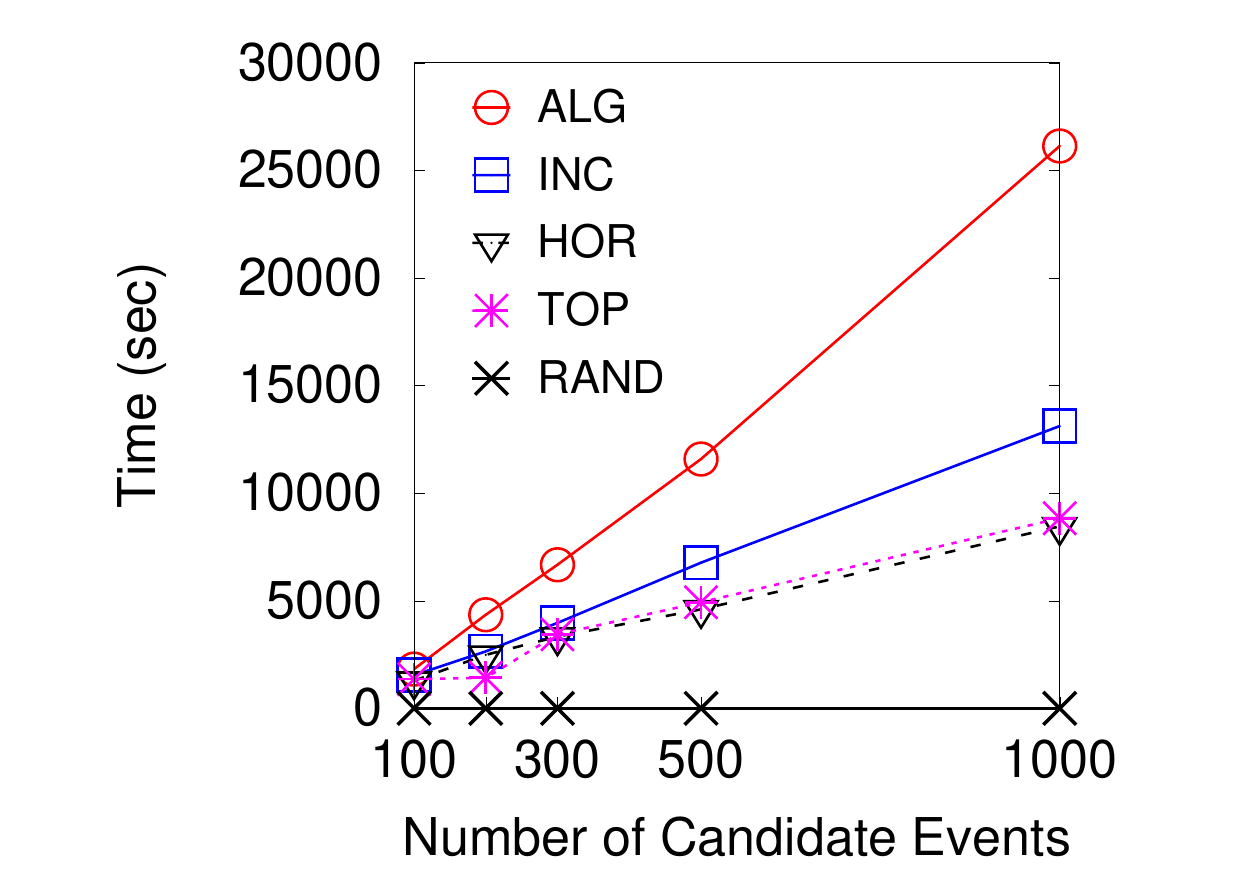}\label{fig:t_YA_e}}\hspace{-0.2cm}
\subfloat[Time (\uni) ]{\includegraphics[height=1.28in]{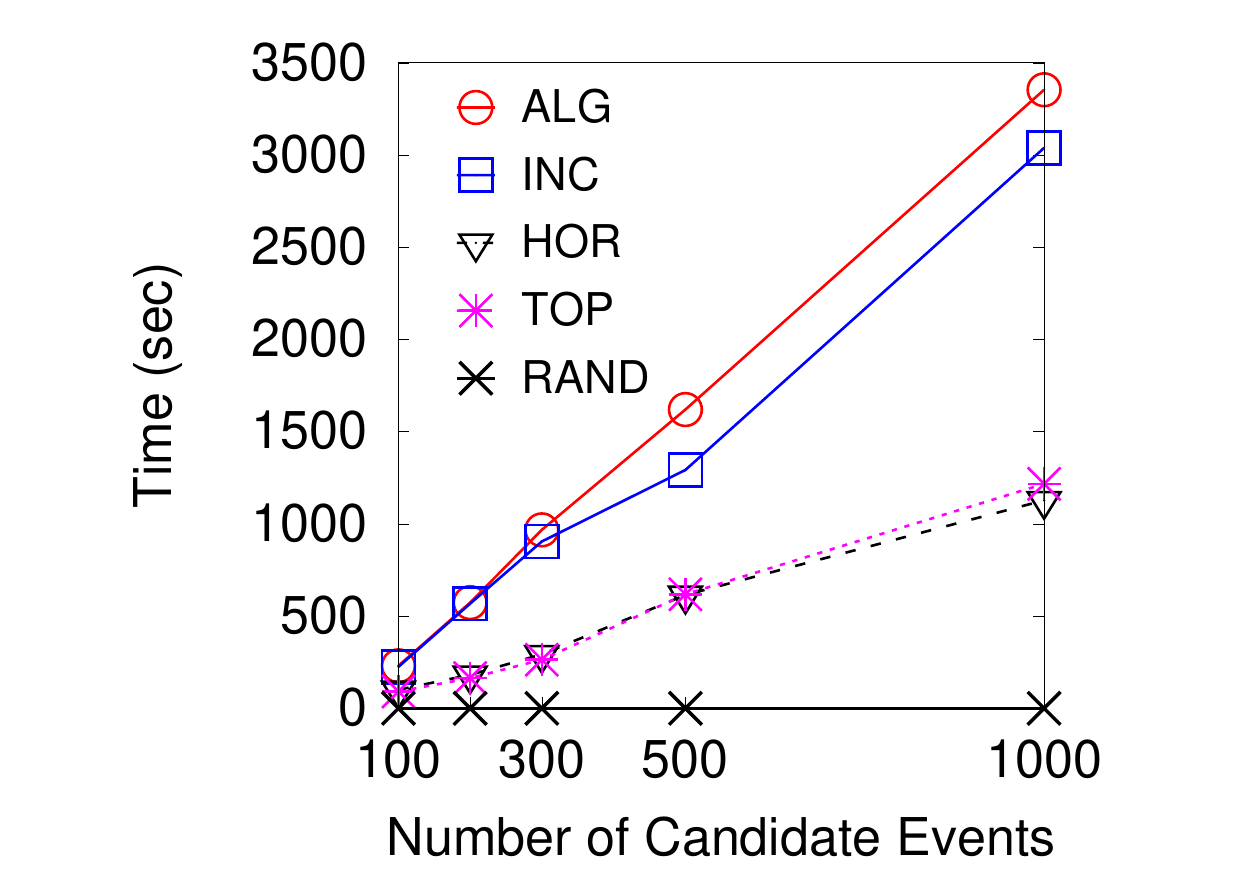}\label{fig:t_IN_e}}\hspace{-0.2cm}
}
\vspace{-8pt}
\caption{Varying the number of candidate events  $|\E|$}
\label{fig:vare}
\vspace{4pt}
\end{figure}

}
{
\begin{figure*}[t]
\centering
\vspace{-12pt}
\hspace{-0.35cm}\mbox{
\subfloat[Utility (\meetupc) ]{\includegraphics[height=1.3in]{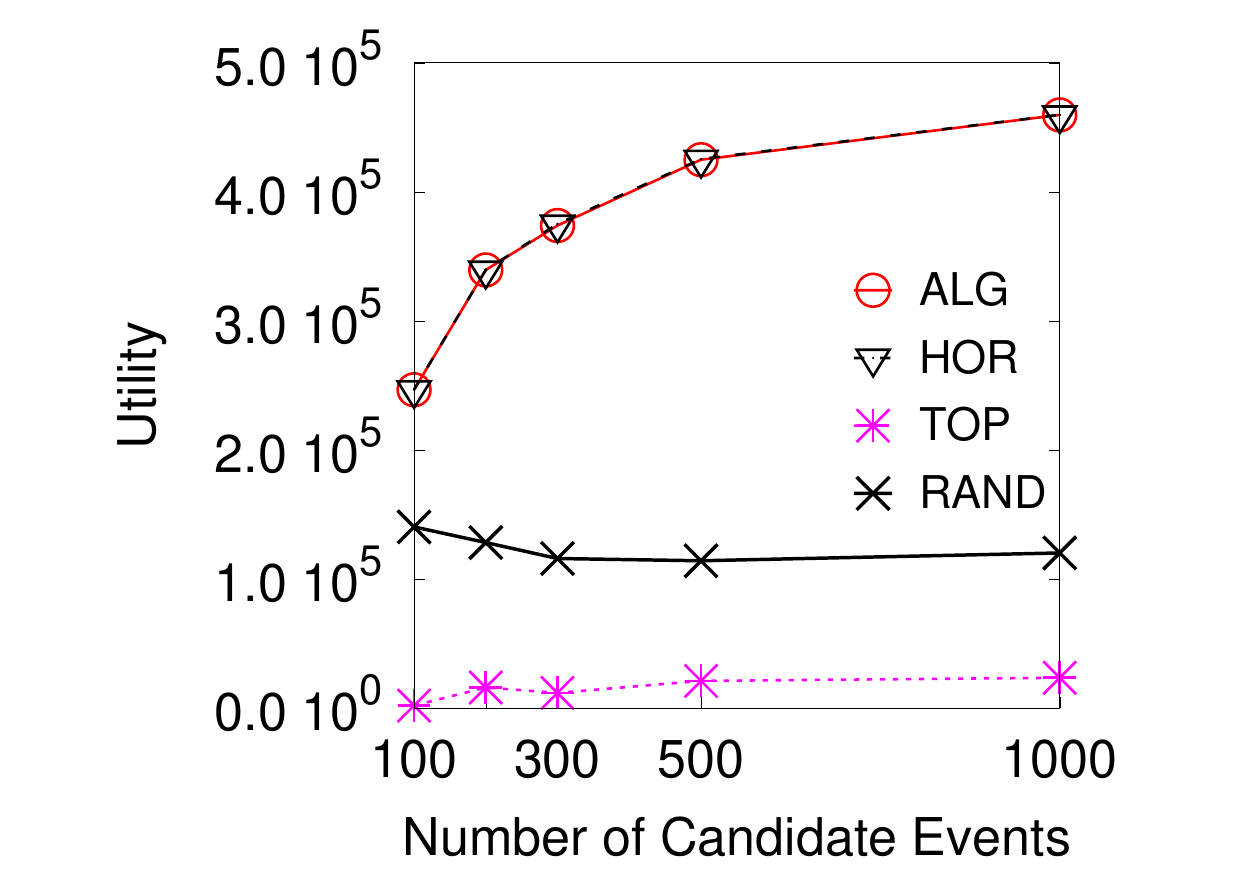}\label{fig:u_CA_e}}\hspace{-0.4cm}
\subfloat[Utility (\music) ]{\includegraphics[height=1.3in]{u_YA_e-eps-converted-to}\label{fig:u_YA_e}}\hspace{-0.4cm}
\subfloat[Utility (\uni) ]{\includegraphics[height=1.3in]{u_IN_e-eps-converted-to}\label{fig:u_IN_e}}\hspace{-0.4cm}
\subfloat[Utility (\zip) ]{\includegraphics[height=1.3in]{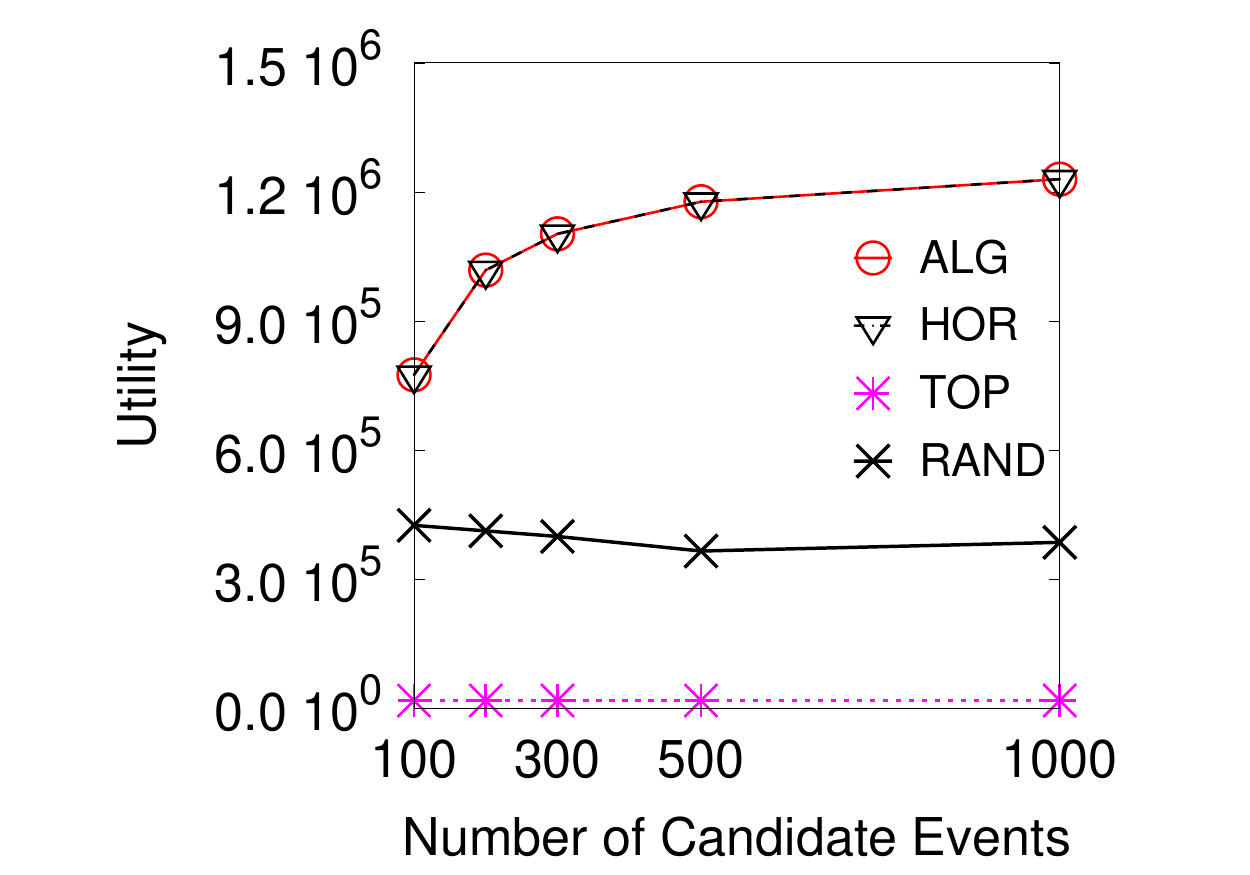}\label{fig:u_ZP_e}}\hspace{0cm}}
\\ \vspace{-5pt}
\hspace{-0.35cm}\mbox{
\subfloat[Computations  (\meetupc) ]{\includegraphics[height=1.3in]{c_CA_e-eps-converted-to}\label{fig:c_CA_e}}\hspace{-0.4cm}
\subfloat[Computations  (\music) ]{\includegraphics[height=1.3in]{c_YA_e-eps-converted-to}\label{fig:c_YA_e}}\hspace{-0.4cm}
\subfloat[Computations (\uni) ]{\includegraphics[height=1.3in]{c_IN_e-eps-converted-to}\label{fig:c_IN_e}}\hspace{-0.4cm}
\subfloat[Computations  (\zip)]{\includegraphics[height=1.3in]{c_ZP_e-eps-converted-to}\label{fig:c_ZP_e}}\hspace{0cm}}
\\ \vspace{-5pt}
\hspace{-0.35cm}\mbox{
\subfloat[Time (\meetupc) ]{\includegraphics[height=1.3in]{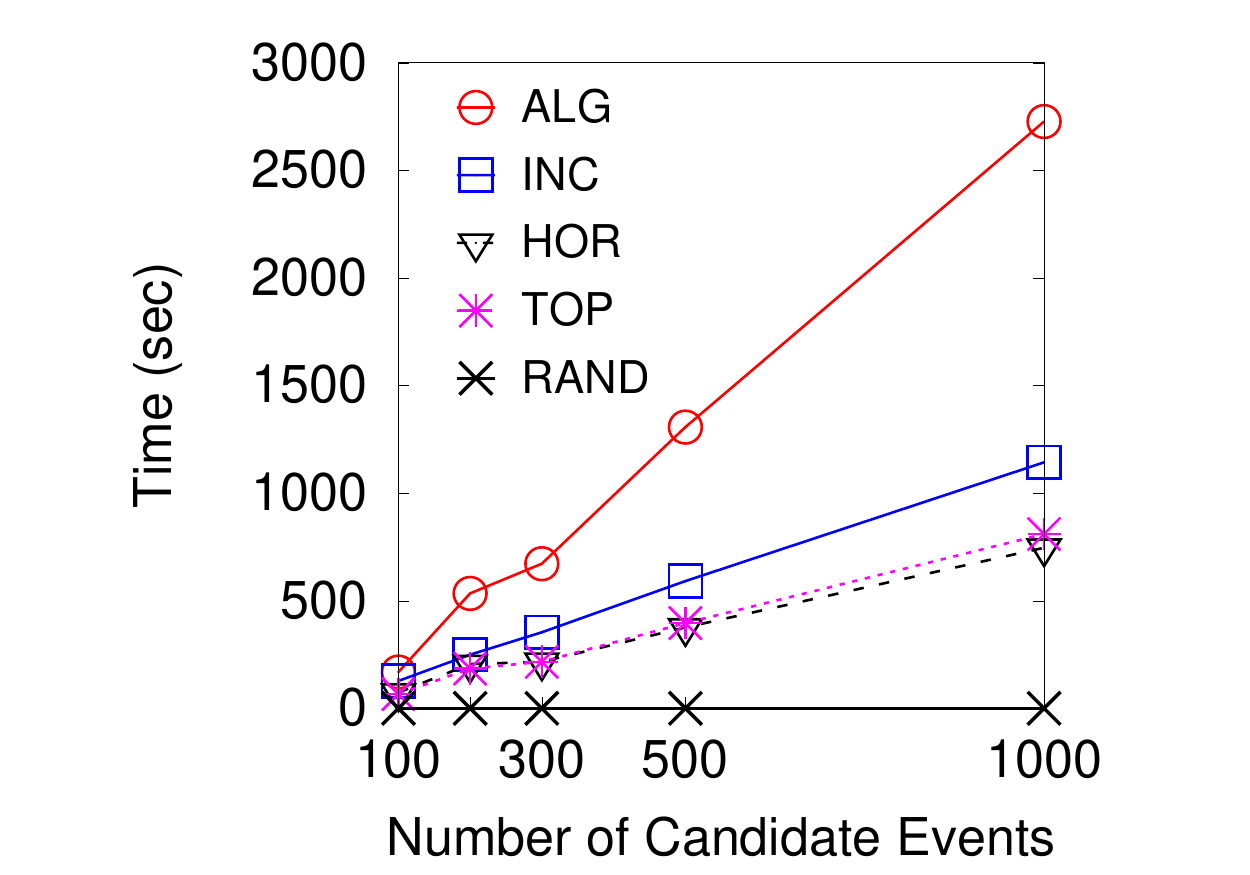}\label{fig:t_CA_e}}\hspace{-0.4cm}
\subfloat[Time (\music) ]{\includegraphics[height=1.3in]{t_YA_e-eps-converted-to}\label{fig:t_YA_e}}\hspace{-0.4cm}
\subfloat[Time (\uni) ]{\includegraphics[height=1.3in]{t_IN_e-eps-converted-to}\label{fig:t_IN_e}}\hspace{-0.4cm}
\subfloat[Time (\zip) ]{\includegraphics[height=1.3in]{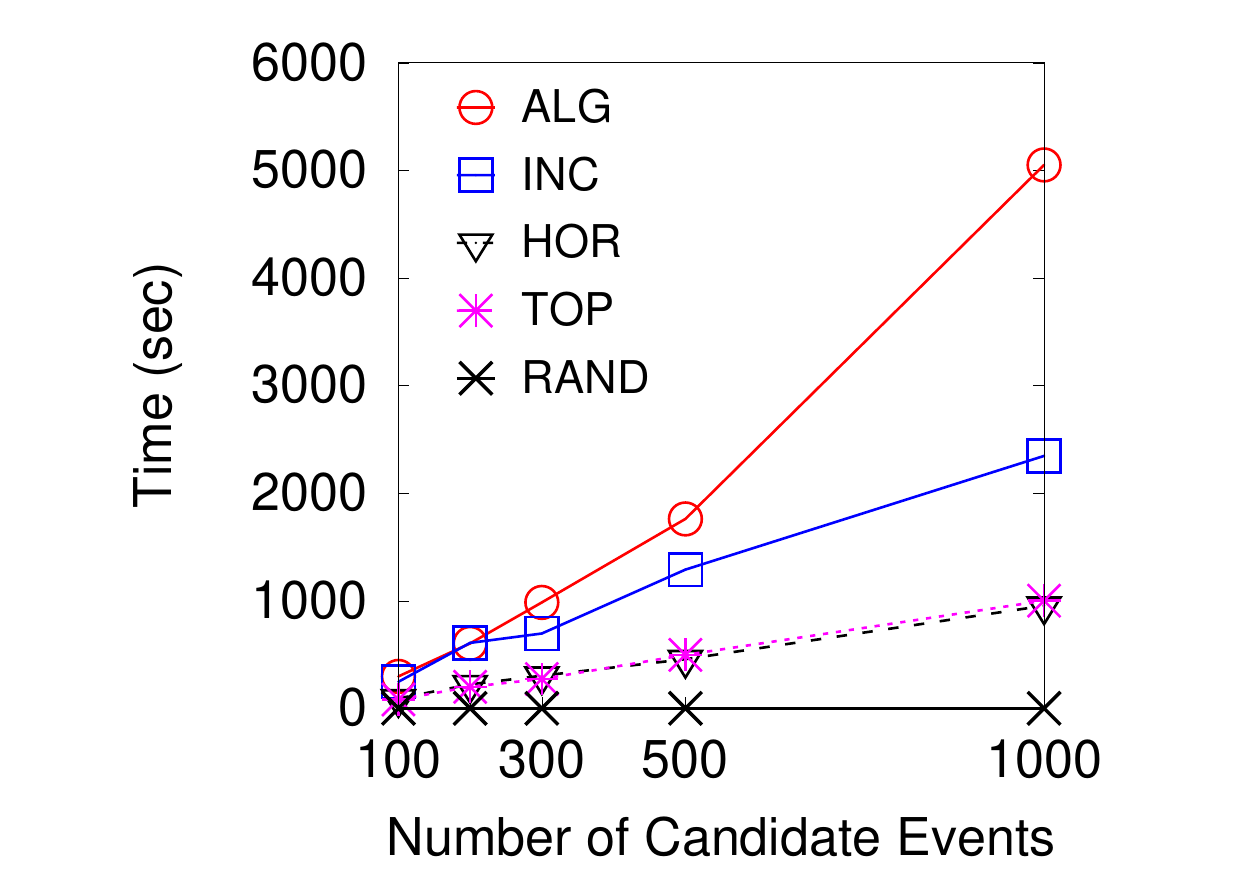}\label{fig:t_ZP_e}}\hspace{0cm}}
\caption{Varying the number of candidate events  $|\E|$}
\label{fig:vare}
\vspace{-9pt}
\end{figure*}
}

\subsubsection{Effect of the Number of Candidate Events}
\hfill \break
We next study the effect of varying the number of candidate events~$|\E|$.
 Note that, in this experiment, since $k<|\T|$,     {\horb is identical to \hor}.
Due to lack of space, in this experiment, the plots for the \meetupc and \zip  are not presented, since  they are similar to \music.

  \stitle{Utility.}
  Also in this experiment   (\mbox{Fig.~\ref{fig:u_YA_e}--\ref{fig:u_IN_e}}) our HOR method has the same utility score as the \bsc in all cases.
  We  observe that  the utility of \bsc and \hor increases with $|\E|$ 
  (with an exception in~\uni). 
  On the other hand,  for \rand it is either stable or is  decreasing. 
This happens since the increase of $|\E|$ results to more candidate assignments.
So, there are more options for the \bsc and \hor methods, while for \rand it is less possible 
to select  ``good'' assignments. 
Notice that, in the \uni case (Fig.~\ref{fig:u_IN_e}), the utility for the non-random methods remains stable. 
The reason is that increasing the number of ``similar'' events (as previously explained) cannot result to better assignments.

 \alt{  
 \stitle{Time.}
Also in this experiment (\mbox{Fig.~\ref{fig:t_YA_e}--\ref{fig:t_IN_e}}), our \hor method outperforms the other, 
with \bnd having noticeably bad performance in \uni,  compared to \hor (as in the  previous experiments).
Further, the difference between   \bsc and our methods increases with $|\E|$, due to the  increasing number of update computations.
Overall, in general \hor  is around 3 to 4 times faster than \bsc, 
and up to  5 times faster in  \zip dataset specifically. 
}
{\sstitleM{Computations}:
As for computations, we should note that, here, 
the  number of both  {initial} and updates computations increases with $|\E|$.
In all cases, \hor outperforms \bnd, while \bsc is the worst. 
Further, the difference between   \bsc and the other methods increases with $|\E|$, due to the  increasing number of update computations.
Additionally, as in the  previous experiments, the bound exploited by \bnd, is less effective in the \uni dataset. 
Overall,  in most cases, \hor performs slightly more than  half computations
that \bsc performs.

 \sstitleM{Time}:
Also in this experiment ((\mbox{Fig.~\ref{fig:t_CA_e}--\ref{fig:t_ZP_e}}), \hor outperforms the other methods (excluding \stat), 
with \bnd having noticeably bad performance in \uni,  compared to \hor. 
Overall, \hor  is around 3 to 5 times faster than \bsc.
  }

 \alt{ 
\begin{figure}[t]
\vspace{-39pt}
\centering
\hspace*{-0.35cm}
\subfloat[Time  \small{($|\T|=150$)}]{\includegraphics[height=1.31in]{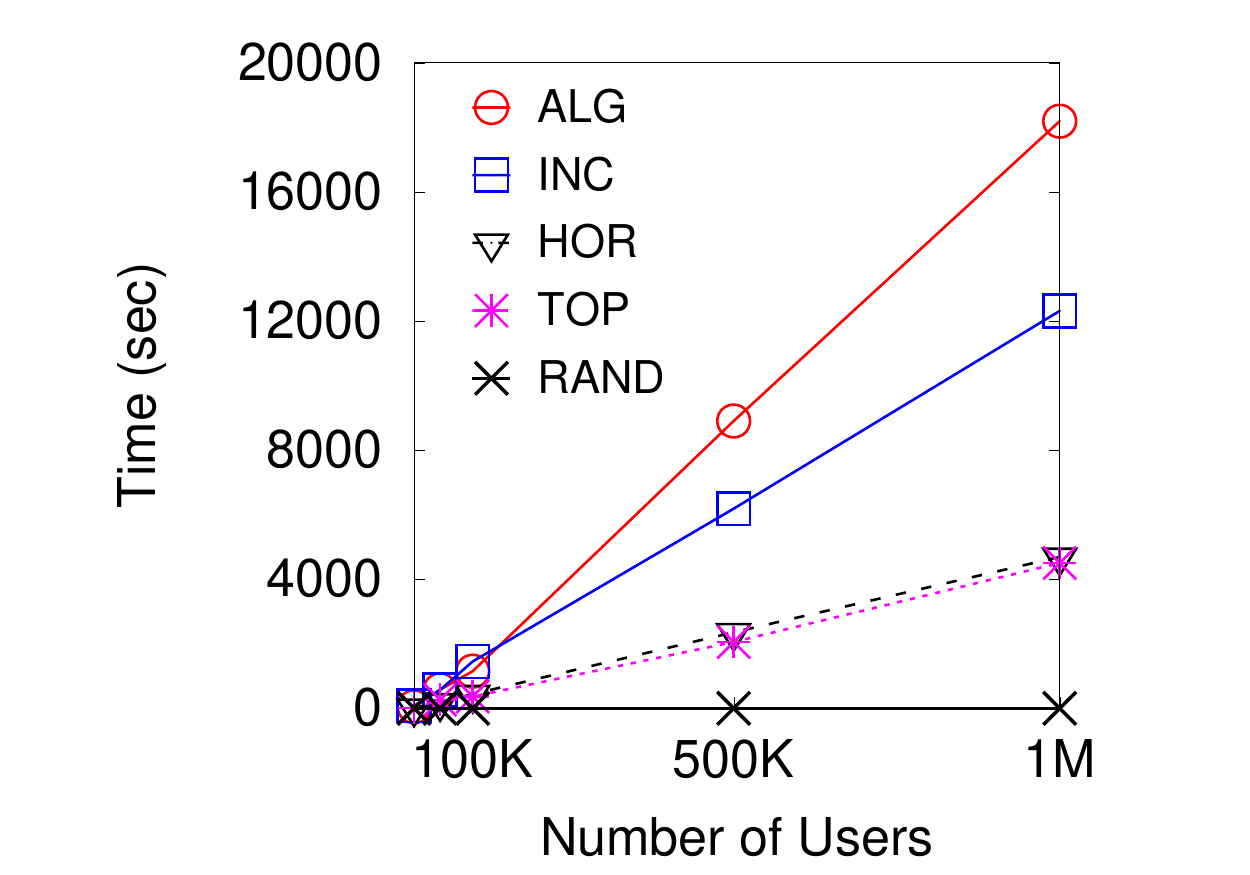}\label{fig:t_IN_u}} \hspace*{-0.45cm}
\subfloat[Time \small{($|\T|=65$)}]{\includegraphics[height=1.31in]{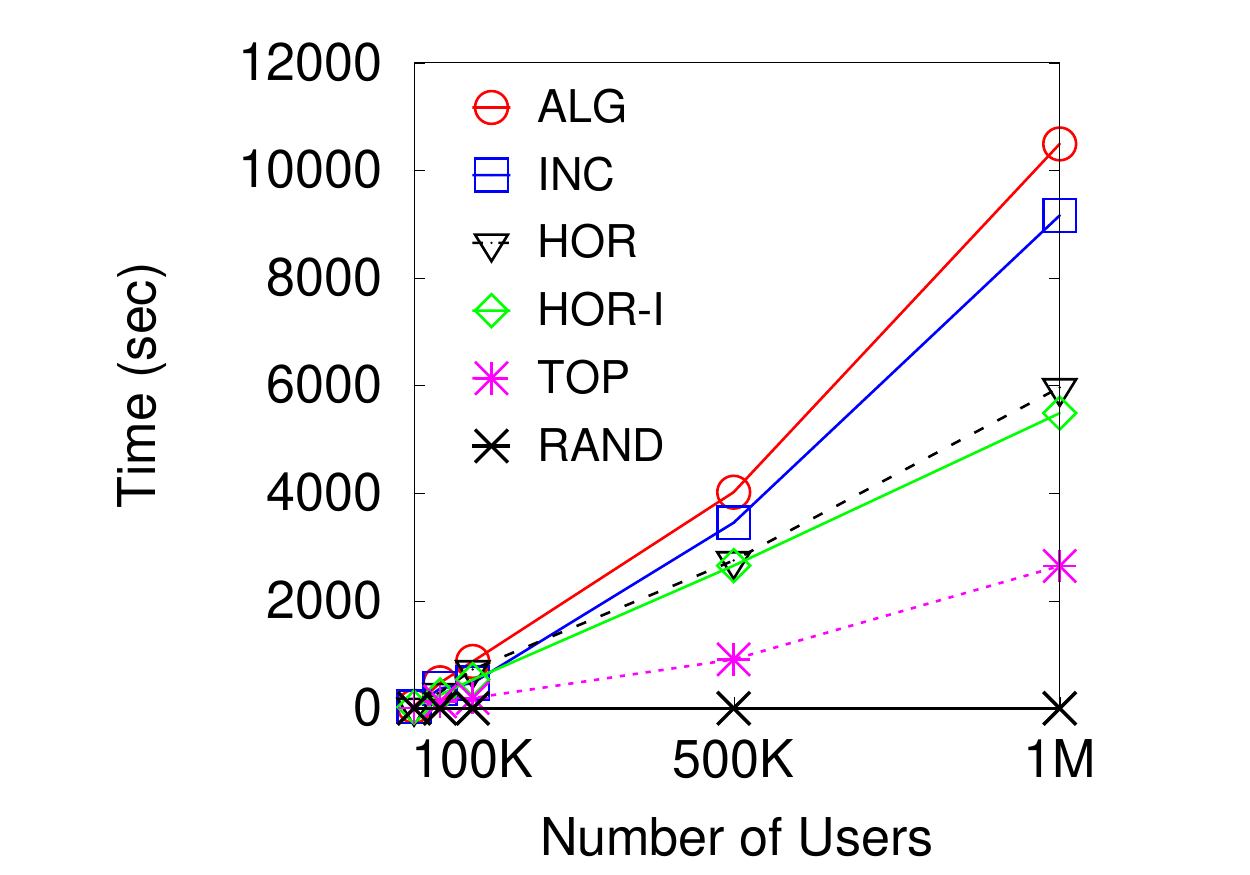}\label{fig:t_IN_u2}}\hspace{0cm}
\vspace{-9pt}
\caption{Varying the number of users $|\U|$ (\normalfont{\uni Dataset}\textbf{)}}
\label{fig:varu}
\vspace{0pt}
\end{figure}
}{
\begin{figure*}[t]
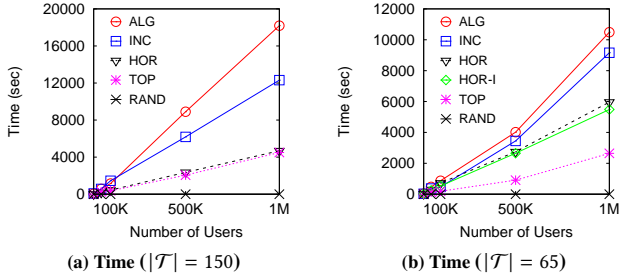

\centering
\vspace{-9pt}
\hspace{-0.35cm}\mbox{
\subfloat[Utility \small{($|\T|=150$)}]{\includegraphics[height=1.3in]{u_IN_u}\label{fig:u_IN_u}}\hspace{-0.2cm}
\subfloat[Time  \small{($|\T|=150$)}]{\includegraphics[height=1.3in]{t_IN_u}\label{fig:t_IN_u}}\hspace{-0.4cm}
 \subfloat[{Computations} \small{($|\T|=65$)}]{\includegraphics[height=1.3in]{c_IN_u2}\label{fig:c_IN_u2}}\hspace{-0.4cm}
\subfloat[Time \small{($|\T|=65$)}]{\includegraphics[height=1.3in]{t_IN_u2}\label{fig:t_IN_u2}}\hspace{0cm}}
\vspace{-5pt}
\caption{Varying the number of users $|\U|$ (\normalfont{\uni Dataset}\textbf{)}}
\label{fig:varu}
\vspace{-0pt}
\end{figure*}
}

\alt{
\vspace*{2pt}
\subsubsection{Effect of the Number of Users}
\hfill \break
We then study the effect of varying the number of users (Fig.~\ref{fig:varu}).
Results for the \zip dataset are omitted since they are similar to the ones reported for \uni. 
Here, the \horb algorithm cannot be defined with the default parameters setting 
(${k=100, \T=150}$). 
Hence, in order to also study \horb, we examine a supplementary experiment (Fig.~\ref{fig:t_IN_u2}), where $\T=65$. 
Note that, this setting ($k=100, \T=65$) corresponds to the  
average case  for the \hor and \horb algorithms (w.r.t.\ the relation between $k$ and $|\T|$;
see Sect.~\ref{sec:horAnal} \& \ref{sec:horbanal}).

In terms of utility (the plot is omitted due to lack of space), as   expected, the utility increases with the number of users. 
The \hor and \bsc methods have the same utility scores in all cases. 
%
Regarding performance, in the first experiment 
(Fig.~\ref{fig:t_IN_u}), \hor performs increasingly better than \bnd and \bsc, 
as the number of users increases.
In the second experiment (Fig.~\ref{fig:t_IN_u2}), 
for larger numbers of users
\linebreak 
(i.e.,~$|\U|>100$K), \bnd performs close to \bsc. 
On the other hand, \hor and \horb outperforms \bnd, with the difference increases with $|\U|$. 
%
Overall, in the first experiment, \hor  is  around 3 to 4 times faster than \bsc;
in the second one, \hor and \horb are  around 2~times faster than \bsc.
}{
\stitle{Effect of the Number of Users.}
We then study the effect of varying the number of users $|\U|$.
The results are presented in Figure~\ref{fig:varu}.
Results for the \zip dataset are omitted since they are similar to the ones reported for \uni. 
 
In this experiment, the \horb algorithm cannot be defined with the default parameters setting (i.e., $k=100, \T=150$). 
Hence, in order to also study \horb, we examine a supplementary experiment, where $\T=65$ (Fig.~\ref{fig:c_IN_u2}~\&~\ref{fig:t_IN_u2}). 
Note that, this setting (i.e., $k=100, \T=65$) corresponds to the  
average case  for the \hor and \horb algorithms (w.r.t.\ the relation between $k$ and $|\T|$;
see Sect.~\ref{sec:horAnal}).

In terms of utility, as   expected, the utility increases with the number of users. 
Additionally,  the \hor and \bsc methods have the same utility scores in all cases. 
%
Regarding performance, in the first experiment 
(Fig.~\ref{fig:t_IN_u2}), \hor performs increasingly better than \bnd and \bsc, 
as the number of users increases.
Further, in the second experiment (Fig.~\ref{fig:c_IN_u2}~\&~\ref{fig:t_IN_u2}), 
for larger numbers of users (i.e.,~$|\U|>100$K), \bnd performs close to \bsc. 
On the other hand, \hor and \horb outperforms \bnd, with the difference increases with $|\U|$. 
Recall that the performance of \bnd and \horb is lower in the \uni dataset.
That why \horb and \bnd  perform slightly better than \hor and \bsc, respectively.
Overall, in the first experiment, \hor  is  around 3 to 4 times faster than \bsc;
in the second one, \hor and \horb are  around 2 times faster than \bsc.
}

\begin{figure}[t]
\centering
\vspace{-39pt}
\hspace{-0.4cm}\mbox{
\subfloat[Utility  vs.\ Locations]{\includegraphics[height=1.28in]{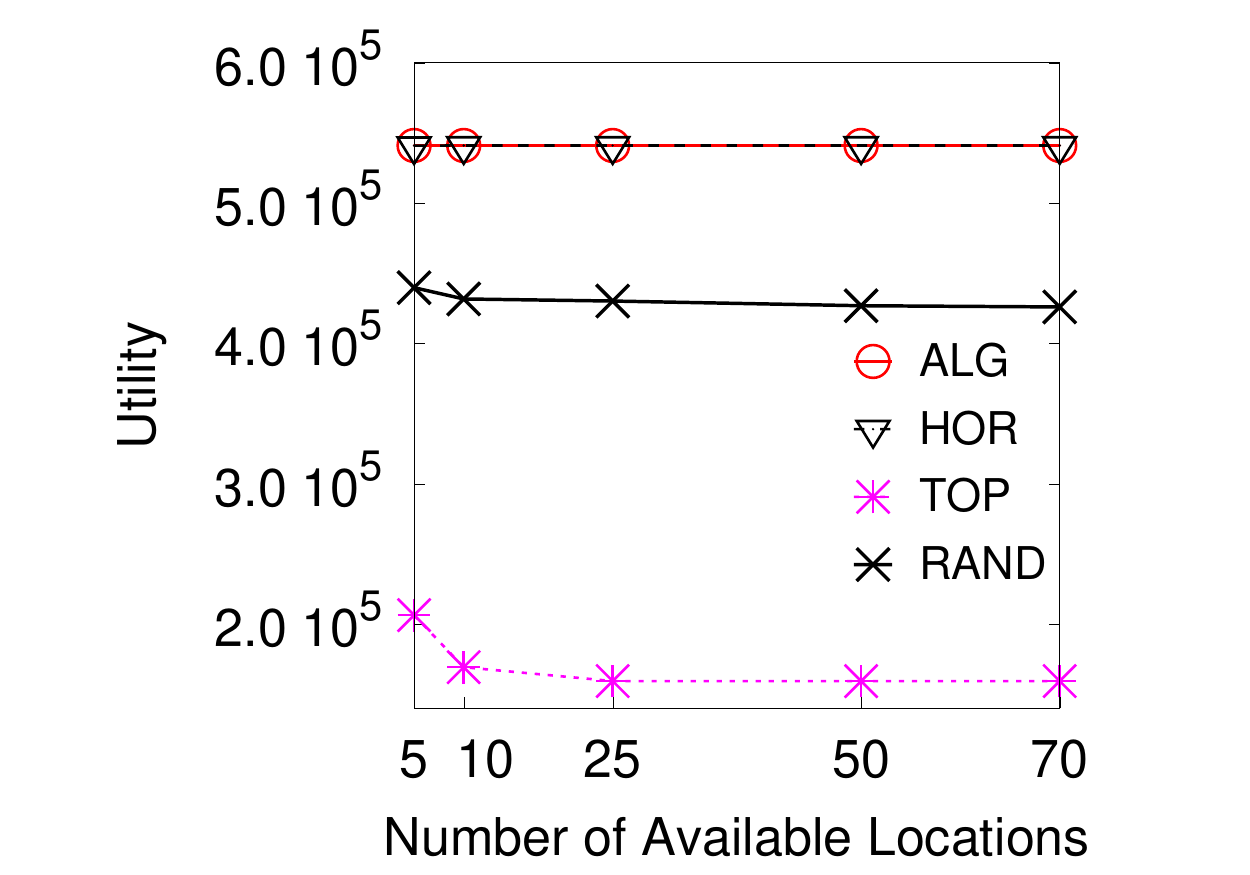}\label{fig:u_IN_l}}\hspace{-0.45cm}
 \subfloat[Time vs.\ Locations]{\includegraphics[height=1.28in]{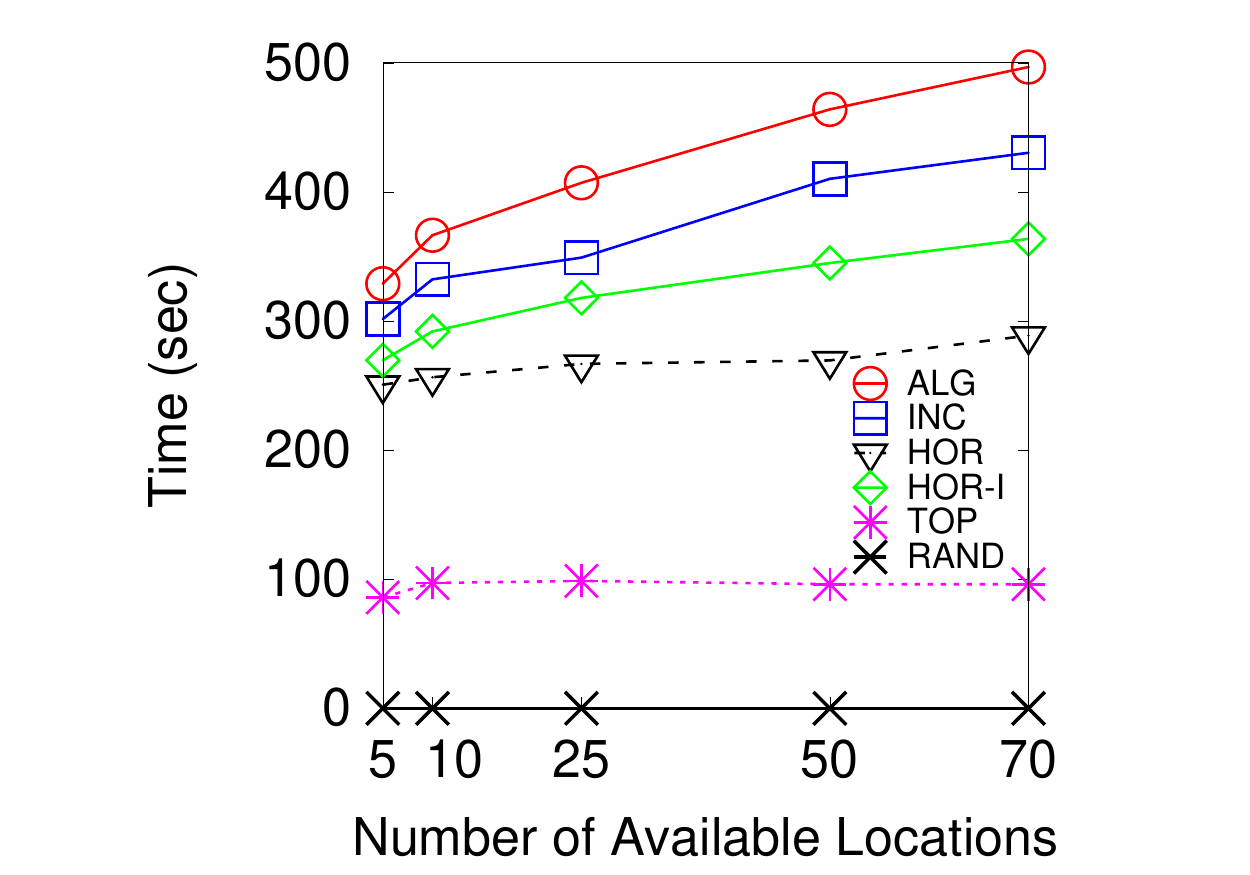}\label{fig:t_IN_l}}\hspace{0cm}}
 \vspace{-6pt}
\caption{Varying the number of  locations (\normalfont{\uni, $|\T|=65$}\textbf{)}}
\label{fig:varla}
\vspace{-0pt}
\end{figure}

\alt{

\subsubsection{Effect of the Number of Available Locations}
\hfill \break
In this experiment we vary the number of available locations of each candidate event (Fig.~\ref{fig:varla}).
 The results correspond to the \uni dataset; though, similar results are reported in all datasets.
 
We can observe, that the utility score (Fig.~\ref{fig:u_IN_l}) remains almost unaffected for the \bsc and \hor methods, while \stat and \rand perform slightly better in 5 locations.
This is expected, since, as the number of locations decreases, the number of feasible assignments decreases too. 
Regarding the execution time (Fig.~\ref{fig:t_IN_l}), in all methods, increases with number of locations.  
This is due to the fact that the number of feasible assignments (as well as the computations) increases too. 
}
{\stitle{Effect of the Number of Locations \& Resources.}
The results presented in the following experiments correspond to the \uni dataset; though, similar results are reported in all datasets.
Finally, similarly to the previous experiment, we set $|\T| = 65$. 
 
Initially, we vary the number of available locations.  
The results are presented in Figures~\ref{fig:u_IN_l}~\&~\ref{fig:t_IN_l}.
We can observe, that the utility score (Fig.~\ref{fig:u_IN_l}) remains almost unaffected for the \bsc and \hor methods, while \stat and \rand perform slightly better in 5 locations.
This is expected, since, as the number of locations decreases, the number of feasible assignments decreases too (i.e., the number of available options).  
Regarding the execution time (Fig.~\ref{fig:t_IN_l}), in all methods, increases with number of locations.  
This is due to the fact that the number of feasible assignments (as well as the number of computations) increases too.

In the next experiment, we vary the number of resources that are required for each event. The results are depicted in  Figures~\ref{fig:u_IN_a}~\&~\ref{fig:t_IN_a}.
In terms of utility,   \bsc and \hor perform almost the same 
in all cases.
Also, as expected, the utility of \rand and \stat increases with the number of required resources.
Further, the execution time decreased with number of resources, due to the decreasing number of feasible assignments.
Note that, we observe that varying the number of available resources $\theta$ does not affect the results.}

\subsubsection{\hor \& \horb Worst Case w.r.t.  $k$ and $|\T|$}
\hfill \break
Here, we consider the setting that corresponds to the worst case
\mbox{w.r.t.\ $k$ and $|\T|$} for the \hor and \horb algorithms (Sect.~\ref{sec:horAnal} \& \ref{sec:horbanal}).
Thus, for $k=100$, the worst case corresponds to $|\T|=99$. 
Fig.~\ref{fig:worstSpace}a presents the execution time for all datasets. 
We can observe that even in the worst case, \horb outperforms all methods 
in all cases (excluding the \stat). Also, in synthetic datasets, 
where the \bnd demonstrates poor performance, \hor is more efficient.

%

%
%

%

\subsubsection{Search Space}
\begin{figure}[t]
\vspace{-20pt}
 {\hspace*{-11pt}  \subfloat[\hor \& \horb worst case]{
 \includegraphics[trim={0mm 10mm 0mm 5mm},height=1.05in]{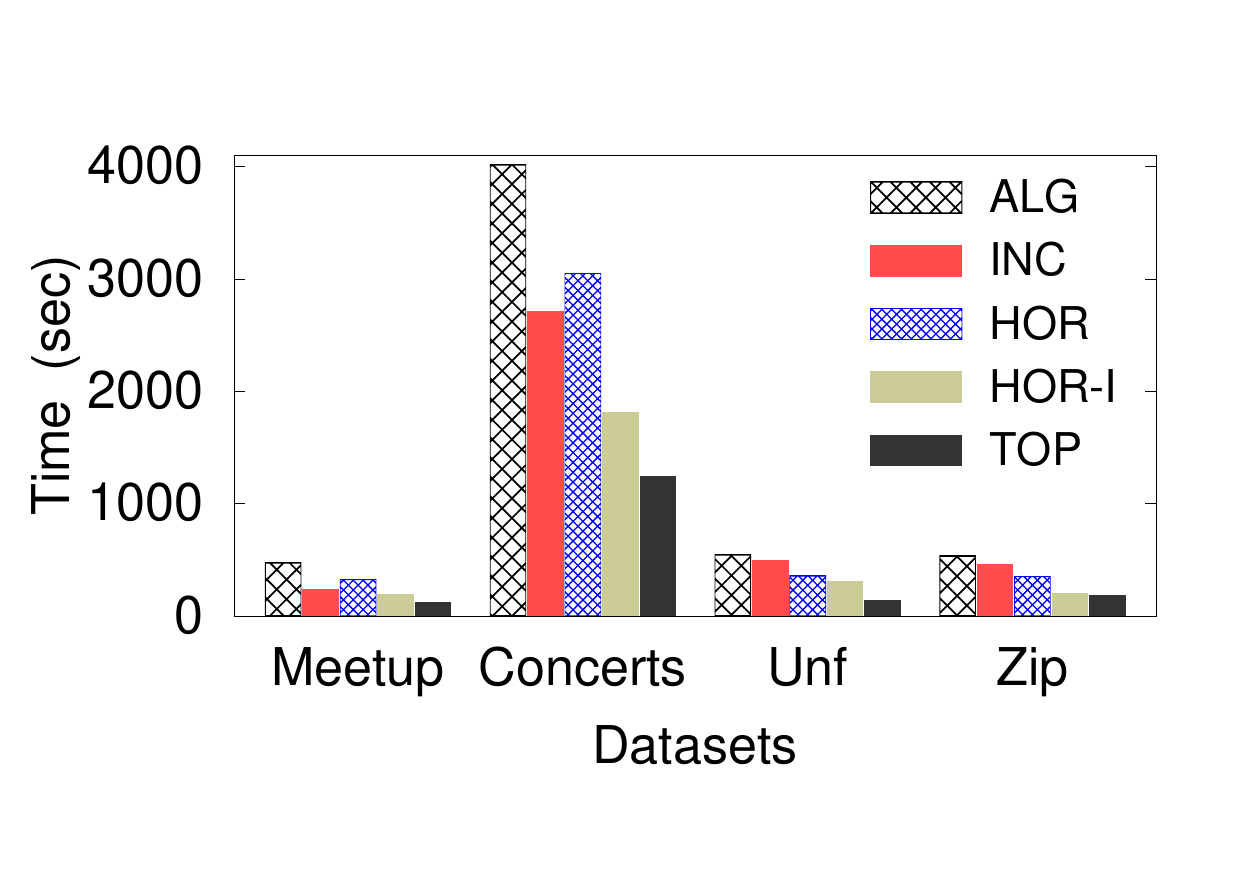}}}
 \hspace{-10pt}
\subfloat[\bsc \& \bnd search space]{
 \includegraphics[trim={0mm 10mm 0mm 5mm},height=1.034in]{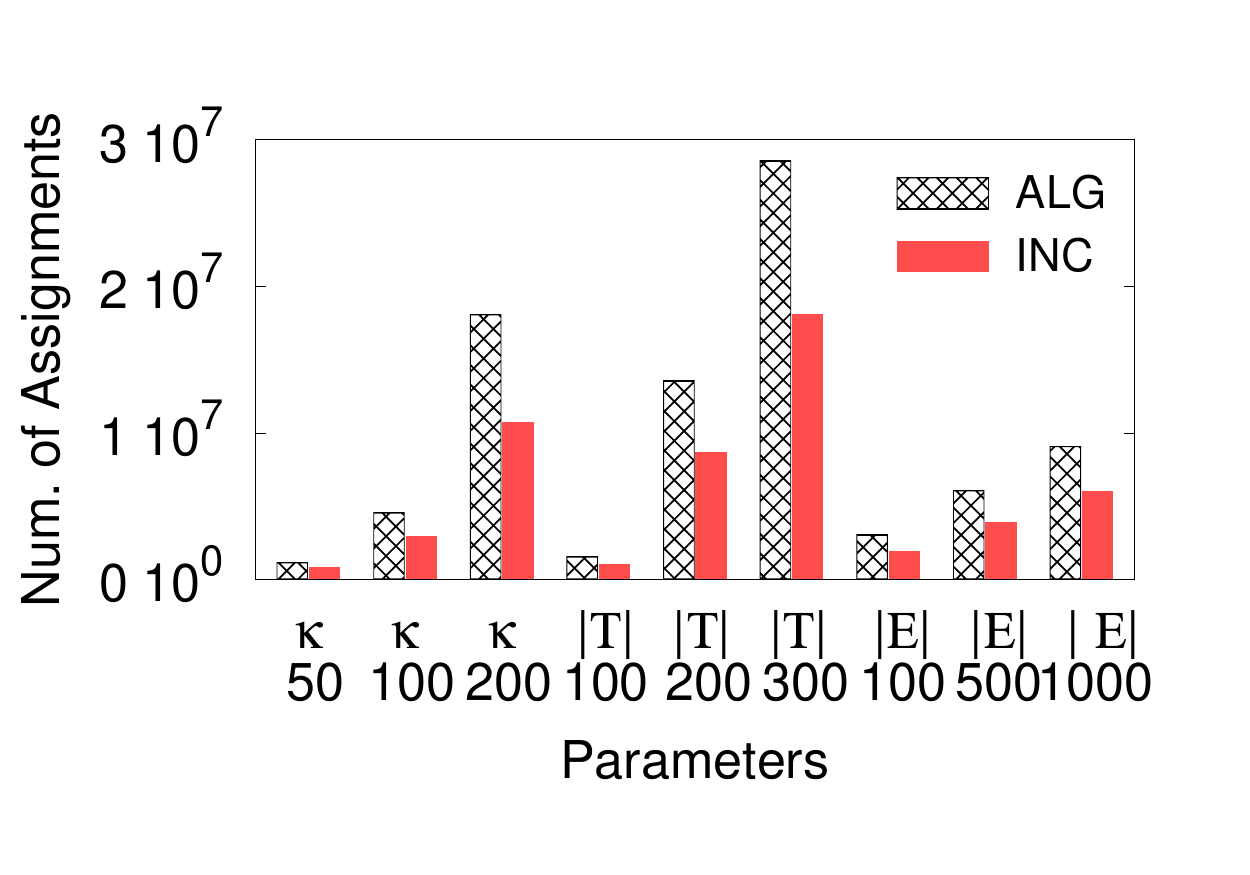}
 \hspace{-15pt}}  
 \vspace{-8pt}
\caption{\hor /-I worst case and \bsc/\bnd search space}
\label{fig:worstSpace}
\vspace{2pt}
\end{figure}

\hfill \break
%
In this experiment (Fig.~\ref{fig:worstSpace}b) we study 
the effectiveness of the proposed assignment organization (Sect.~\ref{sec:interv-org}).
We measure the number of assignments examined by the 
\bsc and our \bnd algorithm,  varying the main parameters ($k$,  $|\T|$, $|\E|$).
In all cases, \bnd accesses noticeable less assignments. 
Also, in each parameter, the differences  between \bnd and \bsc  increases in large 
parameter values.  
Overall,  in most cases, \bnd examines slightly more than  half assignments  that \bsc accesses.

\subsubsection{Summary}\label{ex:summary}
\hfill \break
In what follows, we  summarize our findings: 
$(1)$~\textit{{Between the  datasets used}}, with the exception of \uni, all the methods report similar results. 
In the \uni dataset, the bound-based methods (\bnd \& \horb) 
demonstrate lower performance than in other datasets. 

\hspace*{-7pt}
$(2)$~\textit{{Regarding utility score}}, in all cases, our \hor (and \horb) algorithm achieves almost the same utility score as   \bsc. 
Particularly, in more than $70 \%$ of the performed experiments (including the experiments omitted from the manuscript), \hor and \bsc report the same utility scores,  while in the rest of the cases, the difference in utility is on average $0.008\%$; 
with the largest difference  being $1.3\%$.
Recall that, the \bnd algorithm always return the same solution as \bsc.

\hspace*{-7pt}
$(3)$~\textit{{Comparing}} \bsc \cite{ses} \textit{{with our methods}}, in several cases:  
$(i)$ our \hor and \horb methods are around 5$\times$ faster and perform less than half of  the computations.
$(ii)$ \bnd  is more than 3$\times$ faster, performs less than half of  the computations.

\hspace*{-7pt}
$(4)$~\textit{{Comparing our methods}}:
$(i)$~\horb is always faster than the other methods.
In several cases, \horb is around  3  and 2 times faster than \bnd and \hor, respectively.
$(ii)$~\hor outperforms \bnd in terms of  time and computations, with some exceptions, in cases where
${k > |\T|}$. Overall, in several cases, \hor is around 3$\times$ faster than \bnd.

%
 
%

%


 \section{Related Work}
 \label{sec:rw}
 
 \vspace{-4pt}
 

 \stitle{Event Management \& Mining.}
 The \ses problem studied in this work was recently introduced in \cite{ses}, 
where a simple greedy algorithm was proposed.
Compared to \cite{ses}, here we show that \ses is hard to be approximated
over a factor and we design three efficient and scalable algorithms 
which perform on average half the computations compared the method presented in \cite{ses}  and, in most cases, are 3 to 5 times faster  (more details in Sect.~\ref{sec:intro}).

  Recently, a number of studies have been proposed in the context of 
event-participant planing. 
These works examine the problem of finding assignments between a set of users and a set of pre-scheduled events.
The determined assignments aim to maximize the satisfaction of the  users while satisfying several constraints.
Particularly, \cite{Li2014} assigns one event to each user, based on her interests and social relations. 
%
\cite{She2015a} finds an user-event arrangement by assigning users to events. 
%
The latter work is extended in \cite{She2016}, where the online setting of the problem is examined. 
%
A similar user-event arrangement problem is defined in a more advanced setting \cite{She2015}, 
where more factors are considered (e.g., complex spatio-temporal factors, travel cost).
This work is extended in \cite{ChengYCGW17}, in which 
participation lower bounds on event and  potential changes induced either by  event organizer  or by users (e.g., changes on event location) are considered.
%
In the same context, \cite{Tong2015} tries to maximize the satisfaction of the least satisfied 
user.
%
%
In an online scenario, \cite{She2017} 
exploits the user feedback (i.e., accept or reject the assigned events) 
in order to   adaptively  learn   user interests.
This work tries to maximize the number of accepted assigned events.
Compared to our work, as discussed in Sect.~\ref{sec:intro},  the objective, 
the solution and the setting of our problem substantially differ from the aforementioned approaches.

In a different context, 
\cite{FengCBM14,HHXTG18} attempt to find influential event organizers 
and promoters from online social networks.  
\cite{Romero2017} studies the influence of early respondents in online event scheduling process. 
Further, a number of works  \cite{XuZZXCL15,Zhang2015,DuYMWWG14,Boutsis15} 
analyze several factors form (Event-based) Social Networks data in order to study {user 
attendance} and provide event recommendations. 
Our work studies a different problem compared to the aforementioned approaches. 
However, some of the aforementioned methods can be {exploited in our problem to estimate the  user attendance probability.}

\stitle{Assignment \& Matching Problems.}
The problem studied shares common characteristics 
with the Generalized assignment (GAP) and Multiple knapsack (MKP) problems.
Particularly,  our problem is a 
generalized case of the GAP and MKP problems with identical bin capacities \cite{Martello1990}. 
A major difference of \ses compared to GAP and MKP is that in \ses the 
expected attendance (resp.\ profit) of  assigning an event  (resp.\ item) to an interval (resp.\ bin)
is determined based on the other events assigned to this interval.  
Also, beyond the event  and interval   entities
which are also considered in the aforementioned problems, in
our problem further core entities are involved (e.g., users,   
organizer, competing events).
Additionally, assignment/matching problems (similar to bipartite matching) have been   studied in 
spatial context \cite{TongSDCWX16,UYMM08,LongWYJ13,WongTFX07}. 
In general, the main differences of these works compared to \ses, 
are the same as the  ones that  hold in  GAP and MKP problems (see above).

 \stitle{Recommender Systems.} 
Numerous approaches have been  proposed in the context of location and event recommendations. 
Particularly, several works recommend  events to users \cite{Liu2012,Pham2015,MinkovCLTJ10,YQVHNH18,ZhangWF13}, 
while others offer location-based recommendations
 \cite{0003ZWM15,LiuTGQ2017,YinSCHC13,WangYSCXZ16,YinZCWZH16,LeeKBDCHTV13,LiuLAM13}.
Further, in a more general setting,   approaches have been proposed for recommending  locations or items to groups of people (i.e., {group recommendations}) \cite{YahiaRCDY09,NtoutsiSNK12,QuintarelliRT16,WangLF14,Qian2017,Qi2016,BikakisBS16}. 
Compared to our work, the aforementioned approaches study a different problem, 
that is, recommending objects (e.g., venues, events) to users.



\vspace{-6pt}
\section{Conclusions}
\label{sec:concl}

This paper studied the {\textit{Social Event Scheduling}} (\ses) problem, 
which assigns a set of events to time intervals, so that the number of attendees is maximized.
We showed that \ses is NP-hard to be approximated over a factor, and we proposed three efficient and scalable algorithms. 
The proposed algorithms are evaluated over several  real and synthetic datasets,
outperforming the existing solution three to five times in several of cases. 
 
\vspace{-5pt}

\alt{}
{As future work, there are several variations that can be consider for the \ses problem.
in which new methods have to be designed.
For example, in a profit-based version, each event has a cost and a fee. 
In this scenario, an event cannot be organized if its expected profit 
(which can be computed w.r.t.\ expected attendance and fee) is not over a percentage of its cost.
Further,  users may be associated with a maximum number of attendances 
and/or a budget for a given time period.
In this case, a reasonable approach would consider for each user 
only her top attendances within the number attendances and budget limits.
}
\vspace{-0pt}

%

\bibliographystyle{abbrv}

\normalsize
\alt{
}{
\appendix
\input{appendix}
 }

\end{document}